\documentclass[a4paper]{article}
\usepackage{amsmath}
\usepackage{graphicx}
\usepackage{geometry}
\usepackage{floatrow}
\usepackage{float}
\usepackage{layout}
\usepackage{amssymb} 
\usepackage{multirow}
\usepackage{caption}
\usepackage{subcaption} 
\usepackage{authblk}
\usepackage{hyperref}
\usepackage{indentfirst}
\usepackage{physics}
\usepackage[toc,page]{appendix}
\usepackage{tikz}
\usepackage{cite}
\usetikzlibrary{calc}
\usepackage{relsize}
\tikzset{fontscale/.style = {font=\relsize{#1}}}
\usetikzlibrary{arrows}
\usetikzlibrary{shapes,arrows,shapes.multipart}
\usepackage{circuitikz}

\begin{document}

\newcommand{\beq}{\begin{equation}}
\newcommand{\eeq}{\end{equation}}
\newcommand{\bea}{\begin{eqnarray}}
\newcommand{\eea}{\end{eqnarray}}

\title{\textbf{\huge{Cauchy Slice Holography:\\ A New AdS/CFT Dictionary}}}
\author{\large{Goncalo Araujo-Regado\footnote{ga365@cam.ac.uk} , Rifath Khan\footnote{rifathkhantheo@gmail.com} , Aron C. Wall\footnote{aroncwall@gmail.com}}}
\affil{DAMTP, University of Cambridge} 
\date{\today}

\maketitle

\maketitle

\begin{abstract}

We investigate a new approach to holography in asymptotically AdS spacetimes, in which time rather than space is the emergent dimension.  By making a sufficiently large $T^2$-deformation of a Euclidean CFT, we define a holographic theory that lives on Cauchy slices of the Lorentzian bulk.  (More generally, for an arbitrary Hamiltonian constraint equation that closes, we show how to obtain it by an irrelevant deformation from a CFT with suitable anomalies.)  The partition function of this theory defines a natural map between the bulk canonical quantum gravity theory Hilbert space, and the Hilbert space of the usual (undeformed) boundary CFT.  We argue for the equivalence of the ADM and CFT Hamiltonians.  We also explain how bulk unitarity emerges naturally, even though the boundary theory is not reflection-positive.  This allows us to reformulate the holographic principle in the language of Wheeler-DeWitt canonical quantum gravity.

Along the way, we outline a procedure for obtaining a bulk Hilbert space from the gravitational path integral with Dirichlet boundary conditions.  Following previous conjectures, we postulate that this finite-cutoff gravitational path integral agrees with the $T^2$-deformed theory living on an arbitrary boundary manifold---at least near the semiclassical regime.  However, the $T^2$-deformed theory may be easier to UV complete, in which case it would be natural to take it as the definition of nonperturbative quantum gravity.
\end{abstract}

\tableofcontents

\vspace{-5pt}
\section{Introduction}
\vspace{-5pt}

\subsection{The Unity of Physics}\label{unity}

When you start learning about physics, there seem to be a large number of subfields that all seem quite different from each other.  Later on, you begin to notice the same mathematics popping up in different places.  In particular, there is a very deep structural analogy in physics, which relates mechanics, thermodynamics, and field theory \cite{baez,McCoy:1994zi}.  One possible presentation of this analogy is displayed in the table below:\nopagebreak
\vspace{10pt}
\begin{center}
\begin{tabular}{c|c|c} 
\textbf{Mechanics} & \textbf{Thermodynamics} & \textbf{Field Theory} \\
\hline\hline &&\\
($\vec{x}$, $\vec{p}$) & e.g. ($V$, $P$) & ($J$, $\mathcal{O}$) \\ 
\textit{positions, momenta} & \textit{extensive, intensive}&\textit{sources, operators} \\ &&\\
on-shell action $I$ & free energy $F$ & effective action $S$
\\ &&\\
$dI = p_i\,dx^i$ & $dF = P\,dV +\ldots$ & $dS = \int {\cal O}\,dJ$
\\ &&\\
QM: state $\Psi(x)$ & Stat.$\,$Mech: $Z(V\ldots)$& QFT: $Z[J]$ \\ && 
%\\
%$=\exp (i\int^t_{t_0} {\cal L}/\hbar)$ & $= e^{-\beta F}$ & $= e^{iS/\hbar}\,(\text{or }e^{-S/\hbar})$
%\\ && 
\\
 space at time $t$ & space, in equilibrium & spacetime \\ 
\end{tabular}
\end{center}
\vspace{10pt}
In this analogy, the top row represents a $2N$ dimensional phase space on which there is a naturally defined symplectic structure $\omega$.  Such a phase space naturally arises as the cotangent space of some configuration space of variables, whenever you have a quantity which is extremized (second row).  The variation of this quantity gives a canonical 1-form (from which $\omega$ may be derived by taking the exterior derivative).  After ``quantization'' (or its analogue ``thermalization''), one ends up with a function on an $N$-dimensional configuration space on which $\omega$ vanishes, which includes just \emph{one half} of the phase space (although there are rules for transforming $\Psi$ or $Z$ if you wish to change your basis to a different $N$-dimensional configuration space).\footnote{In the case of statistical mechanics, the entity playing the role of the amplitude is the probability, which is constrained to be positive, unlike the other two columns where the amplitude may be complex.  This is the one truly distinctive feature of Statistical Field Theory, as compared to QFT, which is instead usually constrained to be unitary (or reflection-positive in Euclidean signature).  However, the $T^2$ deformed field  theories considered in this paper are not unitary!}  

The bottom row indicates the geometrical domain on which the symplectic structure is defined.  Strangely, this means that the relation between sources and operators in a $d$ dimensional \emph{spacetime} field theory, are analogous to concepts that involve \emph{space alone} in Thermodynamics or Mechanics.\footnote{It is of course also possible to treat a relativistic field theory as a mechanical system, by moving to a phase space over field variables, but this is not the same as the third column.}  Hence, if we take this analogy seriously as a full-fledged duality, it seems that there is a possibility of an emergent time dimension!

This emergent time dimension is somewhat uninteresting in the case of Thermodynamics, because in that case we normally assume the system is in thermal equilibrium, so all expectation values remain the same as time passes.  On the other hand, if we reinterpret the ($J$, $\mathcal{O}$)'s in a field theory as the $(\vec{x},\vec{p})$'s of a mechanical phase space, then it seems that there is the possibility of a duality which implies an emergent \emph{time dimension} with nontrivial dynamics.  

In the quantum form of this duality, the \emph{partition function} of a field theory in $d$ spatial dimensions would be dual to the \emph{wavefunction} of a QM system which is defined in $d+1$ dimensions.  Such a theory would therefore obey a generalized form of the holographic principle:
\bea\label{Psi=Z}
\Psi[J] = Z[J],
\eea
where on the left side of the equation $J$ plays the role of a configuration-space variable $x$.

The existence of an emergent time dimension is very reminiscent of the AdS/CFT correspondence, in which a $d$-dimensional CFT is dual to a $d+1$-dimension asymptotically AdS quantum gravity theory \cite{Maldacena:1997re, Witten:1998qj, Aharony:1999ti}.  In this paper, we will attempt to make this connection more precise.

As we shall see, in a class of field theories known as $T^2$-deformed theories, this emergent dynamics is naturally described by a \emph{Hamiltonian constraint} ${\cal H}(x)\Psi = 0$, satisfied at each point $x$.  Thus, the $d+1$ dimensional theory is a gravitational theory, even though $Z[J]$ was defined on a fixed background (the metric $g_{ab}$ being one of the field theory sources).  This local constraint equation is inherited from the conformal symmetry of the original, undeformed CFT.\footnote{More generally, in any field theory, the existence of a scalar constraint equation acting on $Z$ may be expected if there is a nontrivial action of a local renormalization group flow on the field theory sources, so long as a CFT lies anywhere in the RG space of the theory.  Assuming the number of local constraint equations cannot change discontinuously, this would imply that \emph{any} QFT which can be continuously deformed to a CFT is dual to a quantum gravitational theory, although not necessarily one with a good semiclassical regime.}  Hence, we are able to reformulate the holographic principle in the language of Wheeler-DeWitt canonical quantum gravity.

As usual in AdS/CFT, in order to have a good semiclassical regime, it is necessary to assume that the field theory has a large number $N$ of color degrees of freedom.  Furthermore, the theory must be in an `t Hooft-like regime \cite{tHooft:1973alw}, in which there exists a subset of ``single-particle'' operators which are weakly coupled (i.e. approximately generalized free fields). The analogy to a bulk phase space works best if we restrict our phase space to linear combinations of single-particle sources $J_\text{s.p.}$ and single-particle operators ${\cal O}_\text{s.p.}$.\footnote{We do not here use the more common term ``single-trace'' since our $T^2$ deformation is double-trace (at leading order) and as a result the single-particle states of the $T^2$ theory correspond to multi-trace operators in the original CFT.}  Higher-particle operators like $J_\text{s.p.}^n {\cal O}_\text{s.p.}^m$ are then analogous to nonlinear product operators of the form $x^n p^m$.  Such operators obviously have to exist in the theory if the phase space is to have an approximately classical limit.  For reasons described in \cite{Lee:2013dln,Lee:2012xba,Shyam:2016zuk}, writing the beta functions in terms of single-trace operators is equivalent to allowing quantum fluctuations over the bulk fields.\footnote{If one includes all operators on equal footing, there is still an analogy between renormalization and symplectic mechanics \cite{Dolan:1994eg}, but the Hamiltonian is linear in the momenta so there are no fluctuation terms.}  

The most important source in the field theory is the background metric $g_{ab}$, and its dual operator is the stress-tensor $T^{ab}$, which is a single-particle operator in the weak-coupling regime.  In the dual bulk interpretation, these correspond to the \emph{spatial} metric and its canonically conjugate ADM momentum $\Pi^{ab}$ respectively.

\subsection{Holography at Finite Cutoff}

In the context of AdS$_3$/CFT$_2$, it was observed that by deforming the CFT by an irrelevant operator called $T \Bar{T}$ which is quadratic in stress tensor, we get a dual gravitational theory satisfying Dirichlet boundary conditions for the metric, on a boundary $\partial \mathcal{M}$ which is now at finite distance from the center of AdS  \cite{McGough:2016lol}.  In the case relevant to the AdS bulk being pure gravity, Zamolodchikov showed that the $T \Bar{T}$ operator has a factorization property, which makes the spectrum of the theory on a cylinder $S_1 \times \mathbb{R}$ exactly solvable \cite{Zamolodchikov:2004ce, Smirnov:2016lqw}.\footnote{This exact solubility is presumably related to the fact that $D = 3$ pure gravity has no local degrees of freedom, since adding matter, e.g. a scalar field, to the $D = 3$ bulk requires additional double trace operators ($\mathcal{O}_\text{s.t.} \mathcal{O}_\text{s.t.}$) in the deformation, which do not have this factorization property.  In this case we must still resort to large $N$. Also the factorization property fails to hold for a curved background metric.}  

Later this deformation was generalized to $d > 2$ dimensions \cite{Taylor:2018xcy,Hartman:2018tkw, Shyam:2018sro, Caputa:2019pam}, in which case it is called a $T^2$ deformation.  In this case, the theory is exactly solvable at leading order in a large $N$ expansion.\footnote{Interestingly, the entanglement entropy in the $T^2$ deformed theory appears to be finite when calculated by the replica trick \cite{Donnelly:2018bef,Banerjee:2019ewu}, although the factorization of the theory across spatial boundaries is not well understood.}

\begin{figure}[h]
\centering
\begin{tikzpicture}[scale=1.5]
\draw (0,0) ellipse (1.25 and 0.3);
\draw (-1.25,0) -- (-1.25,-3.5);
\draw (1.25,-3.5) -- (1.25,0) node [midway] (N) {}; 
\draw (0,-3.5) ellipse (1.25 and 0.3);
\draw (0,0) ellipse (1.25*1.5 and 0.3*1.5);
\draw (-1.25*1.5,0) -- (-1.25*1.5,-3.5);
\draw (1.25*1.5,-3.5) -- (1.25*1.5,0) node [midway] (M) {}; 
\draw (0,-3.5) ellipse (1.25*1.5 and 0.3*1.5);
\fill [gray,opacity=0.2] (-1.25,0) -- (-1.25,-3.5) arc (180:360:1.25 and 0.3) -- (1.25,0) arc (0:180:1.25 and 0.3);
\draw (M) to  (N)[->,>=stealth'];
\node at (M) [label={[xshift=-4.5mm, yshift=-1mm]$\mu$}] {};
\node at (M) [label={[xshift=4mm, yshift=7mm]$\partial\mathcal{B}$}] {};
\node at (N) [label={[xshift=4.5
mm, yshift=7mm]$\partial\mathcal{B}^{(\mu)}$}] {};
\end{tikzpicture}
\caption{\small Illustration of the usual finite-cutoff AdS/CFT duality. The $T^2$ deformed theory lives on a finite radius brick wall, labelled by $\partial\mathcal{B}^{(\mu)}$.}
\label{usual}
\end{figure}
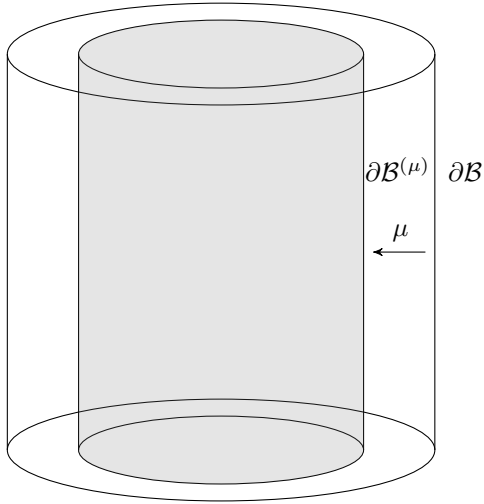

In the usual way of thinking about the $T^2$ theory, one starts with a Lorentzian boundary CFT.  Turning on a finite value of the deformation parameter $\mu$ makes it so the field theory lives on a cylinder at finite sized bulk radius.  On this cylinder, the metric satisfies Dirichlet boundary conditions, so it is a ``brick wall'' on which quantum fluctuations are suppressed, enclosing a finite volume interior (see Figure \ref{usual}).

AdS/CFT is usually thought of as involving an emergent \emph{space} dimension.\footnote{In the best understood versions of the AdS/CFT duality, there are also some number of emergent compact (but large) Kaluza-Klein dimensions that appear in the bulk theory, in order to bring the total number of bulk dimensions to the $D=10$ of string theory or $D = 11$ of M-theory.  In this article we ignore such extra dimensions, and treat the bulk theory as a KK reduced $d+1$ dimensional theory.  This means that, if we consider the case of a pure gravitational theory in AdS$_{d+1}$ (or a theory with a single scalar field as in section \ref{example}), we are considering a model that does not, strictly speaking, have a known CFT dual, although many properties of such field theories are known.  Although such CFTs probably do exist, most likely they require fine-tuning to get a small cosmological constant, especially in the absence of supersymmetry.  We expect that with matter fields included, the tower of KK fields appearing in realistic models can be consistently included using our general approach to the deformation given in section \ref{thegeneraldeformation}.  Alternatively, in regimes where it is self-consistent to restrict to the stress-energy sector of the CFT, it may be possible to apply these methods to the gravitational field only.} The usual approach here is to consider ``sideways'' evolution in the bulk and treat evolution with respect to $z$ as an analogue of the time direction \cite{Hamilton:2005ju,Hamilton:2006az}. Despite the fact that this corresponds to a nonstandard Cauchy problem, one can still make sense of the radial Wheeler-DeWitt evolution \cite{Freidel:2008sh,Hartman:2018tkw,Belin:2020oib}. But how might this relate to an emergent \emph{real} time dimension?

\subsection{Cauchy Slice Holography}

In this paper we will make a different choice and adopt a point of view where the signature of the $d$-dimensional field theory remains Euclidean throughout.  Our perspective is similar to that of Caputa, Kruthoff, and Parrikar \cite{Caputa:2020fbc}; however these authors end their deformation on the $t = 0$ symmetric slice,\footnote{Although these authors did suggest extending the flow further, into Lorentzian signature, in their discussion section.  See also \cite{Kruthoff:2020hsi} where a spatial tensor network is introduced in a somewhat different way to describe the $T^2$ deformation itself.} while we wish to continue it all the way to generic Lorentzian Cauchy slices.

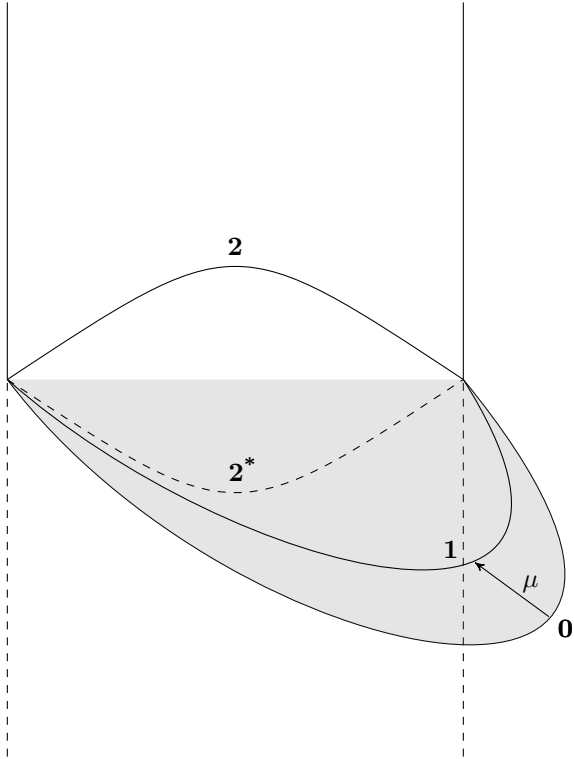
\begin{figure}[h]
\centering
\begin{tikzpicture} 
\def\a{3};
\def\b{5};
\draw (-\a,0) -- (-\a,\b);
\draw[dashed] (-\a,-\b) -- (-\a,0);
\draw (\a,0) -- (\a,\b);
\draw[dashed] (\a,-\b) -- (\a,0);
\draw (-\a,0) ..controls (0,-4) and (11/1.5,-8/1.5).. (\a,0) node[pos=0.69,inner sep=0pt,outer sep=0pt](M){};
\fill [gray,opacity=0.2] (-\a,0) ..controls (0,-4) and (11/1.5,-8/1.5).. (\a,0);
\draw (-\a,0) ..controls (0,2).. (\a,0) node[pos=0.5,inner sep=0pt,outer sep=0pt](p){};
\node at (p) [label={[xshift=0mm, yshift=-1mm]$\textbf{2}$}] {};
\draw[dashed] (-\a,0) ..controls (0,-2).. (\a,0) node[pos=0.5,inner sep=0pt,outer sep=0pt](q){};
\node at (q) [label={[xshift=1mm, yshift=-1mm]$\textbf{2}^\textbf{*}$}] {};
\draw (-\a,0) ..controls (0/1.5,-4/1.5) and (11/2,-8/2).. (\a,0)node[pos=0.65,inner sep=0pt,outer sep=0pt](N){};
\draw (M) to  (N)[->,>=stealth'];
\node at (N) [label={[xshift=7.5mm, yshift=-6.5mm]$\mu$}] {};
\node at (M) [label={[xshift=2mm, yshift=-5mm]$\textbf{0}$}] {};
\node at (N) [label={[xshift=-3mm, yshift=-2mm]$\textbf{1}$}] {};
\node at (M) [label={[xshift=2mm, yshift=-5mm]$\textbf{0}$}] {};
\node at (-2.2*\a,0) {};
\end{tikzpicture}
\caption{\small The evolution of the embedding of the slice $\Sigma$, with a given fixed metric, under the deformation parametrised by $\mu$, in a saddle-point approximation. The illustration shows a (complexified) $\text{AdS}_2$ cross-section of the $\text{AdS}_{d+1}$ vacuum geometry.  The Euclidean CFT boundary is labelled by \textbf{0}, in a conformal frame where its geometry is hyperbolic. Three different stages (\textbf{0}, \textbf{1} and \textbf{2}/$\textbf{2}^\textbf{*}$) of the deformation are shown. Stage $\textbf{0}$ corresponds to no deformation of the CFT. As we perform the $T^2$ deformation the slice moves inwards along the imaginary time direction, while still being embedded in a Euclidean bulk space (stage \textbf{1}). For a sufficiently large deformation there is a phase transition to a Lorentzian saddle-point bulk geometry. By time-reversal symmetry, the slice $\Sigma$ can be embedded in two different ways in the same bulk geometry (with the same intrinsic metric). This corresponds to stages \textbf{2} and $\textbf{2}^\textbf{*}$.  From the perspective of the field theory, this phase spontaneously breaks time-reversal (and therefore $CPT$).}
\label{phase_transition}
\end{figure}

Starting with a \emph{Euclidean} holographic CFT (living at the infinite Euclidean boundary), and then doing a $T^2$ deformation, we can move the slice $\Sigma$ inward along the imaginary time direction.\footnote{Since the $T^2$ deformation is irrelevant, the result depends on the choice of conformal frame for the original CFT metric.}  After a large enough deformation, $\Sigma$ may end up embedded in a Lorentzian signature spacetime (depending on the choice of metric $g_{ab}$ on $\Sigma$).  So in the end the dual holographic theory lives not at spatial infinity, but rather it becomes a field theory living on  a \emph{Cauchy slice} of a Lorentzian AdS spacetime.\footnote{In asymptotically AdS spacetimes, $\Sigma$ is a Cauchy slice iff every complete timelike curve \emph{that does not go to the timelike boundary} passes through $\Sigma$ exactly once.} (Alternatively, we can hold the deformation parameter $\mu$ fixed, in which case the transition to Lorentzian signature occurs when the spatial metric $g_{ab}$ is chosen suitably.)  See Figure \ref{phase_transition}.\footnote{A different way to try to extend to Lorentzian signature metrics, involving constant mean curvature slices, was sketched by \cite{Belin:2020oib}.  However if our proposal is correct this transition already happens in the $g_{ab}$ basis.}

As we shall see, the $T^2$ deformed partition function can be used to define a quantum gravity state on $\Sigma$. Since the Cauchy slice $\Sigma$ has a boundary at infinity, its partition function $Z[\Sigma]$ also defines a state in the usual holographic Lorentzian CFT (the one that lives at timelike spatial infinity). This gives us an (in principle, complete) map between the geometrodynamical degrees of freedom in the bulk, and the CFT degrees of freedom on the boundary. 

In other words, this form of the duality can be thought of a specification of the AdS/CFT dictionary involving arbitrary canonical gravity states in the bulk, including degrees of freedom that lie arbitrarily deep in the bulk.  This includes behind horizons, inside bags of gold \cite{Marolf:2008tx}, etc.  Obviously, this is useful information for purposes of studying the black hole information puzzle \cite{Hawking:1975vcx, Hawking:1976ra, Penington:2019kki, Almheiri:2020cfm, Mathur:2009hf, Raju:2020smc}, which we will discuss further in section \ref{unitarity}.  This picture makes it quite manifest how all degrees of freedom in the bulk must ``flow" outwards in a spacelike direction, so as to be encoded in the boundary.

These time slices $\Sigma$ satisfy Dirichlet boundary conditions, not because there is a physical end-of-time brane on them, but merely to represent the quantum gravity wavefunction in a ``position basis'' in which the metric on $\Sigma$ has a definite value $g_1$.  By picking two such definite metrics, we will show that one can calculate the gravitational inner product $\langle g_2 | g_1 \rangle$ either by means of a bulk gravitational path integral, or by a field theory calculation as a partition function.

\subsection{WDW States}

For some researchers, one of the attractions of defining quantum gravity by means of a holographic duality, is that they don't have to get their hands dirty playing around with quantizing the metric.  Even after much work, geometrodynamics has had limited success resolving fundamental questions such as the information paradox, or what happens at the Planck scale.

That being said, there is no way to express the \emph{equivalence} between geometrodynamics and holography, without defining both sides of the equation. Thus, in this article, we cannot avoid quantizing gravity directly.  In this paper we will mention both canonical and path integral techniques, pointing out potential difficulties and pitfalls as they arise.

In a canonical theory of Quantum Gravity \cite{DeWitt:1967yk}, the state is described by a superposition of metrics, $\Psi[g]$, which satisfies the Hamiltonian and momentum constraints.  In the case of pure GR these are given by:
\begin{equation}\label{ham}
    \mathcal{H}(x)\Psi[g]:=\Bigg\{\frac{16\pi G_N}{\sqrt{g}}:\!\Big(\Pi_{ab}\Pi^{ab}-\frac{1}{d-1}\Pi^2\Big)\!:-\frac{\sqrt{g}}{16\pi G_N}(R-2\Lambda)\Bigg\}\Psi[g]=0,
\end{equation}
\begin{equation}\label{mom}
    \mathcal{D}^a(x)\Psi[g]:=-2(\nabla_b\Pi^{ab})\Psi[g]=0,
\end{equation}
where $g_{ab}(x)$ is the metric on a $d$-dimensional Cauchy slice $\Sigma$, $\Pi^{ab}(x)$ is the canonical momentum conjugate to it:
\begin{equation}\label{Pi}
    \Pi^{ab}(x)=-i \frac{\delta}{\delta g_{ab}(x)},
\end{equation}
and $:\:\,:$ is the normal ordering symbol, meaning that we need to subtract off some divergences{\bf---}we will discuss this in more detail in section \ref{largeN}.  We will henceforth refer to solutions of the above constraints as WDW-states.  From \eqref{Pi} we can deduce the commutator:
\bea\label{com}
[\Pi^{ab}(x),\,g_{cd}(y)]= -i\delta^{ab}_{cd}\,\delta(x-y),
\eea
where $\delta^{ab}_{cd} = \tfrac{1}{2} (\delta^{a}_{c}\delta^{b}_{d} + \delta^{a}_{d}\delta^{b}_{c})$.

Since both WDW-states $\Psi[g]$ and partition functions $Z[g]$ are functionals of a metric, there is a natural analogy between the two concepts, as explained in section \ref{unity}.  We will therefore set about to define a \emph{field theory} whose partition function $Z$ naturally defines a WDW-state:
\bea
Z[g] = \Psi[g].
\eea
In this field theory, $\Pi^{ab}$ plays the role of the stress-tensor operator, while the metric $g_{ab}$ plays the role of the conjugate source.

\subsection{Relation to Tensor Network Models}\label{TN}

Like \cite{Caputa:2020fbc}, this project was conceptually inspired by the goal of coming up with a tensor network model for AdS/CFT states.  A number of holography researchers have constructed various kinds of tensor network models, whose geometry is thought of as being a discrete analogue of hyperbolic space; see e.g. \cite{Swingle:2009bg,Swingle:2012wq,Almheiri:2014lwa,Pastawski:2015qua,Miyaji:2015yva,Hayden:2016cfa,Cotler:2017erl,Kohler:2018kqk,Bao:2018pvs,Caputa:2017yrh,Jafari:2019qns}.  A significant motivator of this line of research has been to construct a model which naturally explains the Ryu-Takayagi formula \cite{Ryu:2006bv} in the bulk: 
\bea
S_\text{CFT}[R] = \min_\gamma \text{Area}[\gamma]/4G_N,
\eea
where $\gamma$ is a codimension-2 bulk surface anchored to the boundary of the CFT region $R$.
Since this formula is only valid on Cauchy slices $\Sigma$ with a time-reversal symmetry (e.g. the $t = 0$ slice of AdS), much of the work on tensor networks is intended as a model only of such slices.

Unfortunately, building a tensor network on a time-reversal symmetric slice $\Sigma_0$ is conceptually untenable as an accurate model of quantum gravity, since it violates the Heisenberg uncertainty principle between $g_{ab}$ and $\Pi^{ab}$.  The characteristic feature of a moment-of-time symmetric slice is that the extrinsic curvature $K_{ab}$ vanishes, and hence $\Pi_{ab} = \sqrt{g}(K_{ab} - g_{ab} K)/8\pi G_N = 0$.  But if we wish for generic surfaces $\gamma \in \Sigma_0$ to have well-defined areas, we also need to have a well-defined metric $g_{ab}$ on $\Sigma_0$.  By the commutator \eqref{com}, this is too much information!

It follows that, if we wish to take tensor networks seriously as a model of quantum space, we necessarily have to be able to use them to describe a generic dynamical Cauchy slice $\Sigma$.  Even if you start with an initial state which is tightly peaked around $K_{ab} = 0$, performing a quantum measurement can take you to another state with large $K_{ab}$.  Even classically, the $K_{ab} = 0$ slice must still satisfy the Hamiltonian constraint.  So even to describe a moment-of-time symmetric slice, we need a formalism in which dynamics is already implicitly present.\footnote{This fact may explain the paradoxical no-go result of \cite{Bao:2018pvs} in which it was impossible to construct a certain tensor network with crossing RT surfaces, using approximate isometries.}

In this article we will show how the $T^2$ deformation provides such a model of holographic space, in which dynamics arises naturally.  Unlike tensor networks, this is a \emph{continuum} model.\footnote{Another example of a continuum approach to holography on a spatial slice is cMERA holography \cite{Nozaki:2012zj,Mollabashi:2013lya,Miyaji:2014mca}.  However, this approach is not rotationally-invariant, and is nonlocal at sub-AdS scales.}  This is technically convenient as it allows us to define a local stress-tensor, and to explore sub-AdS scales.  However, for reasons we shall discuss in sections \ref{largeN} and \ref{UV}, the $T^2$ deformed theory still requires a UV completion in order to describe physics at finite $N$.  (Otherwise we would have just solved an even more famous problem: to provide a nonperturbative definition of quantum gravity!)  It is therefore still possible that space may still be discrete at the Planck scale, in which case the fundamental model of space might be literally a tensor network.\footnote{There is a dynamical version of the holographic entropy, given by \cite{Hubeny:2007xt}: 
$S_\text{HRT}[R] = \min \mathrm{ext}_\gamma \text{Area}[\gamma]/4G_N$.  In the past it has proven difficult to accommodate this formula in a tensor network model for a Cauchy slice $\Sigma'$ that does not pass through the HRT surface, as the minimal area of $\Sigma'$ is always less than the area of the HRT surface \cite{Wall:2012uf}, which gives the boundary CFT entropy.  The resolution of this paradox in the $T^2$ deformed model will be described in more detail in a forthcoming article \cite{Wall_CEB}, but an important clue will be provided in section \ref{exotic} where it is observed that the entropy of $T^2$ in Lorentzian signature sometimes exceeds $\text{Area}/4G_N$ due to the presence of extra states in the spectrum.  In order to be compatible with the HRT formula, a valid dynamical tensor network model must have bond dimensions large enough to include these extra states.  This happens whenever the codimension-2 extrinsic curvature is timelike.}

Since a field theory partition function gives a sum over states whenever it is cut open in the middle, we expect that Cauchy slice holography will implement some form of the surface-state correspondence in which information is attributed to codimension 2 surfaces \cite{Miyaji:2015yva,Sanches:2016sxy} (see also \cite{tHooft:1993dmi,Susskind:1994vu,Crane:1995qj} for precursors of this holographic idea). But in this article we focus on the use of the $T^2$ theory to construct a new form of the holographic AdS/CFT dictionary, which maps bulk WDW states to boundary CFT states.

\subsection{Plan of Paper}

In Section \ref{Hdual}, we study how the constraints of bulk gravity translate into constraints on the holographic field theory meant to describe bulk states. We provide a general prescription for obtaining the correct deformation away from the boundary CFT\footnote{We will generically refer to this deformation as the $T^2$-deformation, as a reference to the usual deformation corresponding to pure Einstein gravity in the bulk.}, which also provides us with a constraint on the CFT anomaly, in the cases in which such an anomaly exists. We also work through the explicit example of a bulk scalar field coupled to gravity in $3+1$-dimensions with a negative cosmological constant.  

In Section \ref{sectionGHP} we propose a generalized version of the holographic principle we are used to from AdS/CFT, taking into account the presence of boundaries at finite distances and times.  

Section \ref{QG Inner Product} is dedicated to a discussion of the bulk gravitational path integral as defining a bulk inner product.  To the extent that this bulk prescription is ambiguous, we propose that the $T^2$ theory could in principle resolve such ambiguities.  In Section \ref{HS} we present a construction of the Hilbert space of solutions to the constraint equations in quantum gravity, based on viewing the path integral method as a linear map from kinematic states to WDW-states.  

In Section \ref{dictionary}, we use the $T^2$ partition function to construct linear maps between the bulk and boundary Hilbert spaces.  That is, given a bulk kinematic Cauchy slice metric $g_{ab}$, we can define the dual CFT state.  Conversely, we can also define a map from the boundary states to solutions of the constraint equations.  (These two maps are not inverses; rather the composition of these two maps{\bf---}assuming the validity of the generalized holographic principle{\bf---}is equivalent to the bulk-to-bulk path integral map defined in Section \ref{HS}.)

%If the latter happens to peak strongly around a particular metric, this would result in the usual notion of the bulk ``dual" geometry. 

In Section \ref{AD} we show the consistency of boundary and bulk time evolution, given the generalized holographic principle.  We also show an equality between the boundary and bulk Hamiltonians, to leading order in large N.  We explain why our dynamics should be compatible with standard approaches to bulk reconstruction.  Finally we discuss the way in which bulk unitarity emerges naturally from the field theory, despite the fact that the $T^2$ spectrum is nonunitary.

Finally, in the discussion \ref{dis} we review our results, and discuss the prospects of UV completing the $T^2$ boundary theory into an ultraviolet finite theory{\bf---}a task which, if completed, might then serve as a satisfactory definition of quantum gravity.  We suggest that it does not actually matter much \emph{how} the theory is UV completed, as different choices are likely to provide dual descriptions of the same physics.\footnote{For readers who wish to get to the main AdS/CFT results as quickly as possible, we suggest the following reading plan: start with \ref{prelim}--\ref{asym} and \ref{form} for an overview of $T^2$ holography, followed by \ref{Lorentzian}, \ref{dirty} to get the gist of our approach to quantum gravity in the bulk.  At this point the key AdS/CFT dictionary results described in sections \ref{dictionary}---\ref{dis} should be accessible.}

\section{The $T^2$ Deformation}\label{Hdual}

\subsection{Preliminaries}\label{prelim}

Consider a holographic CFT which is dual to quantum gravity in an asymptotically-AdS bulk ${\cal B}$.  We assume the low energy bulk corresponds to pure gravity with matter fields, collectively denoted by $\{\Phi\}$, or in other words that the only low-dimension single-trace primary operators in the CFT are the stress tensor $T_{ab}$ and the set $\{\mathcal{O}\}$ which are in correspondence with the bulk matter fields.  The leading order Fefferman-Graham expansion of the metric near the boundary is:
\begin{equation}\label{FG0}
ds^2 = g_{\mu\nu}\,dx^\mu dx^\nu = \frac{g^{\text{bdy}}_{ab} dx^a dx^b + dz^2}{z^2} + ...,
\end{equation}
where $\mu,\nu$ are indices of the $D = d+1$ dimensional AdS bulk spacetime $\cal B$.  A Cauchy slice in this bulk is a noncompact manifold $\Sigma$ with $\partial \Sigma \subset {\partial {\cal B}}$.

We will now evaluate the CFT partition function, not on the usual boundary at infinity $\partial \cal B$, but rather on the Cauchy slice $\Sigma$.  After we deform this CFT partition function by a $T^2$ deformation, it will describe quantum gravity in the bulk. We will choose a convention in which the coupling constant $\mu$ (corresponding to the direction in theory space of this $T^2$ operator) is the only scale in the theory. This is particularly nice when thinking in terms of a Wilsonian picture of RG. Also, any bulk gravity scales will be written in terms of this $\mu$ and a set of dimensionless field theory parameters, collectively denoted by $\{\alpha_i\}$, to fix the ratio between bulk scales.

%We start with a reminder of the properties of the CFT before doing the deformation.  
We pick a \emph{complete commuting set of operators} $\{\chi\}$ acting on the CFT Hilbert space of $\partial\Sigma |_{\partial{\cal B}}$, i.e. on the boundary side of $\partial \Sigma$. Importantly, this is \emph{not} necessarily the same as evaluating the same operators $\{\chi\}$ at $\partial \Sigma|_\Sigma$, because the sharp corner between the two geometries can cause certain operators, such as the stress-tensor, to discontinuously jump across the boundary (as will be discussed explicitly in section \ref{admH=cftH}).  In order to keep our notation clean, without writing junction conditions everywhere, we choose to always think of the Cauchy slice $\Sigma$ as making this transition to $\partial{\cal B}$ \emph{before} we evaluate $\{\chi\}${\bf---}even in situations where two Cauchy slices $\Sigma$ and $\Sigma^{'}$ meet at the same boundary time on $\partial {\cal B}$, with no lapse between them,\footnote{In other words we are implicitly identifying the Hilbert space of $\partial \Sigma$ with that of $\partial {\cal B}|_{\partial \Sigma}$.  Otherwise we would need to explicitly include junction maps ${\cal H}_{\partial {\cal B}|{\partial \Sigma}} \to {\cal H}_{\partial \Sigma}$ everywhere.} as will happen later.  

This defines a basis of states labelled by the set of eigenvalues of $\{\chi\}$\footnote{We will be using the same symbol $\{\chi\}$ to denote both the complete set of operators and the set of eigenvalues used as boundary conditions for the partition function. The context should make this distinction clear.}, while a general state on $\partial\Sigma$ will be a superposition thereof: $\psi[\{\chi\}]$.\footnote{These operators are \emph{not} necessarily local fundamental fields, since those are not usually good operators on a Hilbert space in a strongly coupled theory. The set $\{\chi\}$ can include eigenvalues of nonlocal operators e.g. the total energy or angular momentum.}  %The only local fields we demand our theory to possess are $T^{ab}(x)$ and $\{\mathcal{O}\}$,\textcolor{red}{What do we mean by ${\cal O}$ in this footnote?} so integrals of the form $\int_{\partial\Sigma}T^{ab}$ and $\int_{\partial\Sigma}\mathcal{O}$ are allowed as valid members of $\{\chi\}$.}
The partition function $Z_\text{CFT}[\{J_i\},\{\chi\}]$ is thus a functional of any field theory sources $\{J_i\}$ and of the boundary conditions $\{\chi\}$. The sources include the metric and any other sources which are mapped to bulk matter fields in the low-energy limit. So $\{J_i\}=(g,\{\phi\})$, where $\phi$ is the field theory source which is just a rescaled version of the bulk field $\Phi$ (which is always to be understood as the value of the bulk field induced on the Cauchy slice $\Sigma$). The correct rescaling depends on the nature of the bulk field.

The stress-tensor $T^{ab}$ in any field theory is defined as a variation of the metric:
\begin{equation} 
    T^{ab}(x):=\frac{-2}{\sqrt{g}}\frac{\delta}{\delta g_{ab}(x)} = \frac{-2i}{\sqrt{g}}\Pi^{ab}(x).
\end{equation}
If other sources, collectively denoted by $\{\phi\}$, are switched on in the field theory the corresponding operators, collectively denoted by $\{\mathcal{O}\}$, will be given by linear variation:
\bea
\mathcal{O}(x)=\frac{1}{\sqrt{g}}\frac{\delta}{\delta \phi}=\frac{i}{\sqrt{g}}\Pi_\phi(x).
\eea
Given any set of CFT sources $\{J_i\}$, the generator of diffeomorphism symmetry is
\bea\label{D}
{\cal D}_a := \sum_J \frac{\delta}{\delta \xi^a}\!\Big[{\cal L}_\xi(J) {\cal O}^i \Big] = \sum_J \Big(\partial_a J\cdot\Pi_J \:
+ \!\!\!\!\!\!\!\!\!\!\!\!\!\!\!
\sum_\text{~~~~~~~upper $J$ index}
\!\!\!\!\!\!\!\!\!\!\!\!\!
\partial_b (J^b \cdot \Pi_{Ja})\:
- \!\!\!\!\!\!\!\!\!\!\!\!\!\!
\sum_\text{~~~~~~~lower $J$ index}
\!\!\!\!\!\!\!\!\!\!\!\!\!
\partial_b (J_a \cdot \Pi_J{^b})
\Big),
%= -2\nabla_b \Pi^b_a + \partial_a \phi\,\Pi_\phi,
\eea
where ${\cal L}_\xi(J)$ is the Lie derivative of the tensor $J$, while the generator of Weyl transformations is
\bea\label{W}
{\cal W}:= -\sum_J \Delta_{J}J \cdot \Pi_J, %= 2\Pi-\Delta_\phi \phi \Pi_\phi,
\eea
where ${\cal O}^i$ is the operator conjugate to $J_i$ and $\Delta_{J_i}$ is the conformal weight of the source in the boundary CFT.\footnote{In this expression, we are assuming the sources transform as primaries under a Weyl transformation.} 
A generic CFT is Weyl invariant up to the usual Weyl anomaly $\cal A$:
\begin{equation} \label{euc}
    \big(\mathcal{W}(x)-i\mathcal{A}(x)\big) \, Z^{(\epsilon)}_{CFT}[\{J_i\},\{\chi\}] = 0,
\end{equation}
where $\epsilon$ is a regulator necessary to regulate the logarithmic divergences (which cannot be eliminated by counterterms) arising from the anomaly.
However, this expression does not look like the ADM Hamiltonian constraint, which we need in order to interpret $Z$ as a Wheeler-DeWitt state on $\Sigma$.  Relatedly, the closure relation of the conformal constraint $\widetilde{\cal W} := \mathcal{W}-i\mathcal{A}$ is trivial:
\bea\label{Wclose}
\left[ \widetilde{\mathcal{W}}(x), \widetilde{\mathcal{W}}(y)\right]=0,\label{Wclosure}
\eea
due to the Wess-Zumino consistency condition, whereas the commutator of the Hamiltonian constraint on itself is proportional to the momentum constraint:
\bea\label{Hclose}
\Big[ \mathcal{H}(x), \mathcal{H}(y)\Big]=i\left(\mathcal{D}^a(x)\partial_a^{(x)}-\mathcal{D}^a(y)\partial_a^{(y)}\right)\delta(x-y).
\eea
This encodes the principle of local Lorentz invariance in the bulk, since it implies that when there is a nonzero $\cal H$ gradient (locally a boost) the generator of time translations is shifted by the generator of space translations.\footnote{The action of the Lorentz boost on spatial translations is fixed by the analogous commutator: $[\mathcal{H}(x), {\cal D}_a(y)] = i\partial_a^{(x)}\!({\cal H}(x) \delta(x-y))$.  But this closure relation is an automatic consequence of diffeomorphism symmetry, given that $\mathcal{H}(x)$ is a scalar density on $\Sigma$.}

Hence, to obtain the ADM constraint algebra, we need to deform the theory on the Cauchy slice so that $\widetilde {\cal W}(x)$ becomes ${\cal H}(x)$ and \eqref{Wclose} deforms to \eqref{Hclose}.  (The formula for ${\cal D}_a$ is unaffected so long as the flow is covariant.)  In other words, local Lorentz symmetry in the bulk emerges from local conformal invariance in the boundary, as a deformed WZ consistency condition \cite{Shyam:2017qlr,Shyam:2016zuk}:
\bea\label{LLI}
\text{Conformal Invariance}
 \:\:\underset{\mu}{\Longrightarrow} \:\:\text{Local Lorentz Invariance}.
\eea

\subsection{Asymptotics of the Flow}\label{asym}

Freidel \cite{Freidel:2008sh} showed that any radial WDW state of an asymptotically-AdS quantum gravity theory tends to the following form in the superspace region corresponding to an infinite uniform rescaling of the boundary metric $\tilde{g} = g/\mu^{2/d}$:
\bea\label{asymptotics}
\lim_{\mu\to 0}\Psi\left[\tilde{g}\right] = A_{+} e^{i \text{S}[\tilde{g}] } Z_{+}[\tilde{g}] + A_{-} e^{-i \text{S}[\tilde{g}] } Z_{-}[\tilde{g}],
\eea
where $g$ is the Lorentzian metric on the Dirichlet wall $\partial {\cal B}$, $Z_{\pm}$ look like CFTs\footnote{At least to the extent that they are conformally invariant functionals of the metric, up to the anomaly term.  Holography asserts that, in a UV complete theory of quantum gravity, they should also satisfy the axioms of local QFT.} with opposite anomalies, and $\text{S}[g]$ is an explicit real local action consisting of only relevant terms (\emph{i.e.} terms of dimension at most $d$) which we note to be the holographic counterterms. Furthermore, this local action is the same for all WDW states.  In the $\mu \to 0$ limit, the Dirichlet boundary $\partial {\cal B}$ goes to spatial infinity, due to imposing Dirichlet boundary conditions with infinite volume. %So we colloquially say that this limit is equivalent to taking the boundary $\partial {\cal B}$ ``to infinity". 

The presence of two solutions is a consequence of the radial WDW equation ${\cal H}\Psi = 0$ being of second order.\footnote{Since ${\cal H}(x) = 0$ is a second order differential equation defined at each point $x$, one might think there would be more than 2 solutions if we choose the sign differently for different values of $x$.  However, in the asymptotic limit, such wavefunctions would always violate the diffeomorphism constraint ${\cal D}^a \Psi = 0$ and so are unphysical.  (Here we are assuming that $\partial \cal B$ is connected, otherwise we could pick a separate sign for each connected component.)} Any of the two is independently a solution and so is any linear combination.

If we Wick rotate $Z_+$ and $Z_-$ to Euclidean signature, one of these partition functions (say $Z_-$) corresponds to a state $\Psi$ in which the graviton fluctuations are unnormalizable.  As we wish to define our Lorentzian Cauchy slice theory by analytic continuation from the Euclidean regime, we are therefore forced to restrict our attention to the other one ($Z_+$).\footnote{If we want to do a holographic description of cosmology with $\Lambda>0$, it turns out that it is crucial to consider both branches, as explained in \cite{Araujo-Regado:2022jpj}.}  We believe this choice is closely related to the question of the quantum gravity contour, which we will discuss in section \ref{contour}.

The same behaviour also follows if we add matter fields or higher curvature corrections, since sending the volume of $\partial {\cal B}$ to infinity guarantees that we are in the IR limit of the bulk theory (at least in the neighbourhood the boundary) and so the higher-order corrections to the WDW equation become suppressed in an $M_\text{Pl}$ expansion.  After picking the branch with good Euclidean continuation, we have:
\bea
\lim_{\mu\to 0}\Psi[\tilde{g},\tilde{\Phi}] = e^{i \text{S}[\tilde{g},\tilde{\Phi}] } Z_{+}[\tilde{g},\tilde{\Phi}], %+e^{-i \text{S}[\tilde{g},\tidle{\Phi}] } Z_{-}[\tilde{g},\tidle{\Phi}], 
\label{friedelduo}
\eea
where $\tilde{\Phi} = \mu^{\Delta_\phi/d} \Phi$ is an appropriately rescaled version of the matter field.

Also it is common knowledge in the literature \cite{Heemskerk:2010hk} that taking the boundary to infinity is equivalent to a renormalization group flow of the field theory into the IR. So away from the fixed point, the radial WDW state must be given by a deformation of the CFT by irrelevant (or at most marginal) operators, up to relevant counterterms $\text{S}[g,\Phi]$. %Since the limit $\mu\to0$ probes the same region in superspace as $\det[g]\to\infty$, the WDW solution must again look like the Freidel asymptotics when $\mu\to0$. On a Cauchy slice, $\det[g]$ only tends to infinity close to the boundary. Nevertheless, by this argument the RG flow must be as described above and 
Hence we can reverse the direction of argument and get a quantum gravity state from the CFT partition function by the following flow:
\bea\label{deform}
\Psi^{(\mu)}[g,\Phi] &=&  e^{\text{CT}(\mu)}
      \left( \text{P} \exp  \int_{\epsilon}^{\mu}\frac{d\lambda}{\lambda}\, O(\lambda) \right) Z^{(\epsilon)}_\text{CFT}[g,\phi,\{\chi\}], \\
   \lim_{\mu \to \epsilon}\Psi^{(\mu)}[g,\Phi]   &=& e^{\text{CT}(\epsilon)}  Z^{(\epsilon)}_\text{CFT}[g,\phi,\{\chi\}],
\eea
where $O(\lambda)$ is a spatial integral of local irrelevant and marginal operators, and $\text{CT}(\mu)$ is a counterterm, which is needed for the Hamiltonian constraint equation to take the standard ADM form.\footnote{The solution corresponding to the other Freidel branch, at least formally, be obtained by flipping the sign of the CFT trace anomaly, in which case the sign of the
the relevant part of the counterterms also flips, akin to Freidel, and the deforming operator $O(\lambda)$ will itself be different.} 
%Note that \eqref{deform} is the most general deformation consistent with \eqref{friedelduo} in the IR limit of the RG flow.\footnote{\textcolor{red}{I don't think we are entitled to say this yet since we haven't discussed the dimensional analysis inside of $O(\lambda)$.}}
In section \ref{thegeneraldeformation} we will calculate which $O(\lambda)$ we need so that the Hamiltonian constraint equation is satisfied:
\bea
{\cal H}(x)\Psi^{(1)}[g,\Phi] = 0,
\eea
with ${\cal H}(x)$ an arbitrary constraint that closes properly.

In a $T^2$ deformation, the irrelevant operators in $O(\lambda)$ always includes a term quadratic in the stress tensor.  The fact that the exponential is \emph{path ordered} incorporates the usual rule that, at a given coupling $\lambda$, the definition of the stress-tensor is used is that of the deformed theory with coupling $\lambda$, rather than the stress-tensor of the original CFT.  In other words, if the path ordered exponential is expanded out as a product, then the differential operator $\Pi^{ab} \sim \delta/\delta g_{ab}(x)$ acts on all structures appearing to its right.  (This is equivalent to the more standard $T^2$ notation, in which the deformation to the action is written as a differential equation in $\lambda$.)  A similar remark applies to any other field theory operators ${\cal O} \sim \delta/\delta J$ appearing in $O(\lambda)$.

The existence of marginal operators in $O(\lambda)$ produces log divergences, which require regulation by the lower cutoff $\epsilon$.  We take this to be the same quantity as the UV cutoff $\epsilon$ used to regulate the log divergences in the field theory partition function $Z^{(\epsilon)}_\text{CFT}$, which are associated with the trace anomaly.  These log divergences cancel, so that the overall solution $\Psi^{(\mu)}[g,\Phi]$ is ultimately independent of the regulator $\epsilon$. 

As the simplest example (basically the original $T\overline{T}$ deformation), if you want to obtain pure GR in AdS$_3$ from a $d = 2$ CFT with central charge $c$, $O(\lambda)$ will look like this:

%\begin{equation}
%O(\lambda) = \frac{1}{2}\int d^2x \left[-\frac{\sqrt{g}}{16\pi G_N} R + \frac{(16\pi G_N)\lambda}{\sqrt{g}}\left( :\!\Pi^{ab} \Pi_{ab}\!: - :\!\Pi^2\!: \right)\right]
%\end{equation}

\begin{equation}
O(\lambda) = \frac{1}{2}\int d^2x \left[\frac{c}{24\pi} \sqrt{g}R - \frac{24\pi}{c}\frac{\lambda}{\sqrt{g}}\left( :\!\Pi^{ab} \Pi_{ab}\!: - :\!\Pi^2\!: \right)\right],
\end{equation}
and the counterterm is just a $d = 2$ cosmological constant (not to be confused with $\Lambda$, the bulk $2+1$ cosmological constant):

%\begin{equation}
%\text{CT}(\mu) = \frac{-2}{16\pi G_N}\mu^{-1/2} \int d^2x \sqrt{g},
%\end{equation}

\begin{equation}
\text{CT}(\mu) = -\frac{c}{12\pi}\mu^{-1} \int d^2x \sqrt{g},
\end{equation}
and at the end we obtain a theory satisfying the 2+1 dimensional constraint equations \eqref{ham} and \eqref{mom} with
\begin{equation}
\Lambda = -\frac{1}{\mu} = -\frac{1}{(L_\text{AdS})^2}.
\end{equation}
Furthermore, cancellation of log divergences requires that the central charge of the original CFT satisfy the standard holographic relation:
\begin{equation}\label{c}
c = \frac{3L_\text{AdS}}{2G_N}.
\end{equation}
However, this deformation will not work in $d > 2$ because $R$ looks like a relevant term.  So next we will, extending \cite{Hartman:2018tkw,Taylor:2018xcy}, describe a general procedure for defining the $T^2$ deformation in arbitrary dimension, with general matter fields.

%In the case of Cauchy slice WDW states in AdS, the counterterm $\text{CT}(\mu)$ will turn out to be real.\footnote{Depending on the sign of the cosmological constant and whether we are talking about radial WDW-states or Cauchy slice WDW-states the counterterms will be real or imaginary.}

\subsection{The General Deformation} \label{thegeneraldeformation}

%We now present a general scheme to obtain the required deformation.

Let us start with a Hamiltonian and momentum constraint system describing gravity plus matter, in general bulk dimensions $D = d+1$.  The form of the momentum constraint ${\cal D}^a$ is determined by spatial covariance.  We allow for a general Hamiltonian constraint $\mathcal{H}$ which could be any scalar density satisfying the standard ADM closure condition \ref{Hclose}, which is important enough to type again:\footnote{It should be noted that the standard closure conditions very strongly constrain the form of the theory.  For example, if the only phase space variables are $g_{ab}(x)$ and $\Pi_{ab}(x)$, a series of rigidity theorems \cite{Hojman:1976vp,Kuchar:1974es,Farkas:2010dw,Gomes:2013naa,Gomes:2016vqk} suggest that (modulo field redefinitions, and subject to the constraint that the kinetic term is quadratic) general relativity may be the only allowed theory (but see \cite{Bojowald:2016hgh} for an attempt to deform the constraint algebra).  These theorems require an exact solution to the closure relation.  Presumably higher curvature corrections, such as Lovelock gravity, are allowed in the canonical formulation if one treats them as perturbatively small, and only requires the closure relations to be satisfied order-by-order.}
\bea
\left[ \mathcal{H}(x), \mathcal{H}(y)\right]=i\left(\mathcal{D}^a(x)\partial_a^{(x)}-\mathcal{D}^a(y)\partial_a^{(y)}\right)\delta(x-y).\label{closure}
\eea
It is reasonable not to try to modify the standard closure condition as (modulo redefinitions of the fields and constraints) this must necessarily be satisfied by \emph{any} locally Lorentz invariant bulk theory with a Hamiltonian description.\footnote{It is difficult to slightly deform special relativity while preserving locality. One family of attempts, called ``doubly special relativity'', has been studied in detail (see \cite{Kowalski-Glikman:2004fsz} for a review); however it is not known how to implement this proposal in a local field theory.} 

\paragraph{Irrelevant terms and anomalies:}  Not every choice of $\mathcal{H}$ will be compatible with a given holographic CFT, for as we shall see there are some anomaly matching conditions that need to be satisfied, analogous to \eqref{c}. 
In order to show that $\mathcal{H}$ is compatible, we need to find a canonical transformation:\footnote{We define any similarity transformation of the operators to be a canonical transformation, because in the classical limit ($\hbar \to 0$) this becomes a standard canonical transformation, albeit with complex coefficients.}
\bea\label{canonical}
\tilde{\mathcal{H}}(x) := e^{-\text{CT}} \mathcal{H}(x)  e^{\text{CT}}, \label{arbitraryCT}
\eea
such that $\Tilde{\mathcal{H}}(x)$ has no relevant terms and it contains the CFT trace equation as its marginal part (we will elaborate on how to achieve this at the end of this subsection).\footnote{This canonical transformation can introduce delta function divergences from the reordering of the terms in $\tilde{\mathcal{H}}$ (to put everything in a normal ordered form). These divergences can be eliminated by point-splitting.} That is,
\bea
i\mu^{1/d}\mathcal{\tilde{H}}^{(\mu)}(x)^\text{marg} = \widetilde{{\cal W}}(x) = {\cal W}(x) - i{\cal A}(x).
\eea
The requirement of the above equation will allow us to relate the free parameters in the bulk to the free parameters of the field theory (this will be clear in an example shown in the following section).

%where $\cal{A}$ is the CFT conformal anomaly, and $\cal{W}$ is the Weyl generator on the Cauchy slice: 
%\bea
%{\cal W}:= -\sum_i \Delta_{J^i}J^i {\cal O}_i = 2\Pi-\Delta_\phi \phi \Pi_\phi,
%\eea
%where $J^i$ ranges over all source fields (including the metric), and ${\cal O}_i$ is the corresponding operator and $\Delta_{J^i}$ is the conformal weight of the source in the boundary CFT.\footnote{In this expression, we are assuming the sources transform as primaries under a Weyl transformation.} 

Besides the above marginal couplings, in general $\tilde{\mathcal{H}}(x)$ will also have dimensionful coupling constants in it.  Following \cite{Hartman:2018tkw}, we adopt a convention in which the $T^2$ coupling $\mu$ is the only dimensionful parameter in the flow,\footnote{This is tantamount to adopting a very convenient choice of RG scheme, in which the form of the beta functions is fixed in an \emph{entire neighborhood} of the IR fixed point to exactly equal the leading-order result.  This convention is extremely useful for simplifying the algebra, as without it a path ordered exponential will usually produce infinitely many terms.} which implies that the flow follows an RG trajectory (we will show this more explicitly below).  Hence we absorb all other dimensionful couplings into $\mu$, so as to get a 1-parameter family of Hamiltonian constraints: $\tilde{\mathcal{H}}^{(\mu)}(x)$, which differ only by a rescaling of length. As the deformation parameter $\mu$ is varied, the bulk scales will therefore flow, with the ratios between them being determined by some dimensionless parameters in the $T^2$ theory.\footnote{For an explicit example see the powers of $\lambda$ appearing in \eqref{Xlambda}-\eqref{Xlambda_end} in section \ref{example}, which are fixed by dimensional analysis.}

We define the following local operator $X^{(\mu)}$ (which will turn out to be the operator needed to deform the CFT by an amount $\mu$):
\bea
- i dX^{(\mu)}(x) := i \mu^{1/d} \tilde{\mathcal{H}}^{(\mu)}(x) -  {\cal W}(x).\label{defX}
\eea

\paragraph{Canonical transformation on the algebra:}
As the canonical transformation preserves the commutators, we get the following relation:
\begin{align}
e^{-\text{CT}(\mu)} \left[i\mu^{1/d} \mathcal{H}^{(\mu)}(x),i\mu^{1/d} \mathcal{H}^{(\mu)}(y)\right] e^{\text{CT}(\mu)}
&=\left[i\mu^{1/d}\tilde{\mathcal{H}}^{(\mu)}(x),i\mu^{1/d} \tilde{\mathcal{H}}^{(\mu)}(y)\right]\\
&=\left[{\cal W}(x) - id X^{(\mu)}(x),{\cal W}(y) - id X^{(\mu)}(y)\right]\\
= -id \left( \left[{\cal W}(x),X^{(\mu)}(y)\right] -\left[{\cal W}(y),X^{(\mu)}(x) \right] \right) &+(id)^2\left[X^{(\mu)}(x),X^{(\mu)}(y)\right]. \label{CTofHHcommu}
\end{align}
We define the covariant operator:
\bea
O(\mu) := \int d^dx X^{(\mu)}(x). \label{arbitraryO}
\eea

\paragraph{Conformal classes:}
We split $X^{(\mu)}(x)$ into conformal classes, with each of the terms in the sum having a particular scaling $q_n$ with $\mu$ (determined by dimensional analysis):
\bea
X^{(\mu)}(x) = \sum_{n}  \mu^{q_n} X_n(x), \label{Xexpansion}
\eea
where $X_n(x)$ is a sum of terms with conformal weight $\omega_n$ (i.e. $-2$ for every power of $g_{ab}$ and $\Delta_{\phi}$ for every power of $\phi$, counting $g^{ab}$ or $\Pi^{ab}$ as $+2$ and $\Pi_{\phi}$ as $-\Delta_{\phi}
$) such that:
\bea
\left[
\int d^dy\,{\cal W}(y), X_n(x)\right]
= +i\omega_n X_n(x).
\eea
Because $X_n(x)$ is a scalar density (as can be seen from \eqref{defX}) its mass and conformal dimensions differ by $d$: $\left[X_n\right]=\omega_n+d$, and so we obtain the following universal relation by dimensional analysis: 
\bea
\omega_n =  q_n \, d.
\eea
Note that the operator $X^{(\mu)}(x)$ can be written in terms of its marginal ($\omega_n = 0$) and irrelevant $(\omega_n > 0)$ parts:
\bea
X^{(\mu)}= \frac{\mathcal{A}}{d} + X^{(\mu)}_{\text{irrel}}.
\eea

\paragraph{The power of demanding bulk algebra closure:}
Now we integrate \eqref{CTofHHcommu} with respect to $y$ to get:
\begin{align}
\int d^dy \; e^{-\text{CT}(\mu)} \left[i\lambda^{1/d} \mathcal{H}^{(\lambda)}(x),i\lambda^{1/d} \mathcal{H}^{(\lambda)}(y)\right] e^{\text{CT}(\mu)} &= - id\left(\left[{\cal W}(x),O(\lambda)\right] - \sum_{n} i\omega_n \lambda^{q_n} X_n(x)  \right)\\&\qquad+(id)^2\left[X^{(\lambda)}(x),O(\lambda)\right].&
\end{align}
Next we integrate the above equation with respect to $\log(\lambda)$. However, this integral $\left(\int_0^{\mu} \frac{d\lambda}{\lambda} \lambda^{q_n}\right)$ diverges for $q_n=0$. So we need to regulate the lower bound of the integral by a cutoff parameter $\epsilon$. By rearranging the terms, we get:
\begin{align}
\int_{\epsilon}^{\mu} \frac{d\lambda}{\lambda} \left[{\cal W}(x), O(\lambda)\right] = \frac{1}{2} \left(\mu^{2/d} -\epsilon^{2/d} \right) e^{-\text{CT}(\mu)} \partial_a {\cal D}^a(x) e^{\text{CT}(\mu)}+  i d X^{(\mu)}(x) -  i d X^{(\epsilon)}(x) &\\ 
+id \int_{\epsilon}^{\mu} \frac{d\lambda}{\lambda} \left[X^{(\lambda)}(x),O(\lambda)\right],&\label{relation}
\end{align}
where we used $\int d^dy  \left[\mathcal{H}^{(\lambda)}(x),\mathcal{H}^{(\lambda)}(y)\right]=i\partial_a {\cal D}^a(x)$, which comes from integrating \eqref{closure}.

We now further define a covariant operator from $O(\lambda)$ which is path-ordered in the scale $\lambda$:
\begin{equation}\label{path}
    \left( \text{P} \exp \int_{\epsilon}^\mu\frac{d\lambda}{\lambda}O(\lambda)\right):=\lim_{\delta\to 0}\prod_{n=\epsilon}^{n=\mu/\delta}\left(1+\frac{1}{n}O(n\delta)\right),
\end{equation}
We will eventually be interested in the following ``sandwiched" local operator for the purposes of obtaining WDW states:
\begin{align}
 \left( \text{P} \exp  \int_{\epsilon}^{\mu}\frac{d\lambda}{\lambda}\, O(\lambda) \right)  {\cal W}(x) \left( {\text{P} \exp}-\!\!\int_{\epsilon}^{\mu}\frac{d\lambda}{\lambda}\, O(\lambda) \right)&=  {\cal W}(x) - \int_{\epsilon}^{\mu} \frac{d \lambda}{\lambda} \left[ {\cal W}(x), O(\lambda)\right]  \\ 
 &\!\!\!\!\!\! +\int_{\epsilon}^\mu \frac{d \lambda}{\lambda} \int_{\epsilon}^\lambda \frac{d \lambda'}{\lambda'} \left[ \left[{\cal W}(x), O(\lambda')\right],O(\lambda) \right]  + \ldots, \label{seriesclosing}
\end{align}
where the RHS follows from the Baker-Campbell-Hausdorff formula.

We observe that relation \eqref{relation} generates a neat cascade of cancellations in the infinite series emerging from the closure relation of the Hamiltonian constraint:
\begin{align}
{\cal W}(x)\,- & \int_{\epsilon}^{\mu} \frac{d \lambda}{\lambda} \left[ {\cal W}(x), O(\lambda)\right] + \int_{\epsilon}^\mu \frac{d \lambda}{\lambda} \int_{\epsilon}^\lambda \frac{d \lambda'}{\lambda'} \left[ \left[{\cal W}(x), O(\lambda')\right],O(\lambda) \right]  + \dots \\
&= {\cal W}(x) - id X^{(\mu)}(x) -\frac{1}{2} \mu^{2/d}  e^{-\text{CT}(\mu)} \partial_a  {\cal D}^a(x) e^{\text{CT}(\mu)} \\ &+\left( \text{P} \exp  \int_{\epsilon}^{\mu}\frac{d\lambda}{\lambda}\, O(\lambda) \right) id X^{(\epsilon)}(x) \left( {\text{P} \exp}-\!\!\int_{\epsilon}^{\mu}\frac{d\lambda}{\lambda}\, O(\lambda) \right),  
\end{align}
where the commutator $\partial_a[ {\cal D}^a(x), O(\lambda)] = 0$ by covariance of $O(\lambda)$. Therefore we get:
\begin{align}
 \left( \text{P} \exp  \int_{\epsilon}^{\mu}\frac{d\lambda}{\lambda}\, O(\lambda) \right)  \left\{ {\cal W}(x) - id X^{(\epsilon)}(x)\right\}& \\
 = \Bigg\{ \mathcal{W}(x) - i d X^{(\mu)}(x) -\frac{1}{2} \left(\mu^{2/d}-\epsilon^{2/d} \right) &e^{-\text{CT}(\mu)}\partial_a {\cal D}^a(x) e^{\text{CT}(\mu)} \Bigg\} \left( \text{P} \exp  \int_{\epsilon}^{\mu}\frac{d\lambda}{\lambda}\, O(\lambda) \right), \\
 &= \left\{ {\cal W}(x) - i d X^{(\mu)}(x)  \right\} \left( \text{P} \exp  \int_{\epsilon}^{\mu}\frac{d\lambda}{\lambda}\, O(\lambda) \right) ,
\end{align}
where we have used the momentum constraints in throwing out the $\partial_a {\cal D}^a(x)$ term.\footnote{The closure constraints form an important consistency property on the allowed deformations $O(\lambda)$.  For example, if we replace the standard ADM kinetic term inside $X$ with
\begin{equation}
\Pi_{ab}\Pi^{ab} - \alpha \Pi^2
\end{equation}
with $\alpha \ne \frac{1}{d-1}$, and plug that into \eqref{deformingflow1}, then these cancellations will no longer occur.  The resulting constraint ${\cal H}^{(\mu)}(x)$ at finite $\mu$ will have an infinite series of total derivative terms in it, and in fact the coefficients of these terms fall off slowly enough with the number of derivatives, that it resums to a spatially nonlocal answer!  (The nonlocality of the resulting expression is a consequence of the fact that integrating with respect to $\lambda$ is equivalent to dynamically evolving a finite conformal distance into the bulk.)  Such nonlocal terms are excluded on the hypothesis that the $T^2$ deformed theory has a UV completion which is local down to very short length scales.  A similar argument shows that, if we do not throw away the ${\cal D}^a$ terms, we would get a nonlocal integral over those as well.  This is a necessary consequence of the fact that the algebra of conformal transformations (Weyl plus spatial diffeos) is not locally isomorphic to the ADM constraint algebra.}

\paragraph{The resulting solution:}
So now we can obtain a solution to an arbitrary Hamiltonian and momentum constraints of gravity arbitrarily coupled with arbitrary matter fields, in arbitrary dimensions $d+1$ by deforming an arbitrary $d$-dimensional CFT:
\beq \label{deformingflow1}
 Z^{(\mu)}[g,\phi,\{\chi\}] := e^{\text{CT}(\mu)}
      \left( \text{P} \exp  \int_{\epsilon}^{\mu}\frac{d\lambda}{\lambda}\, O(\lambda) \right) Z_\text{CFT}^{(\epsilon)}[g,\phi,\{\chi\}],
\eeq
where $O(\lambda)$ is defined in \eqref{arbitraryO} and is %\emph{uniquely} 
determined by the choice of starting Hamiltonian constraint $\mathcal{H}^{(\mu)}(x)$ and the choice of canonical transformation $\text{CT}(\mu)$ in \eqref{arbitraryCT}. This is then easily shown to be a WDW state:
\begin{align}
\mathcal{H}^{(\mu)}(x) & Z^{(\mu)}[g,\phi,\{\chi\}] = \mathcal{H}^{(\mu)}(x) e^{\text{CT}(\mu)}
      \left( \text{P} \exp  \int_{\epsilon}^{\mu}\frac{d\lambda}{\lambda}\, O(\lambda) \right) Z_\text{CFT}^{(\epsilon)}[g,\phi,\{\chi\}] \\
      &=  -i\mu^{-1/d} e^{\text{CT}(\mu)} \left({\cal W}(x) - i d X^{(\mu)}(x) 
      \right)
      \left( \text{P} \exp  \int_{\epsilon}^{\mu}\frac{d\lambda}{\lambda}\, O(\lambda) \right) Z_\text{CFT}^{(\epsilon)}[g,\phi,\{\chi\}]  \label{CSeqn}\\
       &=  -i\mu^{-1/d} e^{\text{CT}(\mu)} 
      \left( \text{P} \exp  \int_{\epsilon}^{\mu}\frac{d\lambda}{\lambda}\, O(\lambda) \right) \left({\cal W} (x)-i\mathcal{A}(x)-idX^{(\epsilon)}_{\text{irrel}}(x)\right) Z_\text{CFT}^{(\epsilon)}[g,\phi,\{\chi\}].
\end{align}
Now taking the limit $\epsilon \to 0$ we have that $X^{(\epsilon)}_\text{irrel}(x)\to0$ as a power law and so we learn that the Hamiltonian constraint is satisfied:
\begin{align}
\mathcal{H}^{(\mu)}(x) Z^{(\mu)}[g,\phi,\{\chi\}]&=-i\mu^{-1/d} e^{\text{CT}(\mu)} 
      \left( \text{P} \exp  \int_{\epsilon}^{\mu}\frac{d\lambda}{\lambda}\, O(\lambda) \right) \left({\cal W} (x)-i\mathcal{A}(x)\right)Z_\text{CFT}^{(\epsilon)}[g,\phi,\{\chi\}], \\
      &= 0,\label{zero}
\end{align}
where at the end, we used the trace anomaly equation of the CFT.  This can be viewed as a derivation of what CFT anomaly ${\cal A}(x)$ is required if a holographic correspondence is to hold.

\paragraph{RG flow:} We can now convince ourselves explicitly that the deformation defined by $O(\lambda)$ is really along an RG flow line. Equations \eqref{CSeqn} and \eqref{zero} together show that, modulo the counterterms, the deformed partition function satisfies a Callan-Symanzik equation:
\begin{align}
%    \left(\int d^dx\,{\cal W}(x) - i d \;O(\mu) 
%      \right)
%      \left( \text{P} \exp  \int_{\epsilon}^{\mu}\frac{d\lambda}{\lambda}\, O(\lambda) \right) Z_\text{CFT}^{(\epsilon)}[g,\phi,\{\chi\}]=0,\\
      \left(\int d^dx\,{\cal W}(x) - i d \;\mu\frac{\partial}{\partial\mu} 
      \right)
      \left( \text{P} \exp  \int_{\epsilon}^{\mu}\frac{d\lambda}{\lambda}\, O(\lambda) \right) Z_\text{CFT}^{(\epsilon)}[g,\phi,\{\chi\}]=0,\label{RGflow}
\end{align}
for any finite value of $\mu$ along the flow, where in the first bracket of \eqref{RGflow} the integral of $\cal W$ is just a scale transform (i.e. RG flow).  Hence the entire second bracket of \eqref{RGflow} is nothing other than an integrated RG flow.\footnote{Except for the initial deformation away from the CFT near $\lambda = \epsilon$, to which \eqref{RGflow} doesn't apply, since its derivation is valid only up to terms which vanish as $\epsilon \to 0$.  This is because flowing all the way to an exact CFT requires an infinite amount of RG flow.}

\paragraph{Fixing the counterterms:}
We now give a systematic procedure to fix the counterterms $\text{CT}(\mu)$. First we pick a canonical transformation to exactly generate $\mathcal{W}$, since otherwise we will not obtain the trace anomaly equation at the end. Next we choose a relevant term, say $Y(x)$, that we want to get rid of and perform a further canonical transformation of the form $\sim \int d^dx Y(x)$. This works because:
\bea
e^{-\int d^dy Y(y)}\mathcal{W}(x)e^{\int d^dy Y(y)}&=&\mathcal{W}(x) + \left[\mathcal{W}(x), \int d^dy Y(y)\right]\\
&=&\mathcal{W}(x)+\sim Y(x) + \sim \text{total derivatives}.
\eea 
However, if such a counterterm has already been used to generate the correct form of $\mathcal{W}$, then it might not be possible to eliminate such a relevant term without also modifying $\mathcal{W}$.  This leaves us with only one possibility to eliminate such relevant terms and that is by fixing certain relations between bulk and field theory parameters. This is crucial, since if we do not do this then our starting CFT will not satisfy a well-defined trace anomaly equation, as it would include left-over relevant terms. Therefore, these parameter relations are a direct consequence of the trace anomaly equation of the CFT. Having done that, we proceed with the next relevant terms in an iterative process until we only have marginal and irrelevant terms left, at which point we are in a position to define $X^{(\mu)}(x)$ as in equation \eqref{defX}.\footnote{It is worth noting that there exists another formalism in which one allows relevant terms in $X$, e.g. in pure GR $X$ might for example look identical to the ADM Hamiltonian $\cal H$, even in $d > 2$.  However, the price of doing this is that one must also include a second set of counterterms $e^{CT_2(\epsilon)}$ on the right hand side of the path ordered exponential in \eqref{deformingflow1}.  (Or equivalently, one must agree to throw out all power law divergences in $\epsilon$ when evaluating the $\int_\epsilon^\mu$ integral.)  This formalism has the charm of making the formula for $X$ less dependent on specific dimensions, but it obscures the nature of the RG flow near the CFT.  Since it does not allow for any distinct theories beyond those mentioned in the main text, we do not need to consider it further.}

As long as we restrict attention to CFT sources and fields in the range strictly within the interval $0 < \Delta < d$ (corresponding to $m^2 < 0$) there should be only a finite number of possible relevant terms, equal to the number of possible counterterms, and the process is guaranteed to terminate.  This is what happens in the case of pure Einstein gravity in any dimension.

However, outside this range it could also happen that the newly generated terms are either more relevant or of equal relevancy to the term we were trying to eliminate. This is in fact what will happen for the case of a bulk scalar with $m^2 > 0$. In such cases we would need an infinite series of counterterms to get rid of all the relevant terms.  However, this happens in the regime where we turn on nonrenormalizable sources in the original undeformed holographic QFT.\footnote{This should not be confused with the question of whether the $T^2$ deformation is itself renormalizable.  Since the $T^2$ deformation consists of irrelevant terms, it is always at least superficially nonrenormalizable.} and hence the original QFT (which is no longer conformal) has to be regarded as an effective field theory valid only at low coupling.  Hence, the deformed theory will also only be defined perturbatively in the massive bulk matter fields.  Defining the bulk theory nonperturbatively would presumably require UV completion of the $T^2$ theory.

\subsection{Example: Scalar Field Coupled to Gravity}\label{example}

Consider a scalar field $\Phi$ in $d+1$ dimensions coupled to gravity with the Hamiltonian constraint:
\bea\label{Hscalar}
\mathcal{H}(x)\Psi[g,\Phi]&:=&\Bigg\{\frac{16\pi G_N}{\sqrt{g}}:\!\Big(\Pi_{ab}\Pi^{ab}-\frac{1}{d-1}\Pi^2\Big)\!:-\frac{\sqrt{g}}{16\pi G_N}(R-2\Lambda)\\
&+&\frac{1}{2}\left(\frac{1}{\sqrt{g}}
\!:\!\Pi_\Phi^2\!:
+\sqrt{g}\left(g^{ab}\nabla_a\Phi\nabla_b\Phi+m^2\Phi^2+2V(\Phi,R)\right)\right)\Bigg\}\Psi[g,\Phi]=0, \label{matterHC}
\eea
where $V(\Phi,R)$ is an even potential with arbitrary non-minimal coupling (excluding terms already explicitly in the action).  

By dimensional analysis, the bulk matter field is related to the field theory source via (by dimensional analysis):
\bea\label{Phiphi}
\Phi = \mu^{(1/d)(\Delta_\phi - (d-1)/2)} \phi,
\eea
where $\Delta_\phi$ is the conformal dimension of the field theory source $\phi$.  The Weyl and Diff generators in the CFT are then determined by \eqref{D} and \eqref{W} to be:
\begin{eqnarray}
{\cal D}_a \!\!&=&\!\!\!\! -2\nabla_b \Pi^b_a + \partial_a \phi\,\Pi_\phi,\\
{\cal W} \!\!&=&\phantom{-}2\Pi \:\:-\:\,\Delta_\phi \phi \Pi_\phi.
\end{eqnarray}

\paragraph{Explicit counterterms:} The correct choice of canonical transformation to remove all relevant terms depends on the dimensions $d$ and $\Delta_\phi$.  In order to keep things simple we will consider the following canonical transformation{\bf---}which is valid for a certain range of parameters to be discussed below:
\bea
\text{CT} := \int d^dy \sqrt{g} \left(a + b \ R + c \ \Phi^2 \right). \label{matterCT}
\eea
Under the above canonical transformation, we have:
\bea
\Pi^{ab} \to e^{- \text{CT}} \Pi^{ab} e^{ \text{CT}} &=&  \Pi^{ab} -\frac{i}{2} \sqrt{g} \left(a+ c \ \Phi^2 \right) g^{ab} + i \sqrt{g} \ b \  G^{ab}, \\
\Pi_{\Phi} \to e^{- \text{CT}} \Pi_{\Phi} e^{ \text{CT}} &=& \Pi_{\Phi} -2i \sqrt{g} \ c \ \Phi.
\eea
By performing the canonical transformation of $\mathcal{H}(x)$, we obtain a marginal term of the form:
\bea
16 \pi G_N \frac{i a}{d-1} \Pi(x) - 2ic \Phi \Pi_{\Phi}(x).
\eea
By demanding this to be equal to $-i \mu^{-1/d} \mathcal{W}(x)$, we fix $a$ and $c$:
\bea
a &=& -2\frac{d-1}{16 \pi G_N} \mu^{-1/d},\\
c &=& -\frac{\Delta_{\phi}}{2} \mu^{-1/d}. 
\eea
Now because the cosmological constant term is relevant and we have already used the counterterm $\int d^dx\sqrt{g}$, we need to impose the following relation between the cosmological scale and the deformation coupling:
\bea\label{lambda}
\Lambda = -\frac{d(d-1)}{2}\mu^{-2/d},
\eea
which identifies the scale $\mu^{1/d}$ with $L_{\text{AdS}}$. Similarly, given that for the range of $\Delta_\phi$ in which the counterterm \eqref{matterCT} is valid (as explained below), the mass term is relevant we need to impose the usual holographic mass relation for consistency:
\bea\label{mass}
m^2 = \Delta_{\phi} (\Delta_{\phi}-d) \mu^{-2/d}.
\eea
In order to eliminate the only left-over relevant term, $\sqrt{g}R$, in $\mathcal{H}$ we need to fix $b$:
\bea
b &=& -\frac{1}{(d-2) 16 \pi G_N} \mu^{1/d}.
\eea

The choice of complex sign for $\mu$ determines whether our deformed partition function satisfies the Wheeler-DeWitt equation with positive or negative cosmological constant. In particular,
\begin{eqnarray}
   \Lambda<0 &\longrightarrow& \mu^{1/d}\in\mathbb{R}_+,\\
    \Lambda>0 &\longrightarrow& \mu^{1/d}\in i\mathbb{R}_+.
\end{eqnarray}
Also notice that $m^2$ goes to minus itself in the dS case.
In this paper, we will proceed with the case of $\Lambda<0$, corresponding to AdS/CFT. However it appears that one might also use the same approach to deform the dS/CFT duality \cite{Strominger:2001pn, Anninos:2011ui} to finite time Cauchy slices; although in the dS case, the Euclidean CFT cannot be reflection-positive even prior to the $T^2$ deformation.  (For an application of the formalism to cosmology on closed slices and associated discussions, see \cite{Araujo-Regado:2022jpj}. We also plan another follow-up paper to discuss this case \cite{GRWclosed}.)

One significant difference between the two cases is that for dS deformations, one can probe Lorentzian spacetime even when the parameter $\mu$ is small (measured relative to other length scales set by the choice of metric $g_{ab}$).  But for the AdS case, it is necessary for $\mu$ to be sufficiently large in order to cross a phase transition into Lorentzian signature (see section \ref{exotic}).

\paragraph{Bulk scales:}

As explained earlier, all the bulk scales of the gravity theory can be expressed in terms of the only dimensionful coupling in the $T^2$ theory, $\mu$, alongside a set of dimensionless parameters. We already saw how two of these bulk scales (the AdS scale and the mass of the scalar field) are fixed in this way, in \eqref{lambda} and \eqref{mass}, where $\Delta_\phi$ is the appropriate dimensionless parameter. Similarly we can define another dimensionless parameter, $\alpha$, that fixes the Planck scale relative to the AdS scale:
\bea\label{alpha}
    \alpha &=&\frac{L_\text{AdS}^{d-1}}{16\pi G_N},\\
    \implies 16\pi G_N&=&\frac{1}{\alpha}\mu^{\frac{d-1}{d}}.
\eea

From the field theory perspective $\alpha$ will be related to things like the anomaly coefficient, the stress tensor two-point function coefficient or the coefficient of the divergence in the entanglement entropy. If we are in even dimensions and our theory has a nonzero central charge, then we can fix this ratio of scales at the current level of analysis, by comparing the anomaly term $\mathcal{A}$ directly with the CFT result. 

In odd dimensions, we need to work harder. This ratio can also be calculated from the holographic CFT from e.g. relating the stress-tensor 2-point function to gravitons in the bulk. So if our Cauchy slice theory is to describe a consistent theory of gravity with a holographic dual, we must fix the parameter $\alpha$ to get a consistent local bulk theory.  If we pick the wrong value of $\alpha$, all of the above equations will still hold (unless we fail to satisfy an anomaly matching condition associated with another field theory source), but we expect that the higher n-point functions of the stress-tensor will not agree with a normal semiclassical gravitational bulk with Dirichlet boundary conditions.\footnote{This question is related to the ``fake bulk'' of \cite{Belin:2020oib,fake,Mazenc:2019cfg}.} 

\paragraph{Regime of validity:}

We now discuss the regime of parameters in which the counterterms above are valid.  This requires that $X^{(\mu)}$ contain only irrelevant terms, and that the marginal terms match with anomalies.

The remaining terms in $i \mu^{1/d}\Tilde{\mathcal{H}}^{(\mu)}-\mathcal{W}=-idX^{(\mu)}$ group into conformal classes which are listed in Table \ref{table} along with their conformal weights $\omega$. For a term to be irrelevant, it must have positive conformal weight.

\begin{table}[ht]
\begin{center}
\begin{tabular}{|c|c|c|} 
\hline &&\\
\textbf{Terms in $X^{(\mu)}$} & \textbf{Conformal 
Weight $\omega$ } & \textbf{Condition for $\omega>0$} \\
\hline &&\\
$\frac{1}{\sqrt{g}}\Big(\Pi_{ab}\Pi^{ab}-\frac{1}{d-1}\Pi^2\Big)$ & $d$ & $d>0$  \\ &&\\
$G_{ab} \Pi^{ab} - \frac{1}{d-1} G\Pi$ & $2$ & $2>0$ \\ &&\\
$\sqrt{g}  \left(G^{ab} G_{ab} - \frac{1}{d-1} G^2 \right) $ & $4-d$ &  $d<4$
\\ &&\\
 $\frac{1}{\sqrt{g}}\Pi_\Phi^2$ & $d-2\Delta_{\phi}$ & $\Delta_{\phi}<\frac{d}{2}$ \\ && \\
 $\sqrt{g}g^{ab}\nabla_a\Phi\nabla_b\Phi$ & $2-d+2\Delta_{\phi}$ & $\Delta_{\phi}>\frac{d-2}{2}$ \\ &&
 \\
 $\sqrt{g}\Phi^2$ & $-d+2\Delta_{\phi}$ & $\Delta_{\phi}>\frac{d}{2}$ \\ &&
 \\
 $\Phi^2 \Pi$ & $2\Delta_{\phi}$ & $\Delta_{\phi}>0$ \\ &&
 \\
 $\sqrt{g}\Phi^2 G$ & $2-d+2\Delta_{\phi}$ & $\Delta_{\phi}>\frac{d-2}{2}$ \\ &&
 \\
 $\sqrt{g} \Phi^4$ & $4\Delta_{\phi}-d$ & $\Delta_{\phi}>\frac{d}{4}$ \\ &&
 \\
 \hline 
\end{tabular} 
\caption{\label{relevancy} ‘‘Relevancy" of the terms in $X^{(\mu)}$ for the counterterms chosen in \eqref{matterCT}.}
\label{table}
\end{center}
\end{table}

Notice that the conditions for $\frac{1}{\sqrt{g}}\Pi_\Phi^2$ and $\sqrt{g}\Phi^2$ to be irrelevant are incompatible. That means that depending on the value of $\Delta_\phi$ we will need a different counterterm to cancel whichever one of the two is the relevant term.  

The term $\sqrt{g}  \left(G^{ab} G_{ab} - \frac{1}{d-1} G^2 \right)$ will be relevant for $d>4$, in which case we must include additional counterterms to get rid of it. But it is irrelevant for $d=3$, so we make this choice for the sake of illustration.
Now we can read from the table the allowed range for $\Delta_{\phi}$ such that all the terms in $X^{(\mu)}$ are irrelevant:
\bea
\Delta_{\phi} \in \left(\frac{3}{4},\frac{3}{2}\right).
\eea
Notice that for this range, the mass of the bulk scalar field will be negative but this is allowed in AdS by boundary unitarity:
\bea
m^2 \in \left(-\frac{9}{4}, -\frac{27}{16} \right) L_{\text{AdS}}^{-2}.
\eea

%\newpage
\paragraph{Explicit deformation:}
As a result we have the deformation operator for $d=3$ and the above range of $\Delta_\phi$ to be:\footnote{The need to include the $\Pi_\Phi^2$ term was emphasized in \cite{Kraus:2018xrn}.}
%\textcolor{red}{This deformation is wrong unless we let $16\pi G_N$ vary with $\lambda$ also. The bulk scales also flow along the deformation. If we mean the final $16\pi G_N$, then it is wrong.}
%\bea\label{Xlambda}
%X^{(\lambda)}(x) &=& -\frac{16 \pi G_N}{3}  \lambda^{1/3}  \frac{1}{\sqrt{g}} :\left(\Pi^{ab} \Pi_{ab} - \frac{1}{2} \Pi^2\right): \\ 
%&\;& + i \frac{2}{3}   \lambda^{2/3}   \left(G_{ab} \Pi^{ab} - \frac{1}{2} G\Pi\right) \\ 
%&\;& +\frac{1}{3(16 \pi G_N)} \lambda  \sqrt{g}  \left(G^{ab} G_{ab} - \frac{1}{2} G^2 \right) \\
%&\;& -\frac{1}{3} \lambda^{1/3} \left( \frac{1}{2\sqrt{g}} \Pi_{\Phi}^2  + \frac{\sqrt{g}}{2} \left(g^{ab} \nabla_a \Phi \nabla_b \Phi + \frac{\Delta_{\phi}}{4} R \Phi^2 \right)  +\sqrt{g} V(\Phi,R) \right) \\
%&\;& +i\frac{16\pi G_N}{12} \Delta_{\phi} \Phi^2\Pi -\frac{16\pi G_N}{32} \Delta_{\phi}^2 \lambda^{-1/3} \sqrt{g} \Phi^4.
%\eea
%\textcolor{red}{This is the correct deformation in terms of field theory quantities only, for my definition of $\alpha$. I also feel that we should rescale $X^{(\lambda)}$ by a factor of $d=3$ so that the $T^2$ operator is canonically normalized.}

\bea \label{Xlambda}
X^{(\lambda)}(x) &=& -\frac{\lambda}{3\alpha}   \frac{1}{\sqrt{g}} :\left(\Pi^{ab} \Pi_{ab} - \frac{1}{2} \Pi^2\right): \\ 
&\;& + i \frac{2}{3}   \lambda^{2/3}   \left(G_{ab} \Pi^{ab} - \frac{1}{2} G\Pi\right) \\ 
&\;& +\frac{\alpha}{3} \lambda^{1/3}  \sqrt{g}  \left(G^{ab} G_{ab} - \frac{1}{2} G^2 \right) \\
&\;& -\frac{1}{3} \lambda^{1-2\Delta_\phi/3}  \frac{1}{2\sqrt{g}} \Pi_{\phi}^2 \label{PhiKin} \\ 
&\;& -\frac{1}{3} \lambda^{(2\Delta_\phi-1)/3} \frac{\sqrt{g}}{2}  \left(g^{ab} \nabla_a \phi \nabla_b \phi + \frac{\Delta_{\phi}}{4} R \phi^2 \right)  \\ 
&\;& -\frac{1}{3} \lambda^{1/3} \sqrt{g} V\left(\lambda^{(\Delta_\phi - 1)/3} \phi,R\right)  \\
&\;& +i\frac{\Delta_{\phi}}{12\alpha} \lambda^{2 \Delta_\phi/3}  \phi^2\Pi  -\frac{\Delta_{\phi}^2}{32\alpha}  \lambda^{4\Delta_\phi/3 - 1} \sqrt{g}  \phi^4.
\label{Xlambda_end}
\eea

As shown in section \ref{thegeneraldeformation}, the deformed field theory partition function:
\bea
Z^{(\mu)}[g,\phi,\{\chi\}] := e^{\text{CT}(\mu)}
      \left( \text{P} \exp  \int_{0}^{\mu}\frac{d\lambda}{\lambda}\, O(\lambda) \right) Z_\text{CFT}[g,\phi,\{\chi\}],
\eea
with the $\text{CT}(\mu)$ defined in \eqref{matterCT} and $O(\lambda)$ defined in \eqref{arbitraryO} will satisfy the Hamiltonian constraint \eqref{matterHC}:
\bea
\mathcal{H}^{(\mu)}(x) \ Z^{(\mu)}[g,\phi,\{\chi\}] = 0.
\eea
Note that here we did not need the $\epsilon$ regulator simply because in the range of $\Delta_\phi$ for which this deformation is valid there are no anomalies.

\paragraph{Addressing other ranges:}
If we wish to consider a $\Delta_\phi$ outside the range $(\frac{3}{4}, \frac{3}{2})$, or $d \ne 3$, we need to adjust our deformation accordingly.  For generic values of $0 < \Delta_\phi < \frac{d+2}{2}$ (corresponding to relevant sources in a unitary CFT), it should suffice to simply choose a different set of relevant counterterms.  This will change the set of terms needed in $X^{(\mu)}$.\footnote{We should also be careful about terms in the potential $V$ becoming relevant as we change $\Delta_\phi$. We deal with such terms as described for any relevant term.}  For example, for an ``alternative'' boundary condition scalar with $\frac{d}{2} < \Delta_\phi < \frac{d+2}{2}$, $\Pi_\Phi^2$ is relevant and needs a counterterm, while $\Phi^2$ is irrelevant and can appear in $X^{(\mu)}$. Under a canonical transformation of the form $\sim \int d^dx \frac{1}{\sqrt{g}}\Pi_\Phi^2$ the source field $\Phi$ gets shifted by:
\bea
\Phi(x)\to e^{-\int d^dy \frac{1}{\sqrt{g}}\Pi_\Phi^2}\Phi(x)e^{\int d^dy \frac{1}{\sqrt{g}}\Pi_\Phi^2}&=&\Phi(x)+2i\frac{1}{\sqrt{g}}\Pi_\Phi(x).
\eea
Notice that in this case, when we take the limit $\mu\to\epsilon$, our field theory flows to a different IR fixed-point from $Z_\text{CFT}^{(\epsilon)}[g,\phi]$ due to the presence of this operator-dependent relevant counterterm.
%This is analogous to the case of ``alternative'' quantization in AdS/CFT, which requires a different holographic CFT on the boundary to the case of usual quantization. 

On the other hand, for a measure zero subset of values of $\Delta_\phi$ (e.g. when any of the inequalities in Table \ref{relevancy} are saturated, or any of the similar conditions that arise after doing additional canonical transformations) there will be additional marginal terms in $X^{(\mu)}$, which result in additional anomaly matching conditions to $\cal A$ in the CFT.  For example, when $\Delta_\phi = d/2$ exactly, there might be an anomaly in the CFT proportional to $\phi^2$.  Or when $d = 4$, it is necessary for $\sqrt{g}  \left(G^{ab} G_{ab} - \frac{1}{2} G^2 \right)$ to match with the anomalies in the CFT{\bf---}which implies that the 4d CFT has equal ‘‘central" charges $a = c =\frac{L_\text{AdS}^3}{8G_N}$, as expected in AdS$_5$/CFT$_4$ duality \cite{Henningson:1998ey}.\footnote{These charges are not to be confused with the counterterm coefficients of the rest of the section, which are labelled by the same letters.}

As $\Delta_\phi \to 0$, an increasingly large number of counterterms are necessary, but they all take the form $f(\phi)$ times a finite number of relevant expressions.  On the other hand, for $\Delta_\phi < 0$, there will be an uncontrolled proliferation of counterterms needed, since every expression becomes relevant when multiplied by a high enough power of $\phi$.  (This is related to the fact that the CFT becomes nonrenormalizable when perturbed by the source $\phi$).  It should still be possible to work out the deformation order-by-order in perturbation theory in $\phi$, but we leave the details to future work.

\subsection{Normal Ordering Prescription}\label{largeN}

In order to explain why we need to consider a large $N$ theory, first let us give a more complete description of the normal ordering symbol $:\:\::$ in e.g. \eqref{ham} or \eqref{Hscalar}--\eqref{matterHC}.
For us this symbol actually performs a double duty:
\begin{enumerate}
\item It moves all factors of $\Pi^{ab}$ (with both indices raised) to the right of arbitrary functions of the metric $g_{cd}$, thus removing unpleasant divergences of the form $[\Pi^{ab}(\mathbf{x}), g_{cd}(\mathbf{x})] =-i \delta^{ab}_{cd}\delta(\mathbf{0})$ which arise due to the commutator.\footnote{Note that $\Pi_{ab} = g_{ac} g_{bd} \Pi^{cd}$ has hidden factors of the metric, as does $\Pi = g_{ab} \Pi^{ab}$.}  (This is the same result one would obtain if the divergence is regulated by point-splitting.)
\item It also subtracts off field theory divergences associated with the coincident limit of the two $\Pi^{ab}$ operators as they approach each other in the field theory.  The way this works depends on whether we are at infinite or finite $N$.
\end{enumerate}

\subsubsection*{Infinite $N$:}

In the $N \to \infty$ limit of an 't Hooftian field theory, correlation functions of stress-tensors $T_{ab}$ are dominated by Wick contractions.  It follows that the OPE of two stress-tensors is, up to terms subleading in $1/N$, given by
\begin{equation}\label{Wick}
\Pi^{ab}(\mathbf{x}) \Pi^{cd}(\mathbf{y}) 
= C^{abcd}(\mathbf{x}, \mathbf{y})\phantom{i} +  :\!\Pi^{ab} \Pi^{cd}\!:(\mathbf{x}) + \ldots,
\end{equation}
where $C^{abcd}(\mathbf{x}, \mathbf{y})$ is a c-number which can depend on the background metric and which diverges as $\mathbf{y}\to \mathbf{x}$, while the second term is an operator of dimension $2d$ which is independent of the separation.  Hence, we can define the normal-ordering symbol as:
\begin{equation}\label{NO}
:\!\Pi^{ab} \Pi^{cd}\!:(\mathbf{x}) :=
\lim_{\mathbf{y}\to \mathbf{x}}\left[
\Pi^{ab}(\mathbf{x}) \Pi^{cd}(\mathbf{y}) - C^{abcd}(\mathbf{x}, \mathbf{y})\right].
\end{equation}
This defines $:\!\Pi^{ab} \Pi^{cd}\!:$ (up to a possible curvature ambiguity of weight $2d$).

This property holds even after doing a $T^2$ deformation with finite $\mu$, since in the semiclassical approximation to gravity, fluctuations of the metric are Gaussian.  (Even though the $T^2$-deformed stress-tensor would be multi-trace if written in terms of the original CFT fields.)  Although the theory is no longer conformal, \eqref{Wick} still picks out a unique definition of the $:\!\Pi^{ab} \Pi^{cd}\!:$ operator, up to a c-number correction which can be absorbed into other terms in $\cal H$.

A similar analysis would apply to the kinetic terms associated with matter fields, e.g. $:\!\Pi_\Phi^2\!:$, but for simplicity we consider only the metric in the remainder of this section.

\subsubsection*{Finite $N$, $\mu = 0$:}

The above definition works only at $N = \infty$.  To see why, first consider the case of $N =$ finite, $\mu = 0$.  In such a CFT, there will be additional interaction terms in the stress-tensor OPE, but more importantly, in a generic $d > 2$ CFT, there will be no operator with dimension exactly $2d$.  Instead the dimension of this operator is shifted by some anomalous dimension $\eta(N)$, which would require us to take a limit of the form
\bea
:\!\Pi^{ab} \Pi^{cd}\!:\!(\mathbf{x}) &:=& \lim_{\mathbf{y}\to \mathbf{x}}\left[\frac{1}{|\mathbf{x}-\mathbf{y}|^{\eta}}
\Pi^{ab}(\mathbf{x}) \Pi^{cd}(\mathbf{y}) - 
\hat{C}^{abcd}(\mathbf{x}, \mathbf{y})\right],
\eea
where now $\hat{C}^{abcd}(\mathbf{x}, \mathbf{y})$ can depend, not just on c-numbers, but also on operators of dimension up to $2d + \eta$.  

The anomalous power $\eta$ means that the operator can no longer be specified without introducing an extra length scale into the problem, and we cannot use $\mu$ for this purpose as it vanishes.  Because of this, it will be more convenient to proceed directly to the case of finite $\mu$.  This is acceptable since our deformation is defined in \eqref{deformingflow1} as an integral over finite values of $\mu$.

%If we want to include this $:\!\Pi^{ab} \Pi^{cd}\!:$ operator in our $O(\lambda)$ deformation, we would need an extra fractional power of $\lambda^{\eta/d}$ in that term.  For this reason it is awkward to define the flowing operator at $\lambda = 0$.  However, this is also unnecessary since our definition of the flow involves an integral over positive values of $\lambda$.  We therefore proceed directly to this case.

\subsubsection*{Finite $N$, finite $\mu$:}

Unfortunately, we cannot directly apply the CFT definition of normal ordering at finite values of $\mu$, because at finite $\mu$ the theory is not a CFT in the ultraviolet, and so we do not have access to a well-defined OPE if we take the limit $\mathbf{y} \to \mathbf{x}$ in a careless manner.

We would like to still be able to have a well-defined kinetic operator $G_{abcd}\!:\!\Pi^{ab} \Pi^{cd}\!:\!(\mathbf{x})$ operator at finite $N$ and finite $\mu$; otherwise our theory would not be probing bulk quantum gravity. We will attempt to do this directly for finite value of the deformation parameter $\mu$, using a point-splitting prescription. %In our procedure we start with the Hamiltonian constraint, in which we must define the kinetic operator $G_{abcd}:\!\Pi^{ab} \Pi^{cd}\!:(\mathbf{x})$ as in \eqref{NO}, by subtracting the 2-point function evaluated directly on the $T^2$ deformed theory $Z^{(\mu)}[g,\{\chi\}]$. 
However, we cannot take the point separation close to the Planck length, as this regime is not understood.  The Planck length scales as $l_\text{Pl} \sim \mu^{1/d}/N^p$ for some power $p$ (which depends on the particular AdS/CFT duality).  So we define a dimensionless parameter $\varepsilon$, such that the proper distance $|\mathbf{x}-\mathbf{y}|=\mu^{1/d}\varepsilon$ and $\frac{1}{N^p}\ll\varepsilon\ll 1$. In this way, the separation is much smaller than the AdS scale but much larger than the Planck scale.  Of course this is only possible when $N \gg 1$.

We now consider an OPE expansion of $\Pi^{ab}(\mathbf{x}) \Pi^{cd}(\mathbf{y})$, in order to rewrite it in terms of local operators in the deformed theory. This OPE will in general have relevant, marginal and irrelevant terms. Since $\mu$ is the only scale, the relevancy is entirely determined by the power of $\mu$.  Any relevant terms must be cancelled by the addition of counterterms. The remainder can then be carried through into the definition of ${\cal H}(\mu)$ or (after substituting $\mu \to \lambda$) into the definition of $O(\lambda)$, which will then be a well-defined operator at each value of $\lambda > 0$.\footnote{This definition of $:\!\Pi^{ab} \Pi^{cd}\!:$ becomes singular at $\lambda = 0$, but the elimination of relevant terms ensures that the integral over $\lambda$ values is convergent (up to a possible log divergence, which would cancel with the CFT anomaly as usual).} We can then perform the path-ordered exponential \eqref{deformingflow1} in order to define $Z^{(\mu)}$.\footnote{In the above procedure it is not manifest that the closure relation \eqref{closure} for the Hamiltonian constraint (and hence deformation flow) continues to hold.  Most likely it is necessary to add additional counterterms to eliminate all failures of closure order-by-order in $\varepsilon$.  Presumably this is always possible if the bulk theory has no gravitational anomaly.} 

%eliminated by counterterms (these terms are similar to those included in  $\text{CT}(\mu)$ in section \ref{thegeneraldeformation}.) 

%point-splitted operator of the form $G_{abcd}\Pi^{ab}(\mathbf{x}) \Pi^{cd}(\mathbf{y})$, with $|\mathbf{x}-\mathbf{y}|=\mu^{1/d}\epsilon$.

In order to ensure that the momentum constraint ${\cal D}_a = 0$ is satisfied, it is necessary to ensure that the point-splitting prescription is done in a fully covariant manner.  One possible way to do this, if $g_{ab}$ is sufficiently smooth, is to average over all possible geodesic segments of length $\mu^{1/d}\epsilon$ emanating from $\mathbf{x}$:
\begin{equation}
    G_{abcd}(\mathbf{x})\Pi^{ab}(\mathbf{x}) \int d\Omega\; \tilde{\Pi}^{cd}(\mathbf{y}(\Omega)),
\end{equation}
where $d\Omega$ is the rotationally-invariant measure, $\mathbf{y}(\Omega)$ is the point which is a unit proper distance along a geodesic of length $\mu^{1/d}\epsilon$ at angle $\Omega$,  and $\tilde{\Pi}^{cd}$ is obtained by parallel transporting $\Pi^{cd}$ along the geodesic.\footnote{Recall that our normal ordering prescription takes all metric dependence in the covariant derivative to the left of $\Pi^{ab}(\mathbf{x})$.  This means that the geometry used to define the covariant point-splitting procedures is done using the background metric, prior to the variations associated with $\Pi^{ab}$.}  An alternative, smoother regulator would be to use the heat kernel, in which case the point-split operator would be
\bea
G_{abcd}(\mathbf{x})\Pi^{ab}(\mathbf{x})\int d^d{\mathbf y} \sqrt{g}\; K^{cd}_{ef}(s, {\mathbf x},{\mathbf y}) \Pi^{ef}({\mathbf y}),\qquad 
\sqrt{s} = \mu^{1/d}\varepsilon,
\eea
where $K^{ab}_{cd}(s, {\mathbf x},{\mathbf y})$ is the heat kernel coefficient at Schwinger time $s$ associated with some wave equation which propagates symmetric rank-2 tensors.

There can also be UV divergences in the theory associated with the limit as two or more \emph{distinct} deformation operators (like $:\!\Pi^{ab}\Pi^{cd}\!:$ or $G^{ab}\Pi^{cd}$) approach one another.  These divergences can appear at various powers of $\mu$ and with various curvature couplings.  We assume that any such divergences are also regulated using a cutoff of length $\mu^{1/d}\epsilon$, in order to ensure that 1) the deformation retains the correct dimensional scaling with $\mu$, and 2) the theory is allowed to exhibit locality at sub-AdS scales.

\subsection{Exotic Properties of the Deformed Theory}\label{exotic}
The $T^2$ theory exhibits some exotic phenomena which violate the usual axioms of QFT: nonunitarity and (relatedly) spontaneous CPT violation.  These phenomena should be welcomed as they are crucial for describing the regime in which the Cauchy slice $\Sigma$ is embedded in Lorentzian signature.\footnote{Another exotic property of the Lorentzian signature $T^2$-deformed field theory is that it allows superluminal signalling.  However, we do not analyze this property here since we are primarily interested in the Euclidean signature $T^2$-deformed theory, where this issue does not arise.}

\subsubsection*{Nonunitarity}\label{nonuni}
The Euclidean Wick rotation of a unitary QFT is reflection positive \cite{Osterwalder:1973dx}.  However, the $T^2$ theory is not reflection positive.

One way to see the violation of unitarity is to calculate the energy levels of the $T^2$-deformed theory on a stationary spacetime e.g. a cylinder $S_{d-1} \times \mathbb{R}$, where the sphere has radius $R$ and volume $\Omega_{d-1}R^{d-1}$.  Hartman, Kruthoff, Shaghoulian and Tajdini  \cite{Hartman:2018tkw} did this calculation for states of zero angular momentum, obtaining (in our notation):
\begin{equation}\label{spectrum}
    E(\mu;R)=-2(d-1)\alpha \Omega_{d-1}\frac{R^{d-1}}{\mu}\;\sqrt{1-\frac{\mathcal{E}\mu}{(d-1)\alpha \Omega_{d-1}R^d}+\frac{(d-2)\mu^{2/d}}{R^2}},
\end{equation}
%\begin{equation}
 %   E(\mu;R)=-\frac{(d-1)R^{d-1}\Omega_{d-1}}{2d\mu}\sqrt{1-\frac{4d\mathcal{E}\mu}{(d-1)\Omega_{d-1}R^d}+\frac{2(d-2)d\mu^{2/d}\alpha_d}{R^2}}
%\end{equation}
where the $\mu=0$ boundary condition is $E(\mu=0;R)=\mathcal{E}/R$, with $\mathcal{E}$ being the energy quantum number of the undeformed CFT. This result is only valid in the large-N limit of the theory. The expression above is the result after shifting by the counterterm $\text{CT}(\mu)$, corresponding to the case of pure gravity bulk.

One can see that there is a branch cut when the energy $E = 0$, at some critical value of $\cal E$.  At this critical value, the microcanonical density of states $\rho$ is given by
\bea
\ln \rho(E=0) = S_\text{BH} = \frac{A}{4G_N},
\eea
the black hole entropy.  Higher $\mathcal{E}$ states of the CFT are mapped to imaginary eigenvalues of $E$, for which $S > S_\text{BH}$.   
%At a finite value for the $T^2$ coupling $\mu$, there is a branch cut in the spectrum of the deformed theory.  As a result, there is a maximum CFT energy, beyond which the $T^2$ spectrum becomes complex.  
\cite{McGough:2016lol} argued that these imaginary eigenstates should simply be truncated from the spectrum.  However, this approach is incompatible with the continuation to Lorentzian signature, since it would forbid a timelike sign for the extrinsic curvature $K_{ab}$.  Hence, in order to describe holography on Cauchy slices, it is important to keep these states in the spectrum \cite{Wall_CEB}.  It follows that the theory that lives on the Cauchy slices is inherently nonunitary.\footnote{This type of nonunitarity, associated with a complex Hamiltonian, should not be confused with the totally distinct idea that pure states might evolve to mixed states!  Hence, our holographic model of Cauchy slices cannot be used to argue that information is lost inside black holes, as Hawking originally argued \cite{Hawking:1976ra}.  In fact our model implies the exact opposite: all the information on any Cauchy slice $\Sigma$ ``flows" outward to the boundary at infinity.  See section \ref{unitarity}.}

In our conventions, the field theory always lives on a Euclidean slice $\Sigma$, and the branch cut is associated with changing the signature of the bulk spacetime that $\Sigma$ is embedded in.  (Imaginary $E$ gives a Lorentzian bulk.)\footnote{There is an alternative picture, associated with the radial $T^2$ deformation, in which the bulk spacetime is always Lorentzian.  In this picture, the imaginary values of $E$ correspond to going inside the horizon of a bulk black hole \cite{McGough:2016lol}.  In this picture it is the signature of $\Sigma$ that changes since radial slices inside the black hole are Euclidean.}  See Figure \ref{phase_transition} for an example of this transition (but for a more general $\Sigma$ geometry).

This non-unitarity may seem surprising, but it was argued in \cite{Wall:2021bxi} that the theory of gravitational subregions is necessarily nonunitary, due to the spontaneous breaking of $CPT$ symmetry that is associated with Cauchy slices in Lorentzian spacetime.  Furthermore, dS/CFT dualities, which also have an emergent time direction, are known to be nonunitary \cite{Strominger:2001pn,Anninos:2011ui, Araujo-Regado:2022jpj}.  So we have to learn to live with this.

Fortunately, it turns out that the nonunitarity of the boundary $T^2$ field theory is compatible with unitarity of the bulk theory \ref{unitarity}.  Actually, bulk unitarity emerges in a natural way from the unitarity of the starting CFT, together with the fact that CPT symmetry is not explicitly broken.

\subsubsection*{Spontaneous CPT breaking}\label{CPT}

$CPT$ is however, spontaneously broken in the boundary theory!  Another way of looking at the branch cut in \eqref{spectrum}, is that it corresponds to a transition between two phases with different symmetry properties.  When $\Sigma$ is embedded in a Euclidean bulk, this corresponds to a time-reversal symmetric phase of the gravitational path integral, while a  Lorentzian signature bulk corresponds to a time-reversal violating phase of the gravitational path integral \cite{Wall:2021bxi} (see also \cite{Araujo-Regado:2022jpj} for the analogous discussion in cosmology).  

In fact even the combined $CPT$ symmetry is broken.  In pure gravity, $C$ acts trivially, while $P$ is unbroken by the above construction.  Hence the $T$-violating phase also breaks $CPT$.\footnote{Note that on a given Cauchy slice $\Sigma$, $CPT$ means the same thing for the bulk and boundary theories, even if we add matter.  For $P$ this is obvious. In Euclidean signature, $T$ acts as complex conjugation, which reverses the sign of $\Pi_{ab}$ due to the imaginary sign in \ref{Pi}; so this looks just like time reversal in the bulk.  And $C$ acts trivially if we write the fields in a real basis.}  If we include both possible signs for the imaginary energy eigenvalues (as advocated in \cite{Wall_CEB}), this can be regarded as a form of spontaneous symmetry breaking.  This is of course not possible in a unitary, rotationally-symmetric theory, by the $CPT$ theorem.  But since the energy eigenvalues are imaginary, it is clear the $T^2$ theory is \emph{not} unitary, so this is no contradiction.

The boundary between the two phases is given by a moment-of-time symmetric slice (sometimes called a ``t = 0'' slice), which lies on both the Euclidean and Lorentzian contours.\footnote{However, because of the branch cut, the boundary between the two phases is \emph{not} a standard second order phase transition corresponding to a CFT.  (In fact it is even less smooth than a normal 1st order phase transition.)  The theory on the boundary is highly non-Gaussian, which is related to the fact that, in the gravitational dual, the semiclassical approximation breaks down there, due to the absence of an elliptic boundary condition for the gravitons \cite{Witten:2018lgb}.  In a minisuperspace approximation, the region near the $t = 0$ slice looks like an Airy function $\text{Ai}(x)$, and semiclassical approximation is just the WKB approximation.  This approximation diverges at the transition between the exponential and oscillatory phases, although $\text{Ai}(x)$ itself is perfectly well-defined there.  This may have significant implications for tensor network models of AdS/CFT constructed on the $t = 0$ slice (cf. section \ref{TN}).}

Similar phase transitions ought to occur for non-stationary backgrounds as well, in order to match the dual gravitational path integral, although in such cases it is no longer expedient to analyze them using energy eigenvalues.  Let us be very explicit about the following point:

\medskip

\emph{In order for our proposed AdS/CFT dictionary in this paper to make sense, it is necessary to assume that the $T^2$-deformed theory continues to make sense on the Lorentzian side of this phase transition!}

%\pagebreak[4]

\section{Towards a Generalized Holographic Principle}\label{sectionGHP}

From this point onwards we will be dealing with a pure gravity bulk for simplicity of notation. Matter fields can be added everywhere with minimal modifications.\nopagebreak

\subsection{Form of the Duality}\label{form}

Motivated by the AdS/CFT correspondence \cite{Maldacena:1997re, Witten:1998qj, Aharony:1999ti}, holography at finite cutoff \cite{McGough:2016lol} and the realization that deformed field theories define WDW-states \cite{Hartman:2018tkw, Belin:2020oib, Caputa:2020fbc}, including on Cauchy slices, we now postulate a generalized form of the holographic duality for partition functions with arbitrary boundaries:

%Consider the following: 
%\begin{itemize}
    %\item The set of all manifolds $\{ \mathcal{M} \}$ with the same boundary $\partial \mathcal{M}$.
    %\item We have a $T^2$-deformed field theory living on $\partial \mathcal{M}$ with an arbitrary background complex metric $g$ (e.g. it could be Lorentzian in some parts of $\partial \mathcal{M}$ and Euclidean in some other parts of $\partial \mathcal{M}$, with a smooth complex transition).
 %  \item Let's choose a co-dimension one surface $\Sigma$ in $\partial \mathcal{M}$, such that this surface is fully Euclidean with respect to the induced metric (take $g$ on $\partial \mathcal{M}$ and induce it on $\Sigma$). 
%\end{itemize}
\medskip

\begin{figure}[H]
\centering
\begin{tikzpicture}
\node[inner sep=0] at (-3,0) {\includegraphics[width=.22\textwidth]{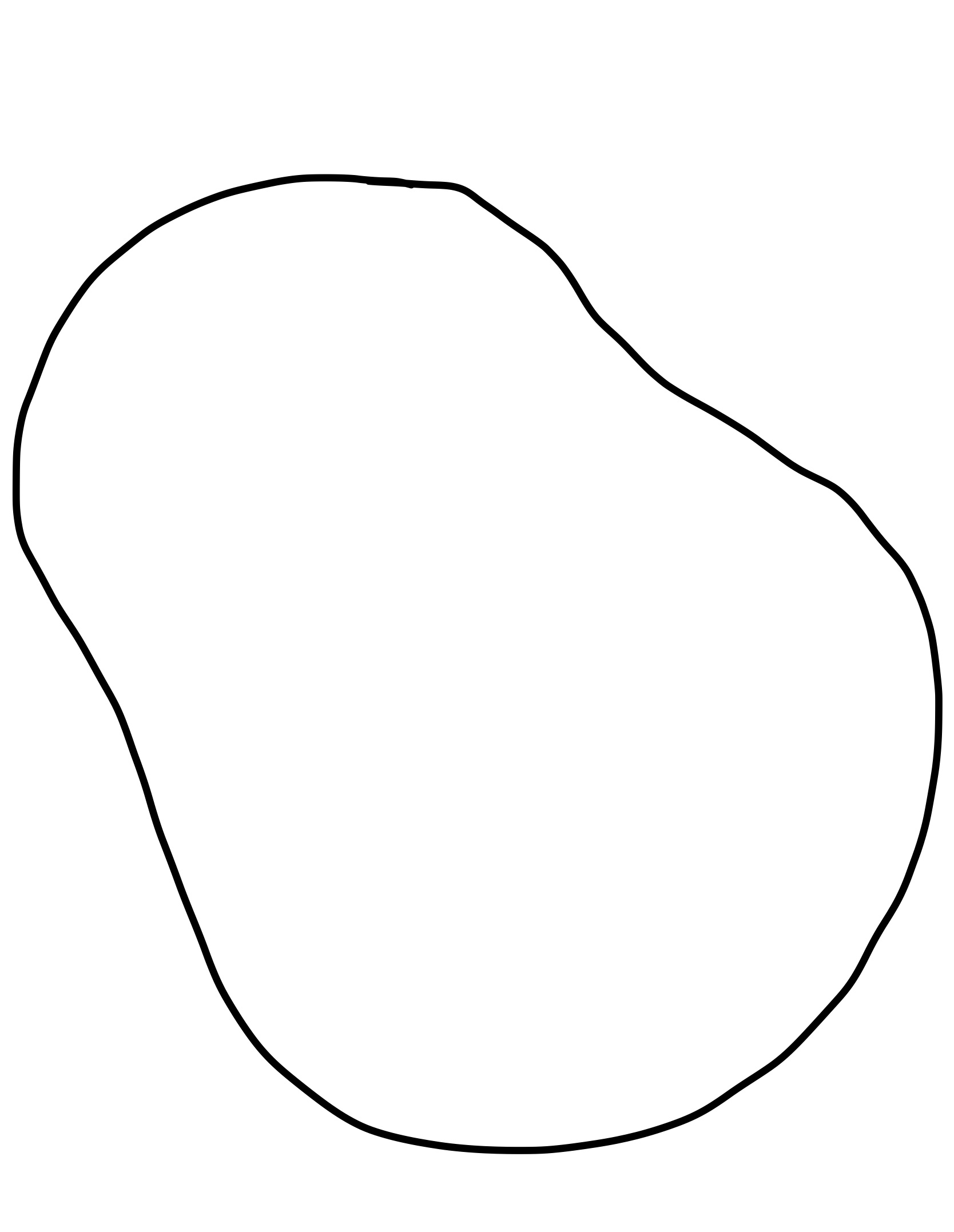}};
\node[inner sep=0] at (3,0) {\includegraphics[width=.22\textwidth]{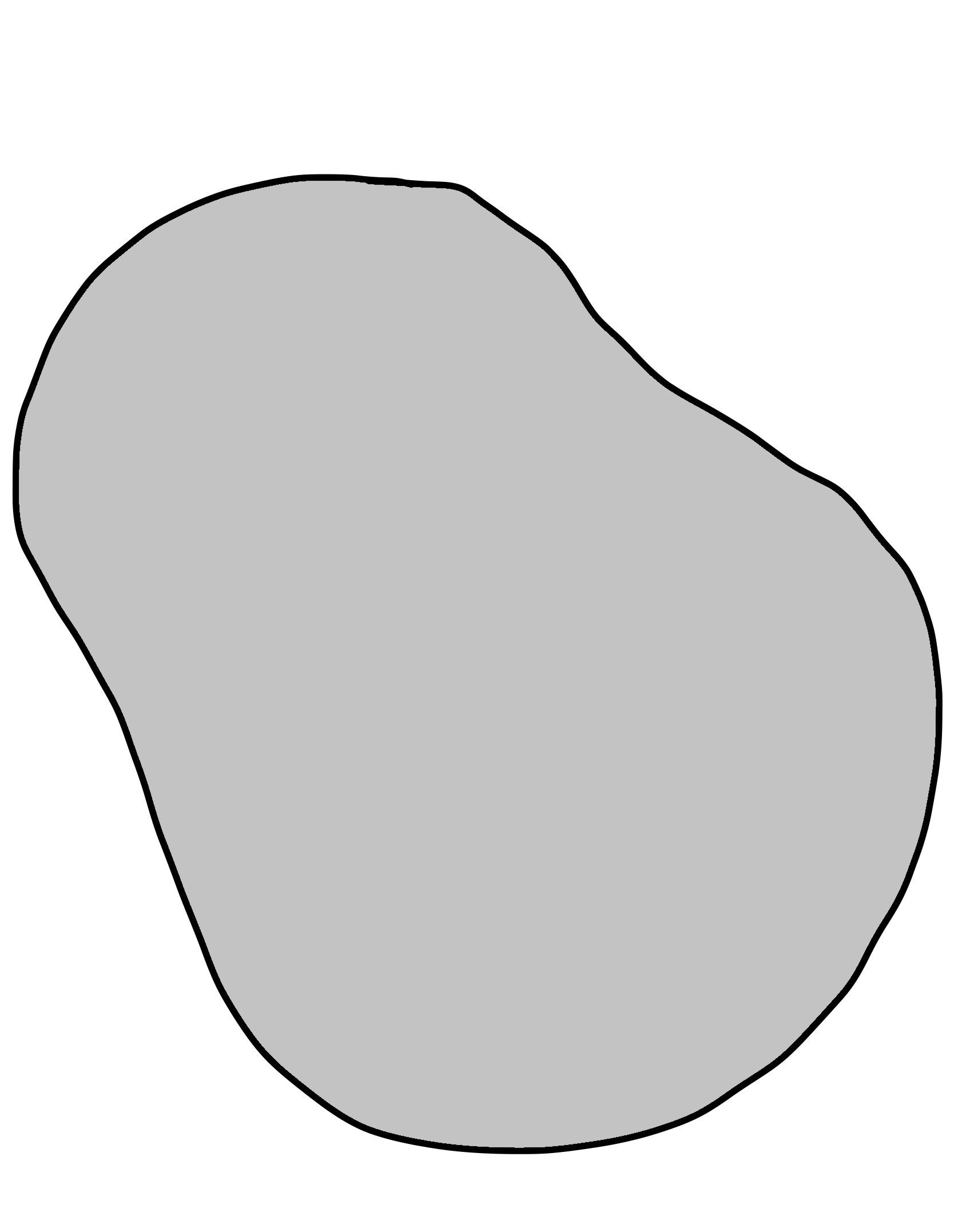}};
\node at (0,0) [fontscale = 5]{$\textbf{=}$};
\node at (3,0.3) [fontscale=1]{Bulk metric};
\node at (3,-.3) [fontscale=3]{$\textbf{g}$};
\node at (-3.5,2) [fontscale=1]{Boundary metric $g$};
\node[align=center] at (-3,-3) {Partition function \\
of the field theory};
\node[align=center] at (3,-3) {Gravitational path integral \\ with Dirichlet boundary conditions \\
$\mathbf{g}|_{\partial\mathcal{M}} = g$};
\draw (-4.05,-1.1) to  (-5,-0.8)[->,>=stealth'];
\node at (-5.5,-0.8) [fontscale=2]{$\partial \mathcal{M}$};
\draw (4,-1.1) to  (5.55,-0.8)[->,>=stealth'];
\node at (6,-0.8) [fontscale=2]{$\mathcal{M}$};
\end{tikzpicture}
\caption{\small The figure on the left represents the partition function of the $T^2$-deformed field theory living on $\partial \mathcal{M}$ (the closed curve) with a background metric $g$. The \textbf{shaded} figure on the right represents the gravitational path integral over the space filling manifold $\mathcal{M}$. The bulk metric $\mathbf{g}$ satisfies the Dirichlet boundary conditions (i.e $\mathbf{g}|_{\partial\mathcal{M}} = g)$.}\label{Duality_hypothesis}
\end{figure}
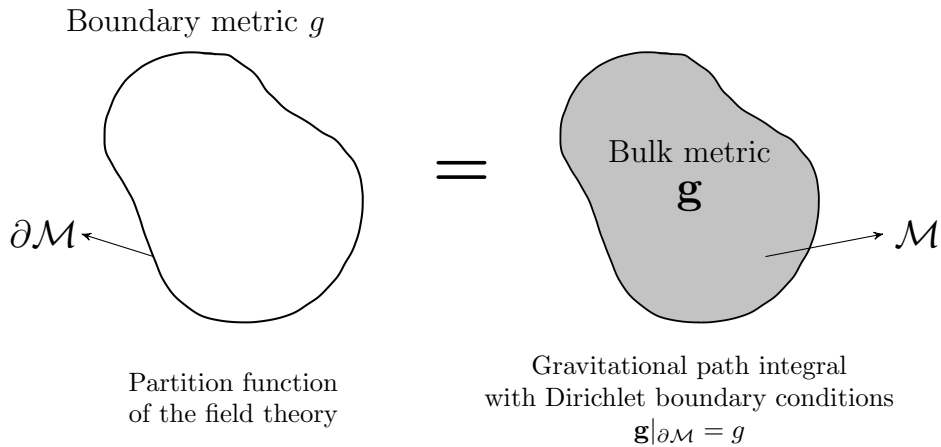

\textbf{The Generalized Holographic Principle (GHP)}: This hypothesis states that the $T^2$-deformed partition function of this boundary theory is equal to the gravitational path integral over space-filling manifolds $(\mathcal{M}, \mathbf{g})$ with Dirichlet boundary conditions on $\partial\mathcal{M}$, as shown in Fig.   \ref{Duality_hypothesis}.  (A more precise definition of the right-hand side will be provided in section \ref{QG Inner Product}.)
\begin{equation}\label{GHP}
    \boxed{Z_{T^2}^{(\partial \mathcal{M})}[g] = 
\sum_\mathcal{M} \int_{\mathbf{g}|_{\partial\mathcal{M}} = g}
\frac{D\textbf{g}}{\text{Diff}(\mathcal{M})}\,e^{ iI_\text{grav}[\textbf{g}]}}
\end{equation}
Here, on the LHS, we have a $T^2$-deformed field theory living on $\partial \mathcal{M}$ with metric $g$.  This could potentially be a complex metric (e.g. it could be Lorentzian in some parts of $\partial \mathcal{M}$ and Euclidean in some other parts of $\partial \mathcal{M}$).

On the RHS, we sum over possible manifolds  $\mathcal{M}$ and bulk metrics $\mathbf{g}$ that are compatible with our boundary conditions{\bf---}albeit with the usual caveats and controversies associated with the gravitational path integral, which are discussed in the remainder of this section.

An initial reason to believe this hypothesis is that we have already shown that the $T^2$ theory obeys the constraint equations of general relativity.  So morally speaking, we are simply requiring that this duality also hold at the level of gravitational path integrals with a single boundary.  However, we do not in this paper claim to show the equivalence between the canonical and path-integral approaches to quantum gravity{\bf---}particularly since the path-integral does more than just project onto states satisfying the constraint equations $\mathcal{H} \Psi = \mathcal{D}_a \Psi = 0$; it also provides a dynamical definition of the inner product between two states, which will be constructed more explicitly in section \ref{QG Inner Product}.\footnote{A careful analysis of the relationship between the $T^2$ deformation and the path integral was done in \cite{Belin:2020oib}, but in their analysis the manifold ${\cal M}$ lies in between $\partial {\cal M}$ and the usual CFT boundary, so it is the ``inverse'' of the usual path integral.  This could be taken as an argument for the GHP, starting with the original form of the AdS/CFT equivalence.}

\subsection{The Need for a Contour Prescription}\label{contour}

What we have said does not quite suffice to fully specify the Generalized Holographic Principle.  For one thing, there are controversies about the correct choice of contour of the integral over metrics in quantum gravity.  As indicated above, we sometimes want to use complex contours, for example those associated with Euclidean signature saddles.  But in the literature, there are (at least!) two divergent philosophies for dealing with these Euclidean saddles:
\begin{enumerate}
\item Hartle and Hawking \cite{Hartle:1983ai} assume that, in cases where a Euclidean saddle is necessary, we always rotate the contour in the direction of the standard Wick rotation $\tau = it$, even in cases where the resulting saddle is exponentially enhanced.  This has the advantage that the horizon entropy is always $S = A/4G$, but sometimes the connection to the Lorentzian picture is unclear.

Fluctuations around such Euclidean saddles are not bounded below, because of the `conformal mode problem', where conformal modes have negative action.  So to calculate the quantum determinant $\Delta_\text{1-loop}$ one must also deal with this, perhaps by rotating the contour of the conformal mode to imaginary values \cite{Gibbons:1978ac}. 

%\textcolor{red}{There is this paper by Mazur and Mottola in which they claim that the conformal mode problem is not there if one deals with the Jacobian factor properly and then do the usual Wick rotation \cite{mazur}. I have only read the introduction and it seems quite important.}

\item Other authors, including Vilenkin \cite{Vilenkin:1984wp,Vilenkin:1986cy} and Feldbrugge, Lehners, \& Turok \cite{Feldbrugge:2017kzv}, have argued that one should instead take the Lorentzian definition of the path integral as fundamental.  In this approach, the contour may be taken to be complex only when, using Picard–Lefschetz theory, one can show that the resulting saddle is a resummation of oscillatory Lorentzian contributions.

Invariably, the saddles that resum an oscillatory path integral are exponentially suppressed, rather than exponentially large.  In some cases, this forces you to take the opposite sign of imaginary time, corresponding to $\tau = -it$.  In such anti-Euclidean saddles, while the conformal modes have positive action, the transverse-traceless modes (corresponding to gravitational waves) and ordinary matter fluctuations would have negative action, leading to inverse Gaussian, unsuppressed fluctuations in the resulting state \cite{Feldbrugge:2017fcc}.\footnote{Cf. \cite{Vilenkin:2018dch} for a proposed solution to this problem.  On the other hand, \cite{Feldbrugge:2017mbc}
argued that, in the context of dS cosmology, there is \emph{no} choice of contour that leads to physically good results!  But even if this no-go result is valid, it would not necessarily rule out a holographic cosmology theory defined on Cauchy slices, since the holographic theory only needs to agree with a semiclassical calculation in the neighborhood of some particular saddle{\bf---}whichever one \emph{it} agrees with{\bf---}and could be quite different from any semiclassical calculation in the far off-shell regime.}  Furthermore, this choice leads to an unphysical value of the Gibbons-Hawking entropy in AdS or dS (corresponding to $-A/4G$ instead of $+A/4G$).
\end{enumerate}

In our opinion, the Hartle-Hawking approach gives results that seem more physically reasonable for defining a ground state.  Since the conformal mode corresponds to a non-propagating degree of freedom, it can always be gauge fixed by a suitable choice of the time variable.  On the other hand, having unnormalizable fluctuations for the \emph{physical} modes around a supposedly dominant saddle, in vacuum dS or AdS, would be disastrous, and of course most of us will want the horizon entropy to have the correct sign.

%The dual $T^2$-deformed theory may be helpful in resolving this controversy.  In cases where there are multiple possible contours, instead of \emph{assuming} which saddle and contour we should use, we could instead simply calculate the $T^2$ partition function and \emph{see} which saddle it corresponds to.  

Of course, the $\mu \to 0$ limit of the $T^2$ field theory corresponds to going back to the original CFT.  And it is generally accepted that in AdS/CFT, a Euclidean CFT is dual to the Hartle-Hawking saddle, since this is the saddle that gives rise to the correct boundary entropy and free energy.  So by continuity, it appears that the $T^2$ field theory must force us to pick the Hartle-Hawking saddle.

For a discussion of the related contour question in the context of $T^2$ deformed cosmology, see \cite{Araujo-Regado:2022jpj}.

\subsection{Challenges to the Validity of the GHP}\label{challenge}

Having said all this, there are also a few potential reasons to doubt the GHP as we've stated it.  Here are some of them:

\paragraph{1. Nonrenormalizability (and other technical problems):}

Obviously, the gravitational path integral is subject to all the usual ambiguities of quantum gravity, including nonrenormalizability of the UV divergences, among other technical issues.  These issues should be resolvable by taking the bulk theory to be a UV complete model, such as (presumably) string theory.

Other potential technical problems include the oscillatory nature of the path integral in Lorentzian signature, the fact that the Euclidean action is unbounded below; questions concerning singularities and topology change, and possible IR issues due to the unrestricted range of time integration.  To define a precise duality at the nonperturbative level, we would obviously need to assume that such issues can be resolved.

\paragraph{2. Factorization Problem:}

If $\partial {\cal M}$ contains 2 disconnected boundary components (i.e. ${\partial \cal M} = B_1 \sqcup B_2$), then any local field theory partition function must factorize:
\begin{equation}\label{Zproduct}
Z[\partial {\cal M}]:= Z[B_1 \sqcup B_2] = Z_1[B_1]Z_2[B_2].    
\end{equation}

In the usual way of doing holography one runs into the problem that it is not at all manifest that the gravitational path integral has this factorization property, since it would appear to include sums over geometries in which the bulk $\cal M$ contains a single connected component, because one or more ``baby universes'' connect the two sides.  For the analogous problem of disconnected CFT boundaries, in some cases one can use curvature conditions to argue \cite{Witten:1999xp} that there are no on-shell saddles violating factorizability \eqref{Zproduct}, but even then one might worry about the contributions from off-shell metrics.\footnote{See \cite{Maldacena:2004rf,Marolf:2021kjc} for further discussion of Euclidean AdS wormholes solutions.}

However, it has also been argued that the sole effect of such baby universes is to shift the value of the allowed counterterms in the local gravitational action \cite{coleman1988black,giddings1988}.  In a bulk theory such as string theory which has no free parameters, this means that baby universes processes cannot really affect anything at all!  Indeed, if one accepts the GHP stated in \eqref{GHP} in complete generality, a baby universe{\bf---}or more generally any closed universe{\bf---}would correspond to a closed partition field theory function $Z$, which can always be evaluated to get a complex number.  This would imply that baby universes cannot in fact transmit any information; in other words any connected geometry in the path integral should be gauge-equivalent to some process in which the two sides are disconnected.  (It might be, however, that the equivalent disconnected geometry requires using some type of end-of-the-world brane, or other structures, which only appear in string theory.)\footnote{Recently there has been interest in relating the gravitational path integral to ensembles of theories with varying coupling constants, because that allows disconnected partition functions not to factorize.  This is of no use with classic forms of AdS/CFT where the CFT coupling constants take definite values.  But if e.g. the pure gravity path integral does not factorize, this could be because it represents a nontrivial ensemble over holographic theories \cite{Saad:2019lba,Stanford:2019vob,Marolf:2020xie,Bousso:2020kmy}.  This is probably why baby universes can play a nontrivial role in the replica wormhole calculations of the Page curve \cite{Penington:2019kki,Almheiri:2019qdq,Engelhardt:2020qpv}.}

%It is to address this issue that we formulated the GHP for connected boundaries only.\textcolor{red}{Is this really likely to be sufficent?  What if the two sides are connected by a tiny wormhole?} For the case of disconnected boundary, we claim that the factorization property holds for any pure WDW-state, where by pure here we mean that there is no entanglement between the two components of the bipartite system. It is only when we consider entangled states (e.g. the state $Z_1[B_1]Z_2[B_2]+Z_1'[B_1]Z_2'[B_2]$) that, if we are to approximate it by a saddle geometry, we will obtain a wormhole connecting the two disconnected boundaries. This is an interesting direction to consider in future work.

\paragraph{3. Information Paradox:} 

Every field theory partition function generates a state of the boundary.  We will define this bulk-to-boundary map more explicitly in section \ref{dictionary}.  

Because of this map, the $T^2$-deformed theory makes it manifest that any information which falls across the horizon of a one-sided black hole (because it lies on a Cauchy slice which is connected to the boundary) is still holographically encoded in the boundary theory.

In this respect it is opposed to the usual geometrodynamics perspective, which in an evaporating black hole regime makes this information escape seem unlikely \cite{Unruh:2017uaw}, but see \cite{Marolf:2008mf,Raju:2019qjq,Chowdhury:2021nxw,Chowdhury:2020hse}.  One way that the information can nevertheless escape, is if the $T^2$ partition function implements additional constraint equations on the spacetime (associated with the finite dimensionality of its Hilbert space at fixed energy) \cite{Wall_CEB} which the purely Lorentzian geometrodynamics theory knows nothing about.\footnote{Although it has been argued that Euclidean path integrals implement some nontrivial additional constraints \cite{Jafferis:2017tiu}.}  If this is true, and if we take the GHP to be valid, then this imposes restrictions on the regime of validity for a semiclassical approximation of the gravitational path integral. For example, this tells us that for an evaporating black hole after the Page time, the semiclassical approximation becomes invalid, and we should instead use the $T^2$ theory to study how the information escapes to the boundary.

\subsection{$T^2$ as the Definition of the Bulk Theory?} \label{defofbulktheory}

In order to deal with all of these problems, we shall adopt the following point of view:  We assume that as long as we evaluate \eqref{GHP} semiclassically, using metrics that are close to dominant classical saddles, that both sides of the duality are well-defined.  In the field theory, this corresponds to a large N expansion.  In that semiclassical regime, this duality could potentially be falsified, or require modifications, if the two sides were found to disagree.  

As an even more ambitious proposal, we suggest that the GHP is likely to be valid at all orders in a 1/N expansion.  This corresponds to an $\hbar G$ expansion in the bulk.

On the other hand, whenever we go \emph{beyond} the regime in which geometrodynamics is likely to be valid, we can take the $T^2$ boundary theory as a \emph{definition} of the gravitational amplitude.\footnote{A similar attitude has sometimes been proposed towards the standard AdS/CFT duality, that the CFT should simply be regarded as the \emph{definition} of quantum gravity.  This is unsatisfying because it gives no way, even in principle, to answer any of the most perplexing questions about quantum gravity, like what happens behind horizons or near singularities.  On the other hand, taking the $T^2$-deformed theory as the definition of quantum gravity seems potentially more fruitful, since the $T^2$-deformed theory is \emph{already} embedded inside the AdS bulk{\bf---}so it doesn't require you to give up on the idea of having a physical picture which is valid deep inside event horizons!} 
Of course, this attitude raises the question of whether the $T^2$-deformed theory \emph{itself} makes sense non-perturbatively.  The fact that the $T^2$ couplings are irrelevant strongly suggests that it may not be, except in those special cases ($d = 2$ or $N = \infty$) where it is exactly solvable.\footnote{Another potential problem is that the $\Pi_{ab}^2$ terms look like inverse diffusion, which is not always well-defined after a finite amount of flow.  It might be that analytically continuing around such obstructions is the origin of the branch cuts in the $T^2$ spectrum.  It would be helpful if someone could do an explicit calculation to confirm or deny this hypothesis.}  

If it is not fully well-defined, then the theory would need to be UV completed somehow, just like its associated bulk quantum gravity theory.  However, it is probably much easier to UV complete a field theory defined on a fixed Euclidean background spacetime, than it is to UV complete a background-independent model of quantum gravity in Lorentizan signature!  We will say more about this issue in the Discussion.

\section{QG Inner Product from the Path Integral} \label{QG Inner Product}

In section 4.1, we start by defining the Lorentzian path integral from a purely gravitational perspective, without making use of the GHP proposed in \eqref{GHP}.  

Section 4.2 describes the importance of integrating over both signs of the Lorentzian lapse $N$ in order to preserve gauge-invariance.  At the end we make a few comments about the possibility of non-Lorentzian contours.  In section 4.3 we discuss the example of maximally symmetric metrics $g$ to illustrate how both Euclidean and Lorentzian contours can arise.

Finally, in section 4.4, we discuss some additional important properties of the amplitude.  It is quite easy to show that the amplitude is linear and hermitian.  We also present an argument that the path integral is positive, at least in a saddle-point approximation, when $\Lambda < 0$.  We also discuss the issue of when the amplitude is finite.  (These properties will all be important in section 5, when we construct the Hilbert space.)

\subsection{Lorentzian Transition Amplitude}\label{Lorentzian}

We will start by defining the Lorentzian transition amplitude (but we will later discuss how things generalize if the contour becomes complex).

We can formally define the Lorentzian transition amplitude as:
\begin{equation}\label{TA}
\langle g_2 \,|\, g_1 \rangle_I =
\sum_\mathcal{M} \int^{\mathbf{g}|_{\Sigma_2} = g_2}_{\mathbf{g}|_{\Sigma_1} = g_1}
\frac{D\textbf{g}}{\text{Diff}(\mathcal{M})}\,e^{ \pm iI_\text{grav}[\textbf{g}]},
\end{equation}
where $\mathbf{g}$ is any $D = d+1$ dimensional Lorentzian metric, on a globally-hyperbolic submanifold ${\cal M} \subset {\cal B}$ of a possible asymptotically-AdS bulk geometry on ${\cal B}$.  Note that the metric $g_{ab}^\text{bdy}$ on $\partial {\cal B}$ is fixed, so the asymptotic behavior of $\mathcal{B}$ is given by a Fefferman-Graham expansion, whose leading order behavior is given by \eqref{FG0}.

$\mathcal{M}$ is the manifold connecting $\Sigma_1$ to $\Sigma_2$.  We require that $\Sigma_1$ and $\Sigma_2$ are anchored to the same boundary Cauchy slice $\Sigma^\text{bdy} \subseteq \partial \cal B$ (i.e. a Cauchy slice of the usual boundary CFT).  $\text{Diff}(\mathcal{M})$ is the group of spacetime diffeomorphisms acting in the interior of ${\cal M}$.\footnote{For purposes of doing explicit calculations it is often convenient to gauge-fix in order to remove the (noncompact) volume integral over the diff group $\text{Diff}(\mathcal{M})$ in the denominator, which necessitates the introduction of Faddeev-Popov ghosts.  Since our present concern is not to manipulate the amplitude into a convenient form for calculation, we will retain the current form, sometimes called ``unitary gauge''.} \footnote{If the topologies of $\Sigma_1$ and $\Sigma_2$ are different in Eq. \eqref{TA}, then this must be interpreted as transition amplitude between different topologies of bulk Cauchy slices, as well as different metrics.}

The gravitational action is:
\begin{eqnarray}\label{Igrav}
I_\text{grav} =
\frac{1}{16\pi G_N}
\int_\mathcal{M} \! d^Dx \sqrt{-\mathbf{g}}\, (\mathbf{R} - 2\Lambda)
\:-\: \frac{1}{8\pi G_N}
\int_{\partial \mathcal{M}} \!
d^dx \sqrt{g}\,K \nonumber \\
\:+\:
\int_\mathcal{M} \mathcal{L}_ \text{RG} + \int_{\partial \mathcal{M}} \mathcal{L}^\text{bdy}_\text{RG}
+\int_{\Sigma^\text{bdy}} \mathcal{L}_\text{lapse}
,
\end{eqnarray}
where $\partial \mathcal{M} = \Sigma_1 \cup \Sigma_2$, $\mathcal{L}_ \text{RG}$ and $\mathcal{L}^\text{bdy}_ \text{RG}$ refer to RG counterterms needed to cancel UV divergences of bulk fluctuations.

$\mathcal{L}_\text{lapse}$ refers to counterterms needed to deal with IR divergences of the action associated with the corner terms between the Cauchy slices $\Sigma_1$ and $\Sigma_2$ as one approaches the asymptotic AdS boundary $\partial {\cal B}$.  In Fefferman-Graham coordinates \eqref{FG0}, we may define a conformally-rescaled relative lapse as $\tilde{N} = zN$, where $N$ is the proper time difference between $\Sigma_1$ and $\Sigma_2$.  There is then a divergence in the spacetime volume (and hence the action) if $\tilde{N} \sim z^{d}$, or any lower power of $z$.  If we consider the class of Cauchy slices which are conformally smooth at $z = 0$, dimensional analysis indicates that only a finite number of IR counterterms will be required.\footnote{We could also restrict attention only to a subset of Cauchy slices whose relative lapse falls off quickly enough that there are no divergences.}  An alternative, somewhat more convenient approach (which we will use in section \ref{admH=cftH}) is to take $\partial \Sigma$ to be a small distance $z = \epsilon$ away from the conformal boundary $\partial {\cal B}$.

$D\mathbf{g}$ is a local covariant measure on the space of metrics, which can be defined formally as the volume induced by the following metric on the space of metrics:\footnote{There is one other possible derivative-free covariant integrand, $\sqrt{-\mathbf{g}}\,
\delta \mathbf{g}_{\mu\nu}\, \delta \mathbf{g}_{\sigma \tau}\,\mathbf{g}^{\mu\nu} \mathbf{g}^{\sigma\tau}$, which would change the measure of the conformal mode.  But since all we care about is the overall volume, this parameter can be absorbed into a rescaling of $C$.}
\begin{equation}\label{metric^2}
d \mathbf{s}^2= C \int_{\cal M} d^Dx\,\sqrt{-\mathbf{g}}\,
\delta \mathbf{g}_{\mu\nu} \,\delta \mathbf{g}_{\sigma \tau}\,\mathbf{g}^{\mu\sigma} \mathbf{g}^{\nu\tau},
\end{equation}
where $C$ is a dimensionful parameter needed to define the path integral.\footnote{$C$ is not actually an independent variable, since rescaling it to a new value $C'$ is equivalent to adding an \emph{imaginary} counterterm to the action $I$ which is proportional to $i \ln(C' / C)$ times a (regulator-dependent) local integral.  Such imaginary terms in the action are not arbitrary, but must be fixed to whatever value is required for unitarity to hold.  (This is one reason why it is simpler to regulate in Euclidean signature, where such measure-dependent counterterms are instead real.  More generally, as discussed in section \ref{unitarity}, imposing $CPT$ symmetry fixes the signs of terms in the action, in a way that guarantees that bulk unitarity cannot be spoiled by local counterterms.)}  This measure $D\mathbf{g}$ is well-defined when considering a finite-dimensional space of metric variations, so it is reasonable to think that the measure can be defined in effective field theory after imposing a UV regulator which cuts off short distance modes.

\subsection{Why we use both Signs for the Lapse}\label{lapse} 

The reason for the $\pm$ sign in front of the gravitational action, is that we are including both signs of the lapse in our gravitational path integral.

A famous paper by Claudio Bunster (formerly Teitelboim) \cite{Teitelboim:1983fh} argues that one ought to restrict the quantum gravitational path integral only to histories that have positive values of the lapse, which is equivalent to saying that $\Sigma_2$ lies to the future of $\Sigma_1$.  Bunster frames this question as a choice between ``causality'' and ``gauge-invariance'', however we respectfully disagree with his interpretation.

Bunster correctly observes that if we restrict the path integral only to positive values of the lapse, that the resulting final state will no longer obey the Hamiltonian constraint equation ${\cal H}(x) \Psi = 0$, violating gauge-invariance.  Normally, the Hamiltonian constraint is guaranteed because it is the Euler-Lagrange equation associated with the variation of the lapse $N(x)$.  But if we integrate over the domain $N(x) \ge 0$, then at the boundary of this domain (where $N(x) = 0$ for some $x$, i.e. $\Sigma_1$ and $\Sigma_2$ touch), we no longer have the ability to freely vary the lapse, hence ${\cal H}(x) = 0$ no longer follows.  (Additionally, the resulting transition amplitude would no longer be Hermitian.)  In our opinion, this is an excellent reason to include both signs of $N(x)$ in the variation. Such a contour over the whole real line (with a small deformation away from the essential singularity at $N=0$) has in fact already been proposed in \cite{DiazDorronsoro:2017hti}.

Bunster resists this conclusion by arguing that if we do include both signs of the lapse in the path integral, the theory will no longer be causal.  We do not share this opinion.  Causality has to do with the propagation of physical signals \emph{within} the manifold $\cal M$ (which is why we demand that $\cal M$ is globally hyperbolic).  It does not have anything to do with the statement that, for two abstractly defined Cauchy slices, $\Sigma_2$ lies to the future of $\Sigma_1$ (a statement that is false in general).  We are merely trying to define a physical quantum gravity inner product $\langle g_2 \,|\, g_1 \rangle_I$ with nice properties, and nothing about this task requires the bra state $\langle g_2 \,|$ to necessarily be to the future of the ket state $|\, g_1 \rangle$.\footnote{If a supernatural being created a whole universe, with some particular initial condition at time $t_0$, and \emph{if} causality holds in the sense that this being's action at $t_0$ only affects times to the future $t \ge t_0$ of that Cauchy slice, then the use of Bunster's ``causal'' propagator might be appropriate.  But this is not the sort of scenario we normally consider in physics.  Usually we assume that there was a past era, prior to the time of whatever experiment is being performed.  For example, in most discussions of AdS/CFT, it is assumed that the AdS spacetime has existed forever.  Even in cosmology, it is usually assumed that the beginning of time was an initial singularity, rather than a Cauchy slice.} In fact, if we restrict to positive lapse solutions, the resulting inner product will no longer be hermitian.

\begin{figure}[h]
\centering
\begin{tikzpicture}
  \draw (0,0) arc (220:320:4 and 3) node[midway,below] [fontscale=2]{$\Sigma_2$} ;
  \draw (0,0) arc (40:140:4 and 3) node[midway,above] [fontscale=2]{$\Sigma_2$};
  \node [xshift=0cm] at (6.6,0) [fontscale=1]{$\partial\Sigma$};
  \node at (0,0)[circle,fill,inner sep=1.5pt]{};
  \node at (-6.1,0)[circle,fill,inner sep=1.5pt]{};
  \node at (6.1,0)[circle,fill,inner sep=1.5pt]{};
  \draw (-6.1,0) -- (6.1,0) node[midway,above] {$\quad \partial R$};
  \draw (-6.1,0) -- (0,0) node[midway,below] [fontscale=2]{$\Sigma_1$};
  \draw (0,0) -- (6.1,0) node[midway,above] [fontscale=2]{$\Sigma_1$};
  
  \end{tikzpicture}
\caption{\small The third case, in which $\Sigma_2$ straddles $\Sigma_1$, is shown. We are using the convention that we fix the time orientation of $\mathcal{M}$ to always be upward so that the time-ordering of the two slices is opposite in the two regions.}\label{straddle}
\end{figure}
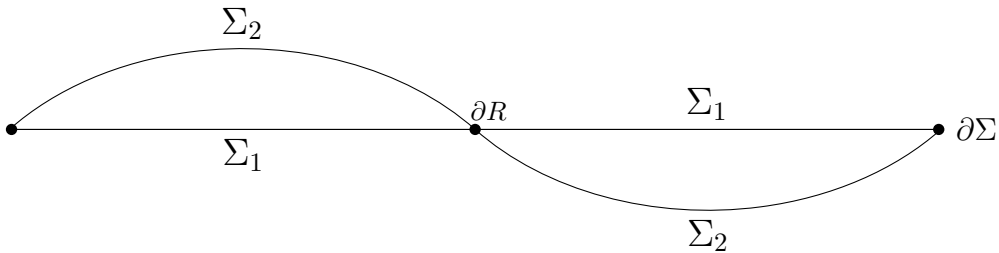

Hence, to obtain a gauge-invariant inner product, we must allow all possible time relationships between the two Cauchy slices.  Since $N(x)$ is a function of spatial position, there are actually three possible cases to consider:\footnote{This classification ignores degenerate cases where $\Sigma_1$ and $\Sigma_2$ entirely coincide in some open region.}
\begin{enumerate}
\item $\Sigma_2$ lies entirely to the future of $\Sigma_1$, \item $\Sigma_2$ lies entirely to the past of $\Sigma_1$, 
\item $\Sigma_2$ straddles $\Sigma_1$ so that it lies partly to the future (in some open spatial region $R$ of $\Sigma_1$) and partly to the past (in the complementary open region $\overline{R}$ of $\Sigma_1$).  (See Fig. \ref{straddle}.)
\end{enumerate}

In portions of $\cal M$ where $\Sigma_2$ is to the future of $\Sigma_1$, one should use the $+i$ sign for the exponential of the action, while where $\Sigma_2$ is to the past, one should use the opposite sign $-i$.\footnote{It is possible to verify that these are the correct sign rules by considering simple toy models in which one takes a direct sum of $n$ non-interacting QM systems, each with its own time variable $t_n$ generated by a Hamiltonian $H_n$, and then imposing the Hamiltonian constraint $H_n \Psi(x) = 0$ on each system separately by defining
\begin{equation}
\langle x_2 \,|\, x_1 \rangle = \prod_n
\int_{-\infty}^{+\infty}  dt_n\, 
\langle x_2 \,| e^{\sum_n iH_nt_n} |\, x_1 \rangle,
\end{equation}
and then switching to the Lagrangian formulation.  Additionally, \cite{Wall:2012fn} checked that these are the correct rules in a discrete (but not holographic) model of spacetime that allows for causal propagation.}

What happens to this argument if we consider quantum gravity along a complex contour which takes us away from Lorentzian signature?  In this case, the lapse $N(x)$ becomes complex.  If the Hartle-Hawking prescription is correct, then we must allow spacetimes with $N(x)$ having sign $+i$ but not $-i$.  However, so long as the integration contour through the space of metrics $\bf g$ is chosen to have no endpoints, the constraint equations will still be satisfied.\footnote{In the Hartle-Hawking no boundary proposal, the sign of the Euclidean imaginary-time lapse $\Delta \tau$ is restricted to be always positive.  This is sort of like a contour endpoint.  However, taking $\tau \to 0$ corresponds to space disappearing entirely (this is the ``no boundary'' part of the no-boundary proposal) so we never violate the Hamiltonian constraint on any given nondegenerate spatial metric $g$.}

%Does this mean that we violate ${\cal H} = 0$ because the range of the Euclidean time is restricted?
%This would be true if we simply integrate our contour along all Euclidean metrics with positive imaginary lapse, but it is not necessarily true if the contour simply passes through a Euclidean saddle.

\subsection{Example: Maximally Symmetric Slices}

As a simple example, suppose that we choose the slices $\Sigma_1$ and $\Sigma_2$ such that their metrics $g_1$ and $g_2$ are homogeneous and isotropic, with (spatial) scalar curvature $R$.  It is important to note that we are \emph{not} considering a minisuperspace cosmology here, as the history $\cal M$ between the two slices not constrained to be isotropic, and thus there will be quantum fluctuations of the anisotropic modes.\footnote{The reason we can impose isotropy on $g_1$ and $g_2$, is simply that these metrics are whatever we want them to be.  We are merely evaluating the full wavefunction on special values of $g$.}  We assume that this isotropy symmetry is not spontaneously broken by the dominant saddle.  Then the ADM momentum is a pure trace: $\Pi_{ab} = g_{ab} \Pi / d$.  Plugging this relation into the Hamiltonian constraint $\eqref{ham}$ we obtain $(R-2\Lambda) \propto -\Pi^2$.  The imposition of this constraint on the slices $\Sigma_1$ or $\Sigma_2$ will determine if the classical geometry that interpolates between them is Euclidean or Lorentzian.

There are three possibilities:
\begin{itemize}
    \item $(R-2\Lambda)<0$. This allows for a real solution for $\Pi^{ab}$, and as such the slice geometry can be embedded in a Lorentzian spacetime.
    \item $(R-2\Lambda)>0$. In this case we get a purely imaginary solution for $\Pi^{ab}$, which corresponds to an embedding in an Euclidean classical spacetime.
    \item $(R-2\Lambda)=0$. This is the transition point between Lorentzian and Euclidean embeddings.  In the case of AdS ($\Lambda < 0$) this transition occurs at negative spatial curvature, when the radius of $\Sigma$ equals that of the bulk spacetime $\cal B$.
\end{itemize}
So we see that the particular state $\ket{g}$ determines whether the saddle-point is Euclidean or Lorentzian in the neighbourhood of the slice.  If we get opposite answers for $\Sigma_1$ and $\Sigma_2$, then the signature of the spacetime must change somewhere in between.\footnote{Although there are many equivalent complex contours that would give such a result, it is nicest to chose a contour in which the metric is everywhere either Euclidean or Lorentzian.}  Whenever there are Lorentzian saddles, we have a choice about whether to take the lapse $N$ to be positive or negative, as argued in \ref{lapse}.  This choice spontaneously breaks $CPT$ symmetry, as discussed in section \ref{CPT}.  The two contributions are complex conjugates.  Both of these saddles have to be included together and so they contribute with a real but oscillatory functional to the transition amplitude \cite{Wall:2021bxi}.

\begin{figure}[h]
\centering
\begin{subfigure}{.47\textwidth}
  \centering
  \vspace{1.05cm}
\begin{tikzpicture}
  \draw (0,0) arc (220:320:4 and 5) node[midway,below] [fontscale=1]{$\Sigma_1,g_1$} ;
  \draw (0,0) arc (140:40:4 and 3.5) node[midway,above] [fontscale=1]{$\Sigma_2,g_2$};
  \node [xshift=0cm] at (6.6,0) [fontscale=1]{$\partial\Sigma$};
  \node at (0,0)[circle,fill,inner sep=1.5pt]{};
  \node at (6.1,0)[circle,fill,inner sep=1.5pt]{};
  \draw [dashed] (0,0) -- (6.1,0) node[midway,above] {$K=0$};
  \end{tikzpicture}
\end{subfigure}
\begin{subfigure}{.47\textwidth}
  \centering
\begin{tikzpicture}
  \draw (0,0) arc (140:40:4 and 5) node[midway,above] [fontscale=1]{$\Sigma_1,g_1$} ;
  \draw (0,0) arc (220:320:4 and 3.5) node[midway,below] [fontscale=1]{$\Sigma_2,g_2$};
  \node [xshift=0cm] at (6.6,0) [fontscale=1]{$\partial\Sigma$};
  \node at (0,0)[circle,fill,inner sep=1.5pt]{};
  \node at (6.1,0)[circle,fill,inner sep=1.5pt]{};
  \draw [dashed] (0,0) -- (6.1,0) node[midway,above] {$K=0$};
  \end{tikzpicture}
\end{subfigure}

\begin{subfigure}{.47\textwidth}
  \centering
  \vspace{2cm}
\begin{tikzpicture}
  \draw (0,0) arc (220:320:4 and 5) node[midway,below] [fontscale=1]{$\Sigma_1,g_1$} ;
  \draw (0,0) arc (220:320:4 and 3.5) node[midway,above] [fontscale=1]{$\Sigma_2,g_2$};
  \node [xshift=0cm] at (6.6,0) [fontscale=1]{$\partial\Sigma$};
  \node at (0,0)[circle,fill,inner sep=1.5pt]{};
  \node at (6.1,0)[circle,fill,inner sep=1.5pt]{};
  \draw [dashed] (0,0) -- (6.1,0) node[midway,above] {$K=0$};
  \end{tikzpicture}
\end{subfigure}
\begin{subfigure}{.47\textwidth}
  \centering
  \vspace{-2cm}
\begin{tikzpicture}
  \draw (0,0) arc (140:40:4 and 5) node[midway,above] [fontscale=1]{$\Sigma_1,g_1$} ;
  \draw (0,0) arc (140:40:4 and 3.5) node[midway,below] [fontscale=1]{$\Sigma_2,g_2$};
  \node [xshift=0cm] at (6.6,0) [fontscale=1]{$\partial\Sigma$};
  \node at (0,0)[circle,fill,inner sep=1.5pt]{};
  \node at (6.1,0)[circle,fill,inner sep=1.5pt]{};
  \draw [dashed] (0,0) -- (6.1,0) node[midway,above] {$K=0$};
  \end{tikzpicture}
\end{subfigure}

\caption{\small For the case of the slices consistent with Lorentzian embedding, we get 4 saddles. The top two ``gibbous'' saddles are time reversals of each other and so they together contribute as a complex conjugate pair to the transition amplitude. Similarly for the bottom two ``crescent'' saddles. In this figure the time orientation is directed upwards and the $K=0$ dashed line is the time symmetric slice.  It can be seen that when $g_1 = g_2$, the crescent saddles go to zero lapse, leading to divergences of the 1-loop quantum determinant $\Delta_\text{crescent}$, which will be discussed in section \ref{prop}.}\label{saddlesfour}
\end{figure}
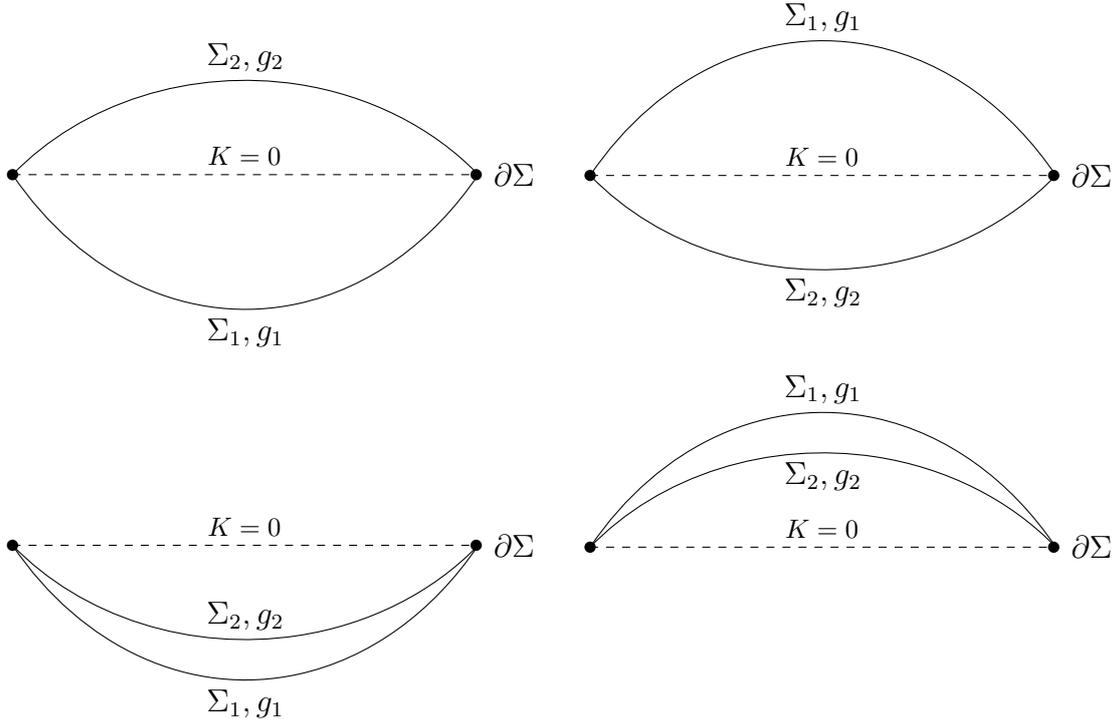

Now if both the slices are consistent with a Lorentzian embedding, then we will have four saddles occurring in two complex conjugate pairs, as shown in Fig \ref{saddlesfour}, arising from all the possible time orientations relative to the time-symmetric slice.

The sum over all 4 saddles gives the amplitude:
\bea\label{4saddles}
\langle g_2 | g_1 \rangle_I = \sum_{\pm} \Delta_\text{gibbous}e^{\pm i (I_1 + I_2)} + \Delta_\text{crescent}e^{\pm i (I_1 - I_2)},
\eea
where $I_n$ is the action for a positive lapse history from the slice $g_n$ to the $K = 0$ slice, and $\Delta$ is the quantum determinant associated with the two possible geometries for $\cal M$ shown in Fig \ref{saddlesfour}, which could be calculated by e.g. heat kernel methods.\footnote{As stated in section \ref{Lorentzian}, these amplitudes are IR divergent and require regulation by moving $\partial \Sigma$ to a finite value of $z$.  In evaluating $I_1$ and $I_2$, we have also ignored in \eqref{4saddles} the Gibbons-Hawking contribution coming from the Euclidean $i\pi$-rotation at $\partial \Sigma$.  This contributes a thermodynamically suggestive (but IR divergent) factor of $e^{\text{Area}[\partial \Sigma]/4G_N}$ to the inner product $\langle g_2 | g_1 \rangle_I$.  Since this factor is the same regardless of the initial and final metrics, it may be ignored when constructing the Hilbert space.}

However, if only one of the slices is consistent with Lorentzian embedding, while the other is embedded in an Euclidean geometry, then there are only two contributing saddles, arising from flipping the time orientation of the Lorentzian piece:
\bea\label{2saddles}
\langle g_2 | g_1 \rangle_I = \sum_{\pm} \Delta_\text{mixed}e^{-I_E \pm iI_L},
\eea
where $I_E$ is the Euclidean part of the action and $I_L$ is the Lorentzian part.  Lastly, if both slices are consistent with an Euclidean embedding, then there is only one dominant saddle,\footnote{The dominant Euclidean saddle is the one where both slices are on opposite sides of the $K = 0$ surface, similar to the top row of Fig. \ref{saddlesfour} but in Euclidean signature.  In addition to having a larger amplitude (not necessarily a decisive point since not every saddle necessarily lies on the physically correct contour) once the Gibbons-Hawking term is taken into account, this saddle is clearly the one which limits to the usual AdS/CFT saddles in the limit that $\mu \to 0$.} contributing with a real and positive functional:
\bea\label{1saddle}
\langle g_2 | g_1 \rangle_I = \Delta_\text{Euclidean}e^{-(I_1 + I_2)},
\eea
where the $I_n$ are now Euclidean signature actions from $g_n$ to the $K = 0$ slice.\footnote{It can be seen that if we neglect the quantum determinants by setting $\Delta = 1$, the resulting norm is nonnegative for arbitrary superpositions of maximally-symmetric metrics.  In fact it is highly degenerate, with the dynamical state space being 1-dimensional.  A more careful check of positivity would therefore require knowing something about the $\Delta$ factors.}

We leave further analysis of this symmetric case to future work.

\subsection{Properties of the Amplitude}\label{prop}

In order for our construction of the Hilbert space in section \ref{HS} to succeed, we will need for our inner product (the previously defined transition amplitude) to satisfy the following four conditions: (a) linearity, (b) conjugate symmetry and (c) positive semi-definiteness, and (d) a subspace on which the inner product is finite.\footnote{See \cite{Wall:2012fn} for discussion of these four requirements in a discrete model of quantum gravity.}

Below we prove the first two properties, and make some arguments concerning positivity and finiteness, leaving more convincing proofs for future work.

\paragraph{Linear.}
Equation \eqref{TA} defines the inner product on the basis elements of the dual space of distributions on functionals of the metric. This space will be later properly defined and called $K^*$. Its action on a general element in $K^*$ is defined from this in a linear fashion. The map is thus linear by construction.

\paragraph{Hermitian.}

We are using a convention in which the time orientation of the space-filling manifold $\cal M$ is fixed. For the inner product $\langle g_2 \,|\, g_1 \rangle_I$ we have the straddle expansion described above, each diagram coming with a specific sign for the exponential of the action. Now consider $\langle g_1 \,|\, g_2 \rangle_I$. This is equivalent to flipping the time orientation of $\cal M$. The effect is to flip the sign of the exponent for each component of the path integral. So we see that $\langle g_2 \,|\, g_1 \rangle_I=\langle g_1 \,|\, g_2 \rangle^*_I$. 

%Moreover, it can be observed from the diagrams that all contributions come in complex-conjugate pairs. Hence, $\langle g_2 \,|\, g_1 \rangle_I=\langle g_1 \,|\, g_2 \rangle_I$ is real.

%\begin{figure}[h]
%\centering
%\includegraphics[width=.7\textwidth]{Figures For Paper.pdf}
%\caption{On the top line, the straddle expansion is shown for $\langle g_2 \,|\, g_1 \rangle_I$. The bottom line is the expansion for $\langle g_1 \,|\, g_2 \rangle_I$. The signs correspond to those to be used in the exponential of the action.}\label{straddle}
%\end{figure}

\paragraph{Positive Semi-definite.}  The inner product cannot be positive definite, because it projects onto states obeying the constraint equations ${\cal H}(x) = 0$ and ${\cal D}_a(x) = 0$, so there ought to be lots of null states.  This is good because it implements the  gauge-invariance of the theory.  So at best, the path integral can be positive semi-definite, so that for any superposition $\Psi$ of spatial metrics $g$:
\begin{equation}\label{pos}
\langle \Psi \,|\, \Psi \rangle_I \ge 0.
\end{equation}
In what follows we will be assuming that this positivity property holds, at least in the context of complete Cauchy slices $\Sigma$ in asymptotically locally AdS spacetimes.\footnote{See \cite{Wall:2021bxi} for an argument that positivity fails for \emph{partial} Cauchy slices which have boundaries in the interior of black holes.}

This property is difficult to show in general, due to the fact that $I$ includes in its definition arbitrary dynamical spacetimes.  In order to prove it, we need to somehow factor out the role of time evolution.

However, an argument for \eqref{pos} can be made that positivity is at least true within the semiclassical regime.  We adopt an argument based on the conformal method of solving the Einstein constraint equations \cite{Isenberg:2007uv, Chruciel2011IntroductionTT, Witten_talk}, in which one switches to an alternative quantum gravity basis $(\gamma_{ab}, K)$ where $\gamma_{ab} = g^{-1/d}g_{ab}$ is the conformal part of the metric, and $K$ is the trace of the extrinsic curvature (below we will choose $K = 0$).

This argument for the equivalence of this basis is only valid for the case of negative cosmological constant ($\Lambda < 0$).\footnote{Perhaps this is related to the fact that the CFTs which appear in dS/CFT are not unitary?}  More precisely, we need to assume that gravity is attractive in all classical saddles relevant for the computation of \eqref{pos}, which would follow if the generic Strong Energy Condition is satisfied:
\begin{equation}\label{SEC}
\textbf{T}_{\hat{t}\hat{t}} + \tfrac{1}{D-2}\textbf{T}\:>\: \frac{\Lambda}{8\pi G_N},
\end{equation}
for all unit timelike vectors $\hat{t}$.\footnote{It might be possible to violate this equation if we turn on an inflating scalar field $\Phi$ deep in the bulk, or by means of a quantum fluctuation.  However, when $\Lambda < 0$, it requires a finite stress-tensor excursion to violate \eqref{SEC} and so this requires a very large quantum fluctuation.  We can neglect this possibility in the semiclassical limit where $G_{N} \to 0$ and quantum fluctuations are linearized.  (This does not preclude us from taking quantum superpositions of different geometries, as long as the inner product \eqref{pos} is dominated by a sum over SEC-satisfying classical saddles.)}

At the classical level, consider the domain of dependence $D[\Sigma]$ of a given Cauchy slice $\Sigma$.  Assuming a plausible form of cosmic censorship, within this domain there always exists a Cauchy slice $\Sigma_{K = 0}$ of maximal spatial volume $V$.  This slice has zero mean extrinsic curvature: $K = \Pi = 0$.  By solving the Raychaudhuri focussing equation for timelike geodesics shot out normal to $\Sigma_{K = 0}$, one can show using the SEC that all other slices $\Sigma'$ have decreasing volume moving away from $\Sigma_{K = 0}$, and thus the $\Sigma_{K = 0}$ slice is also \emph{unique} within each history of $D[\Sigma]$.

Furthermore, the conformal factor of the metric (defined by $g_{ab} = \varphi^{4/(d-2)}\gamma_{ab}$) is determined on-shell by solving the Hamiltonian constraint equation \eqref{ham}.  This equation reduces to the Lichnerowicz equation:
\begin{equation}\label{Lichner}
\tfrac{4(d-1)}{d-2} \nabla^2 \varphi - R[\gamma] \varphi
= -|K|^2_\gamma \varphi^{(2-3d)/(d-2)} - 2\Lambda \varphi^{(d+2)(d-2))},
\end{equation}
where $|K|^2_\gamma := K^{ab}K^{cd}\gamma_{ac}\gamma_{bd}$, with $K^{ab}$ traceless and satisfying the diffeomorphism constraint $\nabla_a K^{ab} = 0$.  It turns out that classically, solutions to \eqref{Lichner} (with $\varphi > 0$) exist and are unique if $\Lambda < 0$.  Hence the Hamiltonian constraint can be eliminated from the phase space together with the conformal part of the metric.  All of this implies that, at the classical level, $(\gamma_{ij}, K^{ij})$ is a well-posed set of Cauchy data (modulo the spatial diffeomorphism gauge symmetry and constraint).

We now suppose that (at least semiclassically) the quantum analogue of these properties also holds.  Because every semiclassical saddle of $D[\Sigma]$ has a $K = 0$ slice in it, it should be possible to use the path integral to write any $|g\rangle$ state as an arbitrary superposition of $|\gamma_{ab}, K = 0\rangle$ states, modulo spatial diffeomorphisms.

Furthermore, at least semiclassically, any two such states $|\gamma_{ab}, K = 0\rangle$ would have positive inner product.  This follows from the uniqueness of $K = 0$ Cauchy slices in the spacetime history.  The only possible amplitudes connecting such $K = 0$ states would therefore be trivial histories with zero lapse.  The sum over such histories are not \emph{quite} trivial, since there can still be a nonzero shift vector $N^a$ relating the two slices.  So in this case, the inner product should reduce to a canonical inner product in which we impose only spatial diffeomorphisms.  The norm of the inner product should look something like this:
\begin{equation}\label{gamma}
\langle \Psi\,|\,\Psi \rangle_\text{spatial} = \int DG 
\int D\gamma\,\Psi^*[\gamma] \Psi[G(\gamma)] \ge 0,
\end{equation}
where $G = \text{Diff}(\Sigma)$, $DG$ is the Haar measure, and $D\gamma$ is some (positive) covariant path integral measure on conformal metrics.\footnote{We assume that conformal invariance is not spoiled by any anomaly coming from the path integral measures $D\gamma$ and $DG$.  If we extend our holographic Cauchy slice hypothesis to this conformal basis, the partition function $Z[\gamma]$ defines a CFT, but in this CFT there is a Liouville-like scalar field corresponding to the conformal mode $\sqrt{g}$.  Presumably this Liouville mode cancels the conformal anomaly.}

The inner product \eqref{gamma} is manifestly nonnegative, since the contribution coming from each leaf of Diff$(G)$ is nonnegative.  Since every $|\,g\rangle$ state is a superposition of $|\,\gamma\rangle$ states, positive semi-definiteness must also hold for linear combinations of  $|\,g\rangle$ states.

\paragraph{Finite Subspace.}  Obviously, all of the above derivations are in vain if the path integral amplitude fails to converge to a well-defined and finite norm for at least some vectors.  Assuming positivity, we are looking for a seminormed subspace of states in which
\begin{equation}
\langle \Psi \,|\, \Psi \rangle_I < \infty.
\end{equation}
What obstacles might there be to constructing such a state?

First of all, in Lorentzian signature, the path integral is oscillatory, and thus extracting well-defined answers may require some prescription for resumming such integrals, perhaps by slightly deforming the complex contour away from the Lorentzian regime (cf. section \ref{contour}).

In general, there might be IR problems coming from the fact that we are summing over histories with arbitrarily large lapse $N$.  However, general relativistic spacetimes have a strong tendency to be unstable, rather than cyclical.  For example, in asymptotically AdS spacetimes we expect (again by the strong energy condition \eqref{SEC}) that in any domain of dependence $D[\Sigma]$ the volume of the universe contracts indefinitely as one departs from the maximal volume slice.  This suggests that for large values of $N$, the wavefunction disperses away from any given metric $g$, resulting a damping of contributions from $|N| \to \infty$.

There could also be UV problems when the lapse $N$ is small.  Our expectation is that the amplitude $\langle g_2 \,|\, g_1 \rangle$ is probably finite for \emph{generic} pairs of metrics $g_1$ and $g_2$, but that in certain coincident limits there are UV problems that arise.  (This is reminiscent of the 2 point function in QFT which is defined at spacelike or timelike separation, but which blows up at null separation.)

For example, if $g_1 = g_2$ in an open region $R \subseteq \Sigma$,\footnote{If there is a matter field $\Phi$, its value will also also need to agree on the two slices in order to allow a zero lapse history with small action.}  then there may exist a classical saddle connecting the two manifolds which has $N = 0$ in $R$.  The quantum determinant $\Delta_\text{1-loop}$ of such a history is likely to diverge badly in this case.  To see this divergence, suppose that the lapse were instead taken to be a small constant $N = \epsilon$.  When quantizing a field with Dirichlet boundary conditions on a narrow strip with width $N = \epsilon$, one expects a Casimir energy contribution to $\Delta_\text{1-loop}$, which by dimensional analysis scales in its most divergent term like:
\begin{equation}
\ln \Delta_\text{1-loop}
\sim \frac{\text{Volume}[R]}{\epsilon^{d}} + \text{subleading}.
\end{equation}
This will produce an essential singularity as $g_1 \to g_2$.\footnote{Given GHP, one might wonder how it is possible for the holographic dual partition function to exhibits a divergence as $g_1 \to g_2$.  It could be that the $T^2$ deformed theory regulates this divergence.  But more likely, the $T^2$-theory is also divergent in such cases, due to a sum over energy eigenvalues with large imaginary energy (cf. section \ref{nonuni}).}  The nature of this singularity in Lorentzian signature (oscillatory, exponentially suppressed, or exponentially enhanced) depends on the dimension $d$ and the matter content of the theory.  Presumably it may be dealt with by %In Lorentzian signature, this contribution is imaginary (i.e. oscillatory as $\epsilon \to 0$) \textcolor{red}{Only when $d =$ odd!  What should we say here?} 
smearing out the metrics $g_1$ and/or $g_2$ slightly, with suitable distributions centered around the chosen metric, using an appropriate choice of contour.\footnote{In QFT, a sufficient condition for (operator-ordered) $N$ point functions to be finite is that all insertions are smeared over time.  This is because the time smearing suppresses all contributions coming from large values of the energy $E$.  Given this fact, one might wonder how it is possible for there to be UV divergences in the WDW amplitude, when they are (by construction) integrated over all possible lapse times $N$, an extreme form of time smearing!  It seems that the ultimate reason for this divergence is that (prior to imposing the constraint) in GR the Hamiltonian is unbounded below due to the conformal mode sign problem.  This makes it so the integral over solutions with $\cal H$ = 0 is a noncompact manifold due to the possibility of $\Pi$ being large.  This suggests that, at least along a Lorentzian contour, slightly smearing the Weyl mode $\sqrt{g}$ should be sufficient to render the amplitude finite.}

Even when $N = 0$ only on a measure zero set (e.g. the straddling history shown in Fig. \ref{straddle}), there may be unique divergences appearing at the corners of the slices.  We assume all such divergences can be dealt with by some combination of renormalization and smearing.

Assuming these properties of the transition amplitude hold, we are now in a position to construct the gravitational Hilbert space.  This construction is purely gravitational, but will be dual by the GHP to the amplitude found by the partition function.

\section{Construction of the Bulk Hilbert Space}\label{HS}

First in \ref{dirty} we give an informal statement of how the Hilbert space may be constructed, omitting some of the more annoying technical details.  Then in \ref{rigorjunky} we will provide somewhat more rigor for those who desire such things. 

Our approach is similar in spirit to previous work constructing the Hilbert space from the gravitational path integral in loop quantum gravity and other diffeomorphism-invariant models \cite{Ashtekar:2000hv,Ashtekar:1995zh,Perez:2001gja,Noui:2004iy,Alesci:2008yf,Freidel:2005qe}.

A reader tempted to skip this section altogether should note that the linear algebra we use in \ref{dirty} to construct the Hilbert space will be an important prerequisite for defining the holographic AdS/CFT dictionary in the following sections.

\subsection{Quick and dirty version}\label{dirty}

Let $K$ be the \emph{kinematic} quantum gravity state space.  We will take $K$ to be a space of functions over spatial metrics $g$:
\begin{equation}
K:= \{\Psi[g]\}.
\end{equation}
To solve canonical quantum gravity, we identify a smaller \emph{dynamical} subspace $C \subset K$ of WDW-states, which are annihilated by the constraint equations $\mathcal{H}$ and ${\cal D}_a$:
\begin{equation}
C:= \{\Psi \in K \:|\: \mathcal{H}\Psi = {\cal D}_a \Psi = 0\}.
\end{equation}
This of course requires that $C$ is within the subspace of $K$ on which $\mathcal{H}$ and $\mathcal{D}_a$ have a well-defined (albeit possibly distributional) action.

In order to interpret $C$ as a Hilbert space (and thus define probabilities), we also need to define an inner product.  States in $C$ are typically not normalized with respect to natural inner products on $K$,\footnote{For example, in minisuperspace cosmology, there is typically a continuous spectrum of $H$ prior to imposing the Hamiltonian constraint.  But then if we view a solution with $H \Psi = 0$ as a wavefunction in the energy representation, it looks like a delta function, which is nonnormalizable with respect to the usual QM inner product.} and thus it is necessary to somehow obtain an inner product from the dynamical evolution itself.

One of the advantages of the path integral formulation of quantum gravity is that this can be done in a very natural way.  The Lorentzian path integral (defined in section \ref{QG Inner Product}) defines a transition amplitude between two metrics $g_1$ and $g_2$. However, the (infinite-dimensional) delta functions $\delta^{\infty}(g - g_1)$ and $\delta^{\infty}(g - g_2)$ are examples of \emph{distributions} on the space of metrics, i.e. they cannot be regarded as vectors in $K$, but rather they span a dual vector space of distributions over spatial metrics:\footnote{If we define a functional integration measure over d-metrics $Dg$, we might be able to use it to convert between functions and distributions, and define an inner product on $K$ in this manner.  However, this inner product would not be the physically relevant one, so it is better not to introduce this extra structure.}
\begin{equation}
K^*:= \{d : K \to \mathbb{C}\}.
\end{equation}

From the discussion in the previous section (\ref{prop}), the path integral formulation of quantum gravity (assuming it is well-defined) should give us a \emph{Hermitian sesquilinear} form on $K^*$, inherited from the path integral:
\begin{eqnarray}\label{I_1}
I: K^* \cross \overline{K^*} &\to& \mathbb{C},
\\
(\delta^{\infty}[g-g_1],\delta^{\infty}[g-g_2]) &\mapsto& \langle g_2 \,|\, g_1 \rangle_I ,
\end{eqnarray}
where the bar represents complex conjugation.  From this we can derive an antilinear map from the space of distributions to the space of functions:
\begin{eqnarray}\label{I_2}
   I: K^* &\to& \overline{K},\\
   d &\mapsto& I(d)=I(d,\cdot),
\end{eqnarray}
where $I(d,\cdot)$ is a linear functional on $\overline{K^*}$,\footnote{When $I$ appears with one argument it should be thought of as the map in \eqref{I_2}, whereas when it appears with two arguments it refers to the map in \eqref{I_1}.} and so it belongs to the original vector space $K$.\footnote{In this (less rigorous) version of the construction, we are conflating $\overline{K}$ with its double dual $\overline{K^{**}}$, even though this is not generally the case for infinite dimensional vector spaces.  We will do better in the next section.} Its action on $\overline{K^*}$ is defined by
\begin{equation}
    (I(d))(v)=I(d,v),\:\:\forall v\in \overline{K^*}.
\end{equation}
Actually, we can make a stronger statement than this, because a diffeomorphism-invariant path integral has the property that its outputs automatically satisfy the constraint equations, hence we can actually assert the stronger relation $\text{Im}_I(K^*)\subseteq\overline{C}$.

From these constraint equations, it also follows (by vector space duality) that there exist some elements of $K^*$ which are ``pure gauge'' and have zero inner product with all other vectors.  Hence, it is also natural to replace $K^*$ with a quotient vector space $Q$, defined by modding out by all such null vectors:
\begin{equation}\label{Q}
Q = \frac{K^*}{K^*_0},
\end{equation}
where $K^*_0$ is the null subspace, consisting of all vectors $v_0$ in $K^*$ such that $\forall w \in \overline{K^*} : I(v_0,w) = 0$.  

Putting these facts together, we actually have a stronger relation:
\begin{eqnarray}
   I: Q &\to& \overline{C},
   \\
   \delta^{\infty}[g-g_1] &\mapsto& \Psi[g]=\langle g \,|\, g_1 \rangle_I.
\end{eqnarray}

Now assuming that the map to $\overline{C}$ is surjective{\bf---}in other words, that there are no additional constraint equations imposed by the path integral besides $\cal H$ and ${\cal D}_a${\bf---}then $\text{Im}_I(K^*) = \overline{C}$ and thus $K^*_0$ can also be characterized as the set of vectors $d \in K^*$ such that $d:C \to \{0\}$.\footnote{If there are any additional constraints coming from the gravitational path integral, as argued in \cite{Jafferis:2017tiu}, then there exists some element $d$ of $K^*_0$ for which $d(C) \ne 0$.  These additional constraints would need to be incorporated into the definition of $C$ in order to proceed further.}  From the same assumption, it also follows that there also exists an inverse map:
\begin{equation}
I^{-1}\!:\, \overline{C} \to Q,
\end{equation}
and a non-degenerate inverse inner product:
\begin{eqnarray}
   I^{-1}\!:\,C \cross \overline{C} &\to& \mathbb{C}, \\
   (c_1,c_2)&\mapsto& \!(I^{-1}(c_2))(c_1)
   \:=\:
   I(I^{-1}(c_1),I^{-1}(c_2)).
\end{eqnarray}
At this point, there is a canonical isomorphism $C \cong \overline{Q}$, and it is therefore possible to get away with identifying the two spaces.

%There is one additional axiom which a physically sensible inner product must satisfy, and that is positivity.  Specifically, we need to know that the inner product $I$ defined on $K^*$ is non-negative, so that for any vector $v \in K^*$:
%\begin{equation}
%I(v, \overline{v}) \ge 0
%\end{equation}

% \textcolor{blue}{What about the example that Aron constructed of a negative-norm state (using the BTZ saddle)? That was for a state in the ``Hilbert space" of a subregion. There is no contradiction with positive-definiteness for the Hilbert space on the whole Cauchy slice.}

The inner product on $Q$ or $C$ is automatically positive-definite (since any nonzero null vectors are included in $K^*_0$ and hence are modded out).  Hence, either $Q$ or $C$ may be regarded as defining the physical Hilbert space of the theory.

Nevertheless, there are some practical advantages to defining states as vectors in $Q$, rather than as vectors in $C$.  Although the two Hilbert spaces are canonically isomorphic, in order to relate them we have to first solve the entire theory, something which is very difficult to do in any realistic model.\footnote{If your model is exactly solvable, it's too simple to be our Universe!}  
Furthermore, until the theory is solved, it is not even possible to write down an explicit formula for any vector in $C$.  On the other hand, it is very easy to write down explicit vectors in $Q$!  Just name any vector in $K^*$, and this is automatically also a vector in $Q$.  (Although if you are very unlucky, it might be the null vector.)

\subsection{Technical details for those who want more rigor} \label{rigorjunky}

The construction in the previous section played fast-and-loose with several features of infinite-dimensional vector spaces.  In this section we will tentatively attempt to fill in some of these details, as would be required in a more rigorous description.  In the process we will make a few slightly arbitrary choices, and it is quite possible that the construction sketched below still requires technical improvements in order to be adequate.

We started out by defining the kinematic quantum gravity state space $K$ as the space of functions $\Psi[g]$ over spatial metrics $g$.  However, if we do not restrict the space of functions in some way, the only allowed elements of $K^*$ will be finite superpositions of delta functions.  This is insufficient since we need continuous functions in $K^*$ to allow for smearing.

To proceed, it is convenient to require the functions in $K$ to be \emph{bounded},\footnote{If boundedness is too strong{\bf---}and it might well be if physically realistic metrics wavefunctions fail to get damped out along certain directions of configuration space{\bf---}we could instead choose a function $f \in K$ with $0 < f < \infty$ and require $\Psi / f$ to be bounded.  This should lead to a similar theory as below.} and also \emph{continuous} with respect to some topology $T_g$ on the space of metrics $g$.  In a minisuperspace model, $T_g$ should look like the usual $\mathbb{R}^n$ topology locally near each point of $g$.  (Most of the technicalities below are required even in this simpler case, if one wishes to avoid assuming any measure \emph{a priori} on the space of metrics.) In full quantum gravity, the space of metrics is infinite-dimensional, and more thought is probably required to determine the most useful choice for $T_g$.  We leave this choice open for the present.

$K$ has a natural norm given by the uniform topology:
\begin{equation}
\forall \Psi \in K, \:||\Psi|| = \sup_g |\Psi(g)|  .
\end{equation}
We can now define $K^*$ as the continuous dual of $K$.  However, in general $K^*$ will be extremely large, mostly because of bizarre nonconstructable distributions, which have no physical relevance.\footnote{As an illustration of the sorts of issues that can arise with the continuous vector space dual in infinite dimensions, consider the Lebesgue spaces $L^p$ on a line $\mathbb{R}$, which is a norm whenever $1 \le p \le \infty$.  Whenever $p$ lies strictly inside this interval, $(L^p)^* = (L^q)$ with $1/p = q$ and hence the double dual is the same as the original space $(L^p)^{**} = L^p$.  (For example $L^2$, being a Hilbert space, is its own double dual.)  But weird things happen in the edge cases.  $(L^1)^* = L^\infty$, where $L^\infty$ is the space of bounded functions (modulo measure zero sets).  Assuming the Axiom of Choice, $(L^\infty)^*$ is a much larger vector space called $ba$ space.  And yet, since there exist models of set theory which deny Choice, in which $(L^\infty)^* = L^1$ \cite{vath1998dual}, it appears that all the additional elements in the double dual $(L^1)^{**}$ are nonconstructive.  Such nonconstructive distributions can have no real relevance to physics; since the outputs of all respectable physics theories should be computable (and hence constructable) at least in principle.}  To cut $K^*$ down to manageable size, we will allow \emph{only} those distributions that can be written in the form of Riemann sums, i.e. limits of countable sums of delta functions.  That is, we accept only distributions of the form:
\begin{equation}
d = \lim_{m\to\infty} \sum_{n = 0}^{\infty}
c_{n,m}\,\delta^\infty(g-g_{n,m}),\qquad n,m\in\mathbb{Z},
\end{equation}
where the coefficients $c_{n,m}$ and metrics $g_{n,m}$ must always be chosen to ensure that the limiting distribution $d[K]$ exists (given the continuity and boundedness of functions in $K$).  Let us call this smaller vector space, the space of \emph{Riemannian distributions} $K^*_R$.  
Since elements in $K^*_R$ are \emph{defined} as limits of delta functions (modulo equivalence in their action on $K$) there is a natural topology on $K^*_R$ defined by demanding continuity with respect to all such limiting sequences.  This justifies the intuitive statement that the space $K^*_R$ is ``spanned'' by delta functions $\delta^\infty(g-g_0)$, in the sense that any continuous functional on $K^*_R$ which vanishes for all delta functions, vanishes everywhere; more generally, it is determined by its value on the delta functions.\footnote{As a corollary, if the topology $T_g$ admits a compact subset $S$ of metrics $g$, the space $K^*_R(S)$ of Riemannian distributions with support on $S$ is itself compact.}  It follows that $(K^*_R)^*$ can be identified with a space of functions $\Psi[g]$, and in fact it is not hard to show that all such functions are bounded and continuous.\footnote{Proof: Suppose for contradiction a function $f \in (K^*_R)^*$ were not bounded.  This means that there is a countably infinite sequence of increasing points $x_n \in g$ such that $|f(x_{n+1})| > 2|f(x_n)|$.  It is now always possible to choose the coefficients of $d = \sum_n c_n \delta(x-x_n) \in K^*_R$ such that $d[f]$ is undefined, e.g. $c_n = 1/n^2$.  

Next suppose that $f$ is not continuous.  This means that there is a convergent sequence of points $x_n$ such that $\lim_{n\to\infty} x_n = x_\infty \in g$ yet $\lim_{n\to\infty} f(x_n) \ne f(x_\infty)$.  But $\lim_{n\to\infty} \delta(x_n)$ converges to $\delta(x_\infty)$ in the natural topology of $K^*_R$ and hence $f$ would not be part of the continuous dual $(K^*_R)^*$.}  (The converse statement, that all elements of $K$ are also elements of $(K^*_R)^*$, is trivial.)  So $(K^*_R)^* = K$.\footnote{Or in other words, $K^*_R$ is a predual of $K$.}  

Another technical problem is that the gravitational inner product $I$ can only converge on, at best, a dense subspace of $K_R^*\cross \overline{K_R^*}$.\footnote{As stated in section \ref{prop}, we are assuming that $\langle g_2 \,|\, g_1 \rangle_I$ can be defined so as to converge for generic pairs $(g_1, g_2)$, and that any infinities arising from nongeneric choices can be dealt with by smearing, at least within some subspace of $K$.} Because of this, a more candid expression than \eqref{I_1} for the gravitational inner product would be:
\begin{eqnarray}
I: K_R^* \cross \overline{K_R^*} &\to& \mathbb{C} \,\cup\, \{\mathbf{undef}\},
\\
(\delta^{\infty}[g-g_1],\delta^{\infty}[g-g_2]) &\mapsto& \langle g_2 \,|\, g_1 \rangle_I,
\end{eqnarray}
where the extra value $\mathbf{undef}$ covers the cases where the inner product is undefined.  In order to get a well-defined linear map, we need to restrict attention to a subspace of normalizable states.  The positivity of $I$ guarantees a maximal such subspace, which may be defined as:
\begin{equation}
    K_{N}^* = \{k\in K_R^*\:|\:I(k,k) < \infty\}.
\end{equation}
Any two such vectors $k_1, k_2 \in K_N^*$ also have finite inner product with each other, by the Cauchy-Schwarz inequality; hence the sesquilinear form becomes an actual map:
\begin{equation}
    I: K_{N}^* \cross \overline{K_N^*}\to \mathbb{C}.
\end{equation}
There may exist some elements $k \in K_N^*$ such that $I(k,K_R^*)$ is either not finite or not continuous on $K_R^*$.  But we may restrict to a smaller subspace which is finite and continuous:
\begin{equation}
K_{f.c.}^* = \{k\in K_N^*\:|\; I(k,\cdot) \in K\}.
\end{equation}
Having done this, we still have antilinear map \eqref{I_2} to the constraint space, but now with a restricted domain:
\begin{equation}
I: K_{f.c.}^* \to K.
\end{equation}
We define the image of this map as the constraint subspace:
\begin{equation}\label{ImC}
\mathcal{C} = \text{Im}_I(K^*_{f.c}) \subseteq C,
\end{equation}
where by making this definition, we allow for the possibility that there might be some solutions to ${\cal H} = {\cal D}_a = 0$ which cannot be produced by the path integral.

Next, we can identify a space of null vectors which have zero inner product with any other vectors:
\begin{equation}
K_0^* = \{k \in K_R^* \:|\: I(k,\cdot) = 0\}.
\end{equation}
Necessarily, such null vectors are \emph{also} elements of $K_N$ and $K_{f.c.}$, by the definitions of those spaces.  Hence we can still define a quotient space as:
\begin{equation}
Q = \frac{K_{f.c.}^*}{K^*_0},
\end{equation}
and thence obtain our physical map:
\begin{equation}
    I:Q\to \overline{\mathcal{C}}.
\end{equation}
This map is automatically surjective due to the restriction to the image in \eqref{ImC}. Hence we obtain an isomorphism between the two vector spaces:
\begin{equation}
    Q \cong \overline{\mathcal{C}}.
\end{equation}
In the end, we obtain an inner product space:\footnote{The whole argument holds for the complex-conjugate vector spaces as well. So we also get the isomorphism $\overline{Q}\cong \mathcal{C}$.}
\begin{equation}
    (\mathcal{C},\,I).
\end{equation}
From the positivity of the inner product $I$, we can induce a norm topology on $\mathcal{C}$ which turns it into a topological inner product vector space.  (This topology is distinct from any of the topologies mentioned previously.)  We now take the topological closure \text{cl}\;$\mathcal{C}$,\footnote{A closure step is needed, even in the simplest QM examples, due to the fact that e.g.~not every element of $L^2$ is actually a function, for example $\Psi(x) = |x|^{-1/3}/(1+x^2)$ is not a continuous function because it blows up at $x = 0$, and yet it is square-integrable.} to finally get the Hilbert space of bulk quantum gravity:
\begin{equation}
    \boxed{\mathcal{H}_{\text{QG}}=(\text{cl}\;\mathcal{C},I)}
\end{equation}
This gives a more precise definition of the bulk quantum gravity Hilbert space.
%\footnote{This path integral construction of quantum gravity Hilbert space is manifestly background independent. The construction of course depends on how well the path integral is defined. If we take the $T^2$ deformed theory as the definition of the path integral as discussed in section \ref{defofbulktheory}, then one views this as the Hilbert space of an effective theory of quantum gravity as the $T^2$ deformation is an irrelevant deformation of the CFT. When this theory is UV completed, one would get the Hilbert space of quantum gravity which is non-perturbative, UV complete and manifestly background independent. See \cite{Ashtekar:2000hv,Ashtekar:1995zh,Freidel:2005qe} for other constructions of Hilbert space for diffeomorphism invariant theories in a manifestly background independent way.}

In the rest of this paper (when discussing holographic maps to the boundary) we will revert to the lower level of rigor of section \ref{dirty} for simplicity.  In other words we do not check that all of our maps give actual proper vectors in the state spaces in question.  This is because it is convenient to refer to entities like metric ket states $|g_1\rangle$, even if they are not really normalizable vectors in the quantum gravity Hilbert space $Q$.  Presumably, everything becomes finite after an appropriate smearing of $|g_1\rangle$.

\section{An Explicit AdS/CFT Dictionary}\label{dictionary}

\begin{figure}[h]
    \centering
    \begin{tikzpicture}[scale=1.5]
\draw (0,0) ellipse (1.25 and 0.3);
\draw (-1.25,0) -- (-1.25,-3.5);
\draw (1.25,-3.5) -- (1.25,0) node[midway,right][fontscale=2]{$\partial\Sigma$}; 
\draw (0,-1.75) ellipse (1.25 and 0.3) node[][fontscale=2]{$\Sigma$} ;
\draw (0,-3.5) ellipse (1.25 and 0.3);
\fill [gray,opacity=0.2] (0,-1.75) ellipse (1.25 and 0.3);
\node at (1.5,-3.5/4) [fontscale=2] {$\partial\mathcal{B}$};
\node at (0,-3.5/4) [fontscale=2]{$\mathcal{B}$};
\end{tikzpicture}
    \caption{\small A Cauchy surface $\partial \Sigma$ of the boundary spacetime $\partial\mathcal{B}$ is shown. The field theory state $\psi[\{\chi\}]$ lives on $\partial \Sigma$. A slice $\Sigma$ whose boundary is $\partial \Sigma$ is also shown. The WDW-state $\Psi[g]$, dual to $\psi[\{\chi\}]$, lives on $\Sigma$.}
    \label{AdStincan}
\end{figure}
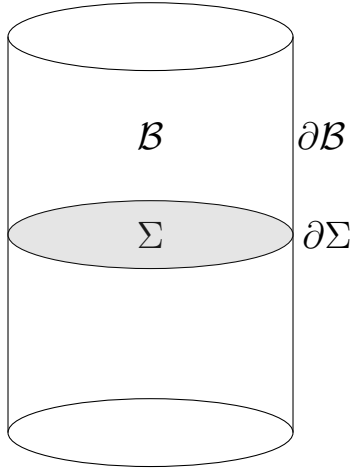

Consider a holographic field theory living on a manifold $\partial\mathcal{B}$ with topology $\mathbb{R}\cross \partial \Sigma$.
%where $N$ is the topology of the \emph{closed} Cauchy slice $\partial\Sigma$, 
%and with background metric $g^\text{bdy}$.
The field theory state $\psi[\{\chi\}]$ lives on $\partial\Sigma$ (see Fig $\ref{AdStincan}$). We now explain how to map this state to a state in the QG Hilbert space. The dual QG state $\Psi[g]$ is to be considered living on some slice $\Sigma$ whose boundary is $\partial\Sigma$. The argument $g$ in $\Psi[g]$ is the metric on this slice $\Sigma$.

Implementing time evolution of the field theory state from $\partial\Sigma_t$ to $\partial\Sigma_{t+\delta t}$, where $t$ is the boundary time, corresponds to some unitary operator acting on $\psi[\{\chi\},t]$. Mapping this new state to a QG state on a slice $\Sigma_{t+\delta t}$, whose boundary is $\partial\Sigma_{t+\delta t}$, we obtain an emergent bulk description interpolating between the two slices $\Sigma_{t}$ and $\Sigma_{t+\delta t}$.

It is crucial \emph{not} to interpret the slices $\Sigma_t$ as being embedded in some prior background bulk manifold $\mathcal{B}$. Rather, we should interpret $\mathcal{B}$ as emerging from the information in the set $\{\Psi[g;t],\;\forall t\}$.\footnote{We allow the possibility for different topologies of $\Sigma$, subject to the  restriction that its boundary must be $\partial\Sigma$. Hence the QG state $\Psi[g;t]$ should be interpreted as a superposition of metrics g on $\Sigma$ and also (implicitly) a superposition of different topologies on $\Sigma$. Henceforth, we will interpret the notion of bulk Cauchy slice $\Sigma$ as encoding these topological features. %In addition, even though the topology of $\Sigma$ should be taken also as an argument of the QG state (the QG state $\Psi[g, \text{Top}(\Sigma);t]$ would give us an amplitude to have the metric $g$ and a particular topology of $\Sigma$ at boundary time t), we will take this as implicitly understood from now on.
\label{superpositionoftopologies}} The only good notion of time evolution is the \emph{boundary} time of $\partial\mathcal{B}$, since time evolution in the bulk is a gauge symmetry (assuming the lapse $N$ falls off fast enough as one approaches the boundary). Any good notion of bulk time should be considered as emerging from the WDW state by  a semiclassical approximation \cite{Kiefer:2021zdq, Halliwell:1987eu, Vilenkin:1988yd, Kiefer:1991ef, Page:1983uc, RKhan}.

We now proceed to construct the isomorphism between the two Hilbert spaces explicitly.  By explicit, we mean that the map can be written simply in terms of the $T^2$ partition function.

%\subsubsection*{\underline{Step 1}: Defining an inner-product preserving map $K^*\to\mathcal{H}_\text{bdy}$} 

\subsection{The Bulk $\to$ Boundary Map}

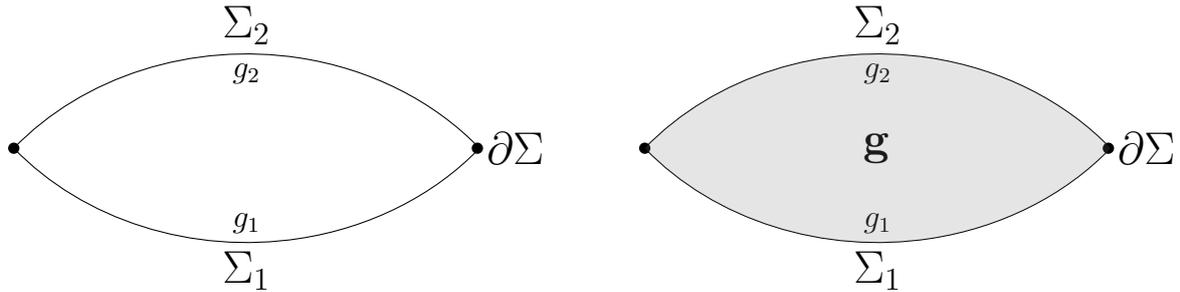
\begin{figure}
\centering
\begin{subfigure}{.47\textwidth}
  \centering
  \begin{tikzpicture}
  \draw (0,0) arc (220:320:4 and 3.5) node[midway,below] [fontscale=3]{$\Sigma_1$} node[midway,above] [fontscale=1]{$g_1$} ;
  \draw (0,0) arc (140:40:4 and 3.5) node[midway,above] [fontscale=3]{$\Sigma_2$} node[midway,below] [fontscale=1]{$g_2$};
  \node [xshift=0cm] at (6.6,0) [fontscale=3]{$\partial\Sigma$};
  \node at (0,0)[circle,fill,inner sep=1.5pt]{};
  \node at (6.1,0)[circle,fill,inner sep=1.5pt]{};
  \end{tikzpicture}
  \caption{The lines represent $d$-dimensional Euclidean manifolds $\Sigma_i$ with metric $g_i$. Their boundaries are glued together at $\partial\Sigma_1=\partial\Sigma_2=\partial\Sigma$. This diagram represents the partition function of the field theory on $\Sigma_1 \cup \Sigma_2$.}
  \label{sametimeqft}
\end{subfigure}
\quad \quad
\begin{subfigure}{.47\textwidth}
  \centering
   \begin{tikzpicture}
  \draw (0,0) arc (220:320:4 and 3.5) node[midway,below] [fontscale=3]{$\Sigma_1$} node[midway,above] [fontscale=1]{$g_1$} ;
  \draw (0,0) arc (140:40:4 and 3.5) node[midway,above] [fontscale=3]{$\Sigma_2$} node[midway,below] [fontscale=1]{$g_2$};
  \node [xshift=0cm] at (6.6,0) [fontscale=3]{$\partial\Sigma$};
  \node at (0,0)[circle,fill,inner sep=1.5pt]{};
  \node at (6.1,0)[circle,fill,inner sep=1.5pt]{};
  \node at (6.1/2,0) [fontscale=3]{$\mathbf{g}$};
  \fill [gray,opacity=0.2] (0,0) arc (220:320:4 and 3.5);
  \fill [gray,opacity=0.2] (0,0) arc (140:40:4 and 3.5);
  \end{tikzpicture}
  \caption{The shaded region represents the gravitational path integral over the space-filling manifolds whose only boundary is $\Sigma_1 \cup \Sigma_2$. The bulk metric $\mathbf{g}$ satisfies Dirichlet boundary condition (i.e. $\mathbf{g}|_{\Sigma_1}=g_1$ and $\mathbf{g}|_{\Sigma_2}=g_2)$. }
  \label{sametimeqg}
\end{subfigure}
\caption{\small By the GHP, the partition function of the field theory (represented by the hollow figure on the left) is equal to the gravitational path integral (represented by the solid figure on the right).}
\label{sametimefig}
\end{figure}

When the deformed partition function $Z^{(\mu)}[g,\{\chi\}]$ (given by \eqref{deformingflow1}) is interpreted as a functional of the boundary set of eigenvalues $\{\chi\}$ (for a fixed $g$), it gives us a map $J$ from the dual vector space of distributions $K^*$ to the boundary Hilbert space $\mathcal{H}_{\text{CFT}}$:
\begin{eqnarray} \label{mapJ}
   J: K^* &\to& \mathcal{H}_{\text{CFT}},
   \\ \label{ketmap}
   \delta^{\infty}[g-g_{0}] &\mapsto& \boxed{\psi[\{\chi\}]=Z[g_0,\{\chi\}]}.
\end{eqnarray}
Using the GHP in equation \eqref{GHP}, one can show that the map $J$ from $K^*$ to $\mathcal{H}_{\text{CFT}}$ is an inner product preserving map. This is because the bulk path integral is equal to the partition function of the field theory on two bulk Cauchy slices $\Sigma_1$ and $\Sigma_2$ whose boundaries are glued together ($\partial \Sigma_1 = \partial \Sigma_2$), with the background metric $g_1$ on $\Sigma_1$ and $g_2$ on $\Sigma_2$ (see Fig. $\ref{sametimefig}$):\footnote{The symbol $d\{\chi\}$ stands for the measure on the space of eigenvalues $\{\chi\}$.  Since we have not specified the basis for the CFT, our notation needs to be interpreted flexibly. If the $\{\chi\}$ basis is discrete, the integral should be replaced with a sum $\sum_{\{\chi\}}$.  If it is a local field in a path integral formulation of the CFT, then we need a functional integration measure $D\{\chi\}$.}

\begin{eqnarray}
     \langle g_2 \,|\, g_1 \rangle_I &=&  I(\delta^{\infty}[g-g_{1}],\delta^{\infty}[g-g_{2}])\\
    &=& \int d\{\chi\}\, \overline{Z}[g_2,\{\chi\}] Z[g_1,\{\chi\}]
    \\
    &=&\int d\{\chi\}\, \, \overline{\psi}_2[\{\chi\}] \, \psi_1[\{\chi\}]
    \\
    &=&  \langle \psi_2 \ | \ \psi_1 \rangle,
\end{eqnarray}
where the bar stands for complex-conjugation.  However, in a time-reversal symmetric theory (such as a holographic dual to pure gravity) we can omit the bar symbol when the metric $g$ is Euclidean (since in Euclidean signature, time-reversal is just complex conjugation).

%(However, $\overline{Z}[g,\{\chi\}] = Z[g,\{\chi\}]$ and $\overline{\psi}[\{\chi\}] = \psi[\{\chi\}]$ for Euclidean metrics in a field theory which is time-reversal symmetric once Wick-rotated to Lorentzian signature,\footnote{This is because the operation of time reversal in Lorentzian signature maps to complex conjugation in Euclidean signature.} and thus we neglect the bars from now onwards.  In the models we are considering, this must at least be true at low energies, since the field theory is dual to pure gravity, but in more general instances of holography the bar will be necessary.)

\subsection{Invertability of the Map} 

Since the map $J$ in Eq. \eqref{mapJ} preserves the inner product on $K^*$ and $\mathcal{H}_{\text{CFT}}$, it maps all the elements in the same equivalence class in $K^*$ to the same element in $\mathcal{H}_{\text{CFT}}$ and so this defines a map from $Q$ to $\mathcal{H}_{\text{CFT}}$, where $Q$ was defined in \eqref{Q}:
\begin{equation}
     J: Q \to \mathcal{H}_{\text{CFT}}.
\end{equation}
We can also show that this map is one-to-one. Suppose two different elements $q_1$, $q_2 \in Q$ have the same dual $\psi \in \mathcal{H}_{\text{CFT}}$. Then by finding a third element $q_0 \in Q$ whose dual is $\psi_0 \in \mathcal{H}_{\text{CFT}} $ such that:
\begin{equation}
 I(q_0,q_1) \ne I(q_0,q_2),
\end{equation}
we get
\begin{equation}
    \langle \psi_0 \ | \ \psi \rangle \ne \langle \psi_0 \ | \ \psi \rangle,
\end{equation}
which is a contradiction. Hence this map is one-to-one. 

It is reasonable to think that that the map $J$ is also surjective if we start with a generic CFT without global symmetries, for which every operator can be generated from OPEs of the stress-tensor $T_{ab}$.  This property should be inherited by $\Pi_{ab}$ in the deformed field theory, since the deformation is itself just a function of the metric and $\Pi_{ab}$.  Hence, one would expect to get every state in $\mathcal{H}_{\text{CFT}}$ by some superposition of bulk Cauchy-slice metrics $g$.  The only obvious reason for this to fail, is if the field theory has a global internal symmetry group ${\cal G}$, in which case the metric $g$, being neutral, cannot generate states charged under ${\cal G}$.  However, in such cases, there should also be a bulk gauge symmetry corresponding to ${\cal G}$, and it is clear that in such cases we ought to have considered additional bulk matter fields besides pure gravity.  On general grounds there should also be bulk objects charged under this symmetry \cite{Harlow:2018jwu}. Then we expect that once such bulk fields are included, the map would become surjective.

Assuming therefore that $J$ is surjective, 
%(if it is not surjective then we take the image of this map $J(Q) = \mathcal{H'}_{\text{bdy}}$ and proceed with this instead), 
we have an isomorphism $Q \leftrightarrow \mathcal{H}_{\text{CFT}}$. Since $Q$ is also isomorphic to $C$\footnote{It is actually $\text{cl}\;\mathcal{C}$ as per the more rigorous subsection \ref{rigorjunky}.}, we have just defined an isomorphism between the boundary CFT Hilbert space and the bulk quantum gravity Hilbert space:
\begin{eqnarray} 
   \mathcal{H}_{\text{QG}} &\longleftrightarrow& \mathcal{H}_{\text{CFT}},
   \\
   \Psi[g]= \langle g \,|\, g_0 \rangle_I &\longleftrightarrow& \psi[\{\chi\}] = Z[g_0,\{\chi\}].\label{map2}
\end{eqnarray}
From this point on, we will refer to Eq. \eqref{map2} as ``state duality".

\subsection{The Boundary $\to$ Bulk Map} 

Unfortunately the map to $C$ is not yet very explicit, since it requires solving the gravitational path integral in the bulk.  Indeed, in the absence of a nonperturbative description of bulk quantum gravity (and hence a UV completion of the path integral), this map from $Q$ to $C$ may not even be fully defined outside of the semiclassical saddle-point approximation.  

Fortunately, we can also define the map $\mathcal{H}_\text{CFT} \to C$, and hence $Q \to C$, in a purely field theoretic manner.  In this way, the holographic theory \emph{determines} the dynamics in the bulk.  From this perspective, the $T^2$ deformation can be taken as the definition of the nonperturbative quantum gravity theory.  Admittedly it is also not yet known how to UV complete the $T^2$ field theory (except for a flat metric in $d = 2$), but since it is a theory with a nondynamical metric $g$, this unsolved problem may be easier to solve than traditional quantum gravity.\footnote{For example, one might identify the $T^2$ deformed theory as the IR regime of an RG flow from some UV field theory, modified by a \emph{relevant} deformation in the UV.  Note that one should not take the \emph{strict} IR limit, because that would give you the boundary CFT without the irrelevant deformation.  If this one little thing can be done, then the result of this section gives a nonperturbative definition of both quantum gravity and the AdS/CFT dictionary.}  

Starting from a field theory state $\ket{\psi}\in \mathcal{H}_{\text{CFT}}$, we construct a quantum gravity state $\ket{\Psi}\in \mathcal{H}_{\text{QG}}$ by integrating over the $\{\chi\}$ variables in the field theory:
%using our basis of WDW states built from the field theory $\{Z[g_0,\{\chi\}]\;|\; \delta^{\infty}[g-g_{0}]\in Q\}$. The coefficients are given by the field theory wavefunctional $\psi[\{\chi\}]$: 
\begin{equation} \label{map1}
  \boxed{\Psi[g]= \int d\{\chi\}\,  Z[g,\{\chi\}] \, \psi[\{\chi\}]}
\end{equation}

Since $Z[g,\{\chi\}]$ satisfies the Hamiltonian and momentum constraints, so does $\Psi[g]$. This defines an explicit map between the Hilbert spaces: $\mathcal{H}_{\text{CFT}}\to\mathcal{H}_{\text{QG}}$, using only field theory concepts.

\subsection{Composition of Maps}

Furthermore, using the GHP, one can show that the map in Eq. \eqref{map1} is compatible with the Hilbert space isomorphism defined by \eqref{map2}; or in other words, holography encodes the \emph{same} bulk dynamics as the gravitational path integral.  

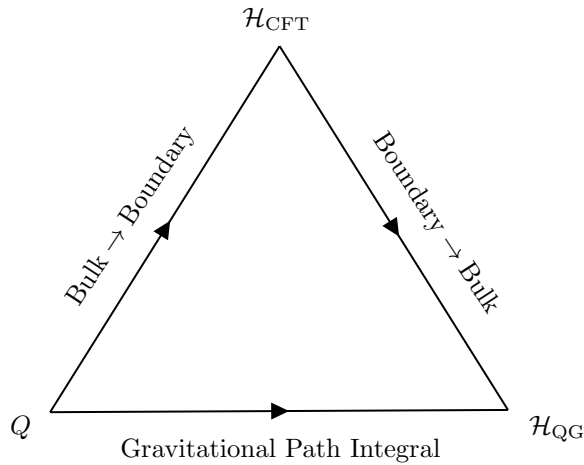
\begin{figure}[h]
\centering

\tikzset{every picture/.style={line width=0.75pt}} %set default line width to 0.75pt        

\begin{tikzpicture}[x=0.75pt,y=0.75pt,yscale=-1,xscale=1]
%uncomment if require: \path (0,431); %set diagram left start at 0, and has height of 431

%Straight Lines [id:da20535770426076494] 
\draw    (321,141) -- (435,324) ;
\draw [shift={(380.64,236.74)}, rotate = 238.08] [fill={rgb, 255:red, 0; green, 0; blue, 0 }  ][line width=0.08]  [draw opacity=0] (8.93,-4.29) -- (0,0) -- (8.93,4.29) -- cycle    ;
%Straight Lines [id:da3348444321680766] 
\draw    (206.43,325.13) -- (321,141) ;
\draw [shift={(266.36,228.82)}, rotate = 121.89] [fill={rgb, 255:red, 0; green, 0; blue, 0 }  ][line width=0.08]  [draw opacity=0] (8.93,-4.29) -- (0,0) -- (8.93,4.29) -- cycle    ;
%Straight Lines [id:da34256361884765996] 
\draw    (206.43,325.13) -- (435,324) ;
\draw [shift={(325.72,324.54)}, rotate = 179.72] [fill={rgb, 255:red, 0; green, 0; blue, 0 }  ][line width=0.08]  [draw opacity=0] (8.93,-4.29) -- (0,0) -- (8.93,4.29) -- cycle    ;

% Text Node
\draw (301,120.4) node [anchor=north west][inner sep=0.75pt]    {$\mathcal{H}_{\text{CFT}}$};
% Text Node
\draw (444,325.4) node [anchor=north west][inner sep=0.75pt]    {$\mathcal{H}_{\text{QG}}$};
% Text Node
\draw (185,325.4) node [anchor=north west][inner sep=0.75pt]    {$Q$};
% Text Node
\draw (240,337) node [anchor=north west][inner sep=0.75pt]   [align=left] {Gravitational Path Integral};
% Text Node
\draw (375.49,180.73) node [anchor=north west][inner sep=0.75pt]  [rotate=-56.86]  {$\text{Boundary}\rightarrow \text{Bulk}$};
% Text Node
\draw (210.92,267.71) node [anchor=north west][inner sep=0.75pt]  [rotate=-303.36]  {$\text{Bulk}\rightarrow \text{Boundary}$};

\end{tikzpicture}

\caption{\small The GHP implies that this triangle diagram commutes, i.e. the composition of the bulk $\to$ boundary map and the boundary $\to$ bulk map, is equivalent to the gravitational path integral, which maps states in the quotient kinematic space $Q$ to the constrained state space ${\cal H}_\text{QG}$.}\label{triangle}
\end{figure}

We show this by starting with a field theory state $\psi_0[\{\chi\}]=Z[g_0,\{\chi\}] \in \mathcal{H}_{\text{CFT}}$.  We map it to $\mathcal{H}_\text{QG}$ using $\eqref{map1}$ to get:
\begin{eqnarray}
   \Psi_0[g] &=& \int d\{\chi\}\,  Z[g,\{\chi\}] \, \psi_0[\{\chi\}],
   \\
   &=& \int d\{\chi\}\,  Z[g,\{\chi\}] \, Z[g_0,\{\chi\}],
   \\
   &=:&Z[g,g_0],
\end{eqnarray}
where $Z[g,g_0]$ (see Fig $\ref{sametimeqft}$) is just the partition function of the field theory on $\Sigma_1 \cup \Sigma_2$ with $\partial\Sigma_1 = \partial\Sigma_2$ and the metric $g$ on $\Sigma_2$ and the metric $g_0$ on $\Sigma_1$.  By the GHP, this equals the gravitational path integral over space-filling manifolds $\mathcal{M}$ whose boundary is $\Sigma_1 \cup \Sigma_2$ (see Fig $\ref{sametimeqg}$):
\begin{equation} \label{sametime}
     \Psi[g]= Z[g,g_0] = \sum_\mathcal{M} \int_{\mathbf{g}|_{\Sigma_1} = g_0}^{\mathbf{g}|_{\Sigma_2} = g}
\frac{D\textbf{g}}{\text{Diff}(\mathcal{M})}\,e^{ \pm iI_\text{grav}[\textbf{g}]}=\langle g \,|\, g_0 \rangle_I,
\end{equation}
which is the same state that one gets from $\psi_0[\{\chi\}]$ using the state duality \eqref{map2} (see Fig. \ref{triangle}).

%\footnote{Recall from equation \eqref{friedelduo} that there are two linearly independent CFTs (of opposite anomaly) which can be deformed into a WDW state. The reader might worry that we could define an alternate map with the same form of \eqref{map1} but using the linearly independent $T^2$ deformed theory. However, Friedel shows that one of the CFT does not have a good semiclassical limit at the boundary $\partial\mathcal{B}$ and so he picks only one. This is the CFT we start with in equation \eqref{friedelduo}. }

%\footnote{Recall from equation \eqref{friedelduo} that there are two linearly independent CFTs (of opposite anomaly) which can be deformed into a WDW state. The reader might worry that we could define an alternate map with the same form of \eqref{map1} but using the linearly independent $T^2$ deformed theory. However such a map would be inconsistent with the GHP. It may be that this is a way of generating states which are in $C$ but not in $\mathcal{H}_\text{QG}.$  Our formalism allows for the possibility of the quantum gravity Hilbert space being a proper subset of the constraint subspace: $\mathcal{H}_{\text{QG}}\subset C$ (cf. \eqref{ImC}).  \textcolor{red}{Keep or remove?}}

Assuming the maps in Fig. \ref{triangle} are surjective, the inverse maps (going opposite to the arrows) are implicitly defined, but we were unable to identify any more explicit expressions.  In particular, we observe that integrating a WDW state $\Psi[g]$ over the spatial metric $g$ does \emph{not} provide a valid bulk to boundary map:
\begin{equation}\label{fakemap}
\psi[\{\chi\}]
\:\ne\:
\int Dg\,\Psi[g]\,Z[g,\{\chi\}] =: \langle Z^* | \Psi \rangle_K
\end{equation}
where on the right hand side, $\langle \cdot | \cdot \rangle_K$ is a \emph{kinematic} norm acting on the kinematic state space $K$ defined in section \ref{HS}, consisting of path integrating the product over all spatial metrics.  This formula fails to be well-defined, because this kinematic inner product does not take into account the momentum and Hamiltonian constraint equations.  But the CFT inner product should be dual to the \emph{dynamical} inner product $\langle \cdot | \cdot \rangle_I$ defined by \eqref{I_1} acting on ${\cal H}_\text{QG}$--- since it is not the gauge-symmetries on either side, but rather only the physical content of the theories, which are dual in AdS/CFT. %This inner product is not the correct one to use on WDW states, which satisfy the constraint equations.  
Therefore \eqref{fakemap} does not properly encode the AdS/CFT duality (and indeed there are divergent factors in the RHS of \eqref{fakemap}, due to spurious integrals over the gauge redundancies.)\footnote{It would be possible to remove the momentum redundancies using the alternative measure $Dg/\text{Diff}(\Sigma)$, but removing the Hamiltonian constraint redundancies would require inverting the gravitational path integral, removing the possibility of a simple expression.}

\section{Asymptotic Dynamics and Unitarity}\label{AD}

In this section, first (\ref{time}) we will show that the bulk time evolution which is given by the Lorentzian path integral is consistent with the boundary time evolution using the GHP. Next (\ref{admH=cftH}) we will show that the ADM Hamiltonian of the bulk is equal to the Hamiltonian of the dual boundary field theory, at least in the large $N$ (classical bulk) limit. In section \ref{compatible} we argue that the dynamics defined by this approach is compatible with the usual bulk reconstruction.  Finally, in section \ref{unitarity} we will argue that the bulk quantum gravity theory is naturally unitary, both with respect to a kinematic norm and with respect to the dynamical norm.

\subsection{Consistency of Time Evolution}\label{time}

%In section $\ref{Wave function of Gravity from a partition function of deformed field theory}$ we started with an Euclidean CFT on an Euclidean manifold $\Sigma$ (Eq. \eqref{euc}) and deformed the theory (Eq. \eqref{deformingflow1}) to get it to satisfy the Hamiltonian constraint (Eq. \eqref{ham}). The Hamiltonian constraint has a particular relative sign between the extrinsic and intrinsic terms depending on the signature of the normal direction. Changing the slice from having timelike to spacelike normal flips this relative sign. For us, this is equivalent to having to pick different dimensionless parameters $(\alpha,\beta)$, as defined in Eqs. \eqref{ct} and \eqref{defo}. Thus, a WDW-state on a timelike slice (as opposed to a spacelike Cauchy slice) would correspond to a deformation of the same type (starting from a Lorentzian CFT), just with a different pair $(\alpha,\beta)$. This modification leads to exactly the deformation proposed in \cite{hartman} for finite-radius holography with $\Lambda<0$ and it corresponds to the field theory we will put on the timelike boundary in the argument that follows.

In section \ref{dictionary} we showed how to map field theory states to quantum gravity states. We also know how to time evolve the field theory states, using the CFT Hamiltonian. With these, we can figure out how to time evolve the quantum gravity bulk states with respect to the asymptotic boundary time $t$. Let us call this type of bulk evolution the \emph{inherited evolution}, because it is inherited from the boundary CFT.  For consistency, the inherited time evolution must be given by the gravitational path integral with an appropriate timelike boundary. We will show precisely this now.

\begin{figure}[ht]
\centering
\begin{subfigure}{.47\textwidth}
  \centering
  \begin{tikzpicture}[scale=1.5]
\draw (0,0) ellipse (1.25 and 0.3)  node[] [fontscale=1.5]{$g$} ;
\draw (-1.25,0) -- (-1.25,-3.5);
\draw (1.25,-3.5) -- (1.25,0) node[midway,right][fontscale=2]{$\partial\mathcal{B}|^t_0$}; 
\draw (0,-3.5) ellipse (1.25 and 0.3) node[] [fontscale=1.5]{$g_0$};
\node at (0,.5) [fontscale=2]{$\Sigma_t$};
\node at (0,-4) [fontscale=2]{$\Sigma_0$};
\node at (1.37,0) [fontscale=1]{$t$};
\node at (1.62,-3.5) [fontscale=1]{$t=0$};
\end{tikzpicture}
  \caption{This hollow figure represents the partition function $Z[g,t;g^\text{bdy};g_0,0]$ (in Eq. $\eqref{difftimepartition}$) of the field theory on $\partial\mathcal{M}=\Sigma_t \cup \partial\mathcal{B}|^t_0 \cup \Sigma_0$. The metric on $\Sigma_0$ is $g_0$, on $\Sigma_t$ is $g$ and on $\partial\mathcal{B}|^t_0$ is $g^{bdy}$.}
  \label{Differenttimeqft}
\end{subfigure}
\quad \quad
\begin{subfigure}{.47\textwidth}
  \centering
  \begin{tikzpicture}[scale=1.5]
\draw (0,0) ellipse (1.25 and 0.3)  node[] [fontscale=1.5]{$g$} ;
\draw (-1.25,0) -- (-1.25,-3.5);
\draw (1.25,-3.5) -- (1.25,0) node[midway,right][fontscale=2]{$\partial\mathcal{B}|^t_0$}; 
\draw (0,-3.5) ellipse (1.25 and 0.3) node[] [fontscale=1.5]{$g_0$};
\node at (0,.5) [fontscale=2]{$\Sigma_t$};
\node at (0,-4) [fontscale=2]{$\Sigma_0$};
\node at (1.37,0) [fontscale=1]{$t$};
\node at (1.62,-3.5) [fontscale=1]{$t=0$};
\node at (0,-3.5/2) [fontscale=2] {$\mathbf{g}$};
\fill [gray,opacity=0.2] (-1.25,0) -- (-1.25,-3.5) arc (180:360:1.25 and 0.3) -- (1.25,0) arc (0:180:1.25 and 0.3);
\end{tikzpicture}
  \caption{The shaded region represents the gravitational path integral (in Eq. $\eqref{timeevolved}$) over the space-filling manifolds whose boundary is $\partial\mathcal{M}=\Sigma_t \cup \partial\mathcal{B}|^t_0 \cup \Sigma_0$. The bulk metric $\mathbf{g}$ satisfies Dirichlet boundary condition (i.e. $\mathbf{g}|_{\Sigma_0}=g_0$ , $\mathbf{g}|_{\Sigma_t}=g$ and $\mathbf{g}|_{\partial\mathcal{B}|^t_0}=g^\text{bdy}$). }
  \label{Differenttimeqg}
\end{subfigure}
\caption{\small By the GHP, the partition function of the field theory (represented by the hollow figure on the left) is equal to the gravitational path integral (represented by the solid figure on the right). The two Cauchy slices $\Sigma_0$ and $\Sigma_t$ are separated from each other by the boundary time $t$.}
\label{Differenttime}
\end{figure}
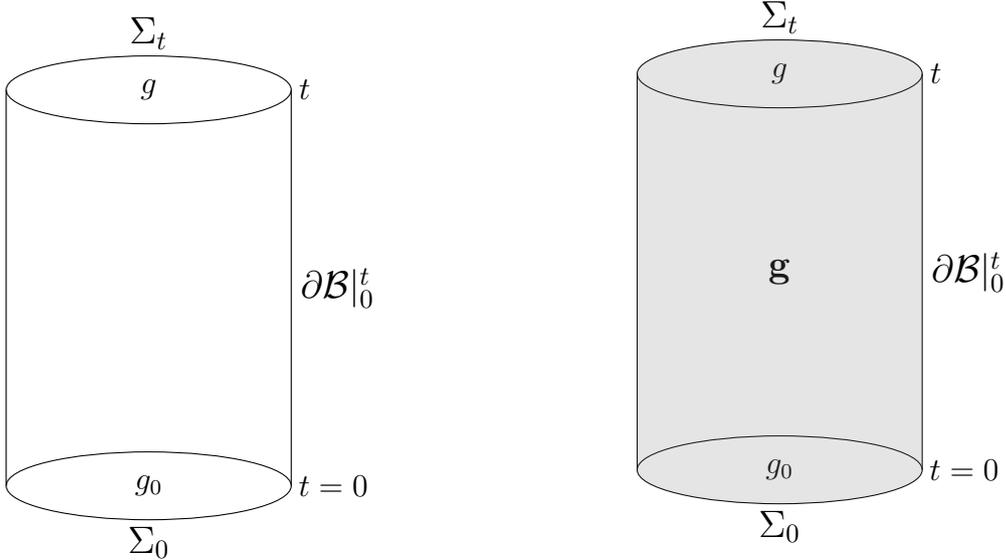

At time $t=0$ let the bulk be in a metric ket state $|\, g_0 \rangle$,
%\begin{equation}
 %  % C \ni \Psi_0[g] =  \langle g %\,|\, g_0 \rangle_I,
%    \langle g \,|\, g_0 %\rangle_I = \Psi_0[g] \in %\mathcal{H}_{\text{QG}},
%\end{equation}
which corresponds to $\delta^{\infty}[g-g_0]$ in $K^*$, which we assume corresponds to a nonzero (albeit probably non-normalizable) vector in the physical Hilbert space $Q = \mathcal{H}_{\text{QG}}$. To time evolve this state, we first take its dual in the boundary Hilbert space (using Eq. $\eqref{map2}$):
\begin{equation}
    \psi_0[\{\chi\}]= Z[g_0,\{\chi\}].
\end{equation}

Let $Z_\text{CFT}[g^\text{bdy}, \{\chi,\tilde{\chi}\}]^t_0$ represent the field theory path integral on the truncated boundary cylinder $\partial\mathcal{B}|^t_0 =[0,t]\cross \partial\Sigma$ (see Fig $\ref{Differenttimeqft}$) with background metric $g^\text{bdy}$ and boundary conditions $\{\tilde{\chi}\}$ on $\partial\Sigma_0$ and $\{\chi\}$ on $\partial\Sigma$. 
%Also let $Z[g_0,\{\tilde{\chi}\}]$ be the corresponding field theory path integral on the $bulk$ Cauchy slice $\Sigma_0$ with boundary $\partial\Sigma_0$. 
Using this cylinder, we can evolve our field theory state at time 0 to get another state at time $t$:
\begin{eqnarray}
    \psi[\{\chi\},t] &=& \int d\{\tilde{\chi}\}\,\braket{\{\chi\},t\,\,}{\phantom{\sum}\!\!\!\!\!\!\!\{\tilde{\chi}\},0}
    %\left\langle\{\chi\},t\,|\,\{\tilde{\chi}\},0\right\rangle\,
    \psi_0[\{\tilde{\chi}\}] ,
    \\
    &=& \int d\{\tilde{\chi}\}\,Z_\text{CFT}\!\left[g^\text{bdy}, \{\chi,\tilde{\chi}\}\right]^t_0\, Z[g_0,\{\tilde{\chi}\}].
\end{eqnarray}

Now we map this field theory state at time $t$ to a quantum gravity state (using Eq. $\eqref{map1}$):
\begin{eqnarray}
   \Psi[g,t] &=& \int d\{\chi\}\,  Z[g,\{\chi\}]\, \psi[\{\chi\},t],
   \\
   &=&\int d\{\chi\}\,  \int d\{\tilde{\chi}\}\,\, Z[g,\{\chi\}]\, Z_\text{CFT}\!\left[g^\text{bdy}, \{\chi,\tilde{\chi}\}\right]^t_0\, Z[g_0,\{\tilde{\chi}\}],
   \\
   &=& Z_{\partial\mathcal{M}}\!\left[g|g^\text{bdy}|g_0\right]^t_0,\label{difftimepartition}
\end{eqnarray}
where $Z_{\partial\mathcal{M}}[g|g^\text{bdy}|g_0]^t_0$ is the partition function on the closed cylinder $\partial \mathcal{M} = \Sigma_t \cup \partial\mathcal{B}|^t_0 \cup \Sigma_0$, which glues $Z_\text{CFT}$ on the side to the deformed partition function $Z^{(\mu)}$ on the top and bottom. The metric on $\Sigma_t$ is $g$ and the metric on $\Sigma_0$ is $g_0$.\footnote{Note that in order to continuously glue the components of this manifold we need the background metrics to match at the gluing points. Thus, we require $\lim_{z\to{0}}(z^2g)|_{\partial\Sigma_t}=g^\text{bdy}|_{\partial\Sigma_t}$ and $\lim_{z\to 0}(z^2g_0)|_{\partial\Sigma_0}=g^\text{bdy}|_{\partial\Sigma_0}$, where the metric rescaling can be seen from the asymptotic expansion in \eqref{FG0}.} It is important to note that in order for this gluing to be well-defined, we need the Hilbert spaces of the CFT and $T^2$-deformed theories to match at the corners. This requires the CFT to be shifted by infinite counterterms, which is to be understood from now on every time we write $Z_\text{CFT}$. Alternatively, if we regulate the problem by bringing the boundary $\partial\mathcal{B}|^t_0 $ to a small but finite distance, we would need a $T^2$-deformed theory on the timelike boundary itself, so that the whole boundary cylinder $\partial\mathcal{M}$ has a smoothly-joining $T^2$ theory on it, albeit with a change of signature at the corners. It is only in such a context that the GHP was even defined. By then taking the limit of the sides of the cylinder going to infinity, we recover the CFT on the timelike portion of the boundary, up to the counterterms just mentioned. 

By the GHP, $Z_{\partial\mathcal{M}}[g|g^\text{bdy}|g_0]^t_0$ is then equal to the gravitational path integral over the space-filling manifolds, whose boundary is $\partial\mathcal{M}$ (see Fig $\ref{Differenttimeqg}$), so 
\begin{equation}
\Psi[g,t] = \sum_\mathcal{M} \int_{\mathbf{g}|_{\Sigma_0}=g_0}^{\mathbf{g}|_{\Sigma_t}=g}
\frac{D\textbf{g}}{\text{Diff}(\mathcal{M})}\,e^{ \pm iI_\text{grav}[\textbf{g}]},\;\;\;\;\;(\mathbf{g}|_{\partial\mathcal{B}|^t_0}=g^\text{bdy}). \label{timeevolved}
\end{equation}
Although Eq. $\eqref{timeevolved}$ looks similar to Eq. $\eqref{sametime}$, the difference is that in Eq. $\eqref{sametime}$ the two Cauchy slices have the same boundary (corresponding to the same boundary time), while in Eq. $\eqref{timeevolved}$ the two Cauchy slices $\Sigma_t$ and $\Sigma_0$ are separated from each other by the boundary time $t$.\footnote{Recall that there were straddles in the Lorentzian path integral \eqref{TA} when the boundaries of the two Cauchy slices matched. But if there is a finite boundary time between $\partial\Sigma_0$ and $\partial\Sigma_t$, then the straddles will be restricted to an open subregion of the Cauchy slices away from the boundary (as these Cauchy slices are always spacelike or at most null hypersurfaces). For a sufficiently large boundary time evolution, there could be no straddles in the bulk. } The inherited time evolved state in Eq. $\eqref{timeevolved}$ is the same as the expected gravitational path integral (with the appropriate boundary conditions). Hence consistency is shown between the bulk and boundary time evolution.

%It is interesting to consider the limiting case of taking the timelike boundary $\partial\mathcal{M}$ to infinite distance, by which we mean sending $\text{Vol}(\partial\Sigma)\to\infty$. Performing an infinite Weyl rescaling of the metric $g$, while keeping $\mu$ fixed, is equivalent to keeping the metric fixed, while sending $\mu\to0$. By looking at the form of the deformed field theory in \eqref{deformingflow1} (and eqns \eqref{ct}, \eqref{defo}), we can clearly see that:
%\begin{equation}
   % \lim_{\mu\to0}Z^{(\mu)}[g,\{\chi\}]= \lim_{\mu\to0}\left(e^{\text{CT}(\mu)}\right)Z_{CFT}[g,\{\chi\}]
%\end{equation}
%\textcolor{red}{Should connect with Freidel \cite{freidel}.}
%Thus, we recover the boundary CFT corresponding to the following asymptotically-AdS duality:
%\begin{equation}
  %  e^{\text{CT}}Z_{CFT} \approx e^{-I_{grav}[\textbf{g}]}
%\end{equation}
%The presence of the counterterms is a consequence of having defined the bulk gravity action to be $I_{grav}=I_{EH}+I_{GH}$ \emph{without} the holographic counterterms. This was done so as to obtain the Hamiltonian constraint in its canonical form. This differs from the ``simpler" form of the duality, in \eqref{ads/cft}, which gives a Hamiltonian operator modified by a canonical transformation implemented by the usual holographic counterterms.
%\textcolor{red}{Hey Aron : Can you let us know if this holographic counterterm argument is correct. Also the holographic counterterm has exactly the same form as our \text{CT}. This gives us confidence that this is the right deformation. Because ... we will explain in detail in call.}

\subsection{ADM Hamiltonian $=$ CFT Hamiltonian }\label{admH=cftH}

In the previous section, we were assuming the GHP in order to prove that the bulk and boundary time evolution is the same.  In this section we demonstrate a key consistency check: that the CFT Hamiltonian equals the bulk ADM Hamiltonian.  Recall that in the canonical formulation, the ADM Hamiltonian is given by
\bea
\text{H}_\mathrm{ADM} =
\int_\Sigma d^dx\,\Big(N{\cal H}(x) + N^a{\cal D}_{a}(x)\Big) + \int_{\partial \Sigma} 
d^{d-1}x\,{\cal Q}.
\eea
where ${\cal Q}$ is the ADM flux out at infinity.  Since the first term vanishes in the dynamical Hilbert space, in this section we identify $\text{H}_\mathrm{ADM}$ with the boundary term, written in a form that we will derive in the Appendix.

Since the ADM Hamiltonian can be identified as an operator in the $T^2$ Cauchy slice field theory, this equality can be derived on the field theory side of the duality, without recourse to the GHP.  Since the ADM Hamiltonian generates bulk evolution in $\cal H_\text{QG}$, this confirms the result of the previous section.

Our derivation below relies on the factorization property of the $T^2$-operator. Even in $d=2$ this is only known to hold for the case of a flat background metric \cite{Zamolodchikov:2004ce}. For general curved backgrounds (and higher dimension) we will imagine ourselves to be in a large-$N$ t'Hooftian limit of the starting CFT, as a means of obtaining factorizability.\footnote{If there are other ways to make the $T^2$-operator factorize, other than through t'Hooftian field theories, then such theories could also be good candidates. But we are only aware of the large $N$ method, which has in addition been explicitly verified to give a dual bulk description in the known AdS/CFT dualities.} Thus (as discussed in section \ref{largeN}) we work below at leading order in an $1/N$ expansion, although we expect that our conclusion continues to hold at all orders in $1/N$.

Consider a $T^2$-deformed field theory living on a Euclidean manifold on which there is a crease, \emph{i.e.} a discontinuous first derivative of the background metric. This curvature singularity will induce a discontinuous jump in the stress-energy tensor $T^{ab}$. In a general field theory, computing this jump may be difficult or even undefined (if there are additional divergences at the corner).  However, the $T^2$-deformed field theory satisfies a Hamiltonian constraint equation (as well as the usual conservation equation), which allows us to derive a finite answer for the junction condition at the crease. 

\begin{figure}[h]
\centering

\begin{tikzpicture}

\node[inner sep=0] at (0,0) {\includegraphics[width=.9\textwidth]{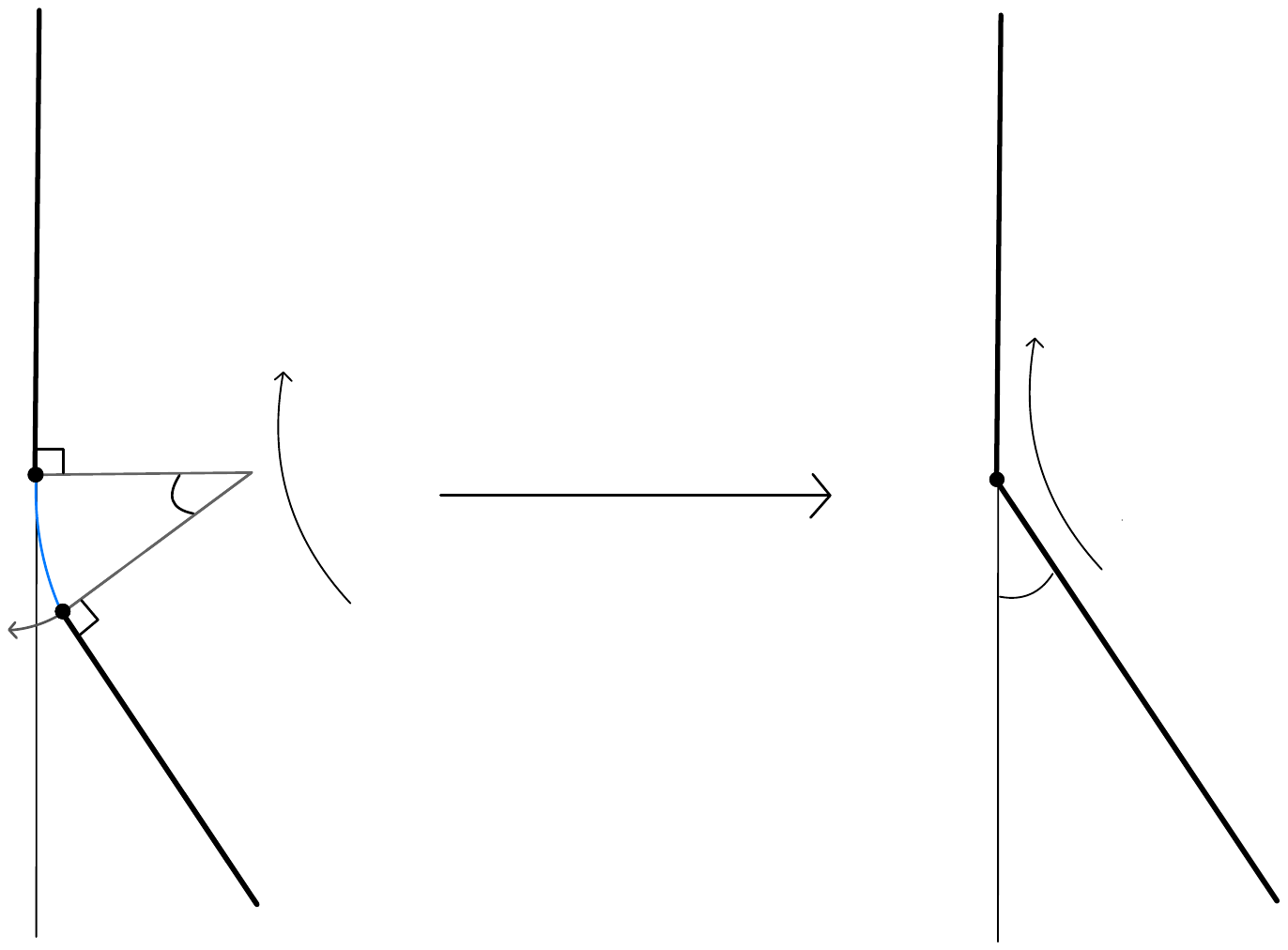}};
\node at (0.3,0.1) [fontscale = 1]{$R \to 0$};
\node at (4.3,0.8) [fontscale=1]{$\tau$};
\node at (-2.6,0.8) [fontscale=1]{$\tau$};
\node at (3.2,0) [fontscale=1]{$\partial \Sigma$};
\node at (-5.7,0) [fontscale=1]{$\partial \Sigma$};

\node at (-4.2,0.3) [fontscale=1]{$R$};
\node at (-4,-1) [fontscale=1]{$R$};
\node at (-4.2,-0.25) [fontscale=1]{$\delta\alpha$};
\node at (4.1,-1.4) [fontscale=1]{$\delta\alpha$};

\node[align=center] at (-3.4,-3) [fontscale=2]{$\Sigma$};
\node[align=center] at (6.1,-3) [fontscale=2]{$\Sigma$};

\node at (-6,-0.8) [fontscale=1]{$\partial \Sigma_{\text{smooth}}$};
\node at (-6,-1.5) [fontscale=1]{$\tau=0$};
\node at (-5.8,3) [fontscale=1]{$\partial \mathcal{B}^{(+)}$};
\node at (3,3) [fontscale=1]{$\partial \mathcal{B}^{(+)}$};

\end{tikzpicture}
\caption{\small The $T^2$-deformed field theory lives on the Euclidean manifold $\partial\mathcal{B}^{(+)}\cup\Sigma$, as shown on the right. The angle is taken to be infinitesimal at this point of the analysis. Shown is the neighbourhood of the junction. The coordinate $\tau$ measures proper distance on both sides of $\partial\Sigma$ and takes value $\tau=0$ at the junction. The left figure shows a smooth geometry with a cap $\partial\Sigma_\text{smooth}$ which is an arc of a circle of radius $R$. The limit $R\to 0$, while keeping the angle $\delta\alpha$ fixed, recovers the final infinitesimal kink. }\label{junction_fig}
\end{figure}

We can imagine embedding this manifold into a one dimension higher flat Euclidean space, in such a way that the metric induced on it is the background metric of the field theory. From this perspective, the kink corresponds to an angle between the two sides of the junction, as shown in the right panel of Figure \ref{junction_fig}.

We can restrict to the following scenario, relevant to AdS/CFT. Take the boundary of the spacetime $\partial\mathcal{B}$ and consider its junction $\partial\Sigma$ with a bulk Cauchy slice $\Sigma$. $\partial\Sigma$ divides $\partial\mathcal{B}$ into two semi-infinite cylinders which we call $\partial\mathcal{B}^{(+)}$ and $\partial\mathcal{B}^{(-)}$.  Considering the junction conditions between $\Sigma$ and either of these (say $\partial\mathcal{B}^{(+)}$) will imply that $\text{H}_\text{QFT}=\text{H}_\text{ADM}$. 

In order to regulate this problem we bring the AdS boundary $\partial\mathcal{B}$ to finite (but very large) distance, in which case it is known that the dual field theory living on $\partial\mathcal{B}$ becomes the $T^2$-deformed version of the original CFT on a very large cylinder; in fact it is the same theory that lives on $\Sigma$ (but in Lorentzian signature if the boundary is Lorentzian). The dual gravity theory in $\mathcal{B}$ is one with Dirichlet boundary conditions at $\partial\mathcal{B}$.  
%We can carry out an expansion of the bulk metric in Fefferman-Graham gauge in the neighbourhood of $\partial\mathcal{B}$, in which corrections arise due to metric fluctuations in the bulk. But all we will need is this asymptotic behaviour of the metric induced on $\Sigma$ near the junction. 
In the end, we will take the limit of the boundary going to infinite proper distance, in which case the theory living on $\partial\mathcal{B}$ becomes the original CFT, up to a counterterm.

We now derive these junction conditions for the case of a Euclidean boundary $\partial\mathcal{B}$. The AdS/CFT result will then follow by analytic continuation back to Lorentzian signature.

%For concreteness, we will be working in $d=3$ dimensions.  \textcolor{red}{Is this restriction necessary?} 

Let the field theory live on the d-dimensional manifold $\partial\mathcal{B}^{(+)}\cup\Sigma$ and let the junction be a $\tau=0$ surface. We can perform an ADM foliation in which the metric in the neighbourhood of the junction is:

\begin{equation}
    ds^2 = d\tau^2 + g_{ij} dx^i dx^j. 
\end{equation}

Consider a $T^2$-deformed field theory on $\partial\mathcal{B}^{(+)}\cup\Sigma$ which satisfies the following relations at every point\footnote{Notice the relative sign is positive and this is because we will eventually think of this being embedded in the euclidean bulk. In this section we are using the convention in which $\Pi^{ab}$ takes real values, although our result should also apply in Lorentzian signature.}:
\begin{equation}
    \frac{16\pi G_N}{\sqrt{g}}\Big(\Pi_{ab}\Pi^{ab}-\frac{1}{d-1}\Pi^2\Big)+\frac{\sqrt{g}}{16\pi G_N}(R-2\Lambda)=0,
\end{equation}
\begin{equation}
    \nabla_b\Pi^{ab}=0.
\end{equation}
We require the metric to be differentiable on either side of the junction, but its derivative along the $\tau$ direction ($\partial_{\tau} g_{ij}$) is discontinuous at the junction. This makes the Christoffel symbols ($\Gamma^\tau_{ij} , \Gamma^i_{\tau j}$) have discontinuities across $\tau=0$. Consequently, the curvature scalar will diverge like $\sim\delta(\tau)$. For the relations to keep holding everywhere we will need compensating divergences in the stress-tensor.  We will denote any weights of Dirac delta divergences by a superscript: $A=A^{(\delta)}\delta(\tau)+\text{finite terms}$, for any given quantity $A$.

In order to compute the jump for a finite angle, let us first solve for the situation of infinitesimal angle $\delta\alpha$. In this case, all the weights go as $A^{(\delta)}\sim \delta\alpha$, to leading order.  The constraints above lead to the following requirements relating the coefficients of the divergent terms:\pagebreak[4]
\begin{eqnarray}
   \Pi^{i(\delta)}_i&=&\Pi^{j(\delta)}_j \ \ \forall i,j=1,...,d-1, \\
   2 \delta\alpha\;\Pi^\tau_\tau&=&\frac{\sqrt{\text{det}(g)}}{16\pi G_N}R^{(\delta)},\label{junk1}\\
   \Delta\Pi^{\tau}_\tau&=&\frac{\sqrt{\text{det(g)}}}{16\pi G_N}\delta\alpha \sum_{i=1}^{d-1}\Gamma^i_{\tau i} \label{junk2},
\end{eqnarray}
where the symbol $\Delta$ stands for the jump of a quantity across the crease, $\Delta A=A^{(+)}-A^{(-)}$. We further introduce the notation $A^{(+)}:=A|_{\tau=0^+}$ and $A^{(-)}:=A|_{\tau=0^-}$. Also $\text{det}(g)$ is the value of the d-dimensional metric determinant evaluated at $\tau=0$. The equations \eqref{junk1} and \eqref{junk2} are valid only to leading order in $\delta\alpha$. To derive them the following argument was needed to relate the divergence of the ``pressure" terms of $\Pi^{ab}$ to the angle $\delta\alpha$.

We imagine obtaining the infinitesimal angle kink at $\partial\Sigma$ by first considering a smooth interpolating manifold $\partial\Sigma_\text{smooth}$ between $\partial\mathcal{B}^{(+)}$ and $\Sigma$. We take the proper size of this smooth cap to be infinitesimal. We can imagine embedding this manifold into a bulk space of one higher dimension, which we can take to be flat in the neighbourhood of this infinitesimal cap.\footnote{We take the scale of this smoothing geometry to be much smaller than any curvature scale of the bulk space.} Thus, usual Euclidean geometry applies and we take the bulk metric to be:
\begin{equation}
    ds^2=dr^2+r^2d\phi^2+g_{ij}dx^idx^j,
\end{equation}
where we have defined the angular variable $\phi=\tau/r$ and we take $R$ to be the proper radius of the arc of circle $\partial\Sigma_\text{smooth}$ in the bulk geometry. The centre of the circle is at $r=0$. The kinked geometry with folding angle $\delta\alpha$ is then recovered in the limit of $R\to0$, while keeping the angle spanned by the arc fixed. With this setup one can obtain that the divergence of the stress-tensor diagonal components is related to the embedding angle via:
\begin{equation}\label{theta}
    \Pi^{i(\delta)}_i=\frac{\sqrt{\text{det(g)}}}{16\pi G_N}\;\delta\alpha,\; \forall i.
\end{equation}

The trace of the extrinsic curvature of the junction $\partial\Sigma$  as embedded in $\Sigma$ (or $\partial\mathcal{B}^{(+)}$) is $K^{(-)}$ ($K^{(+)}$). In terms of the intrinsic geometry on $\Sigma$ ($\partial\mathcal{B}^{(+)}$) this can be written as:
\begin{equation}
    K^{(\pm)}=\sum_{i=1}^{d-1}\Gamma_{\tau i}^{i\;(\pm)}.
\end{equation}
Because we are taking the angle to be infinitesimal we can write, to leading order in $\delta\alpha$, $\Delta A=A^{(-)}(0)-A^{(-)}(\delta\alpha)=:-\delta A$.  

\begin{figure}[h]
\centering
\begin{tikzpicture}

\node[inner sep=0] at (0,0) {\includegraphics[width=.9\textwidth]{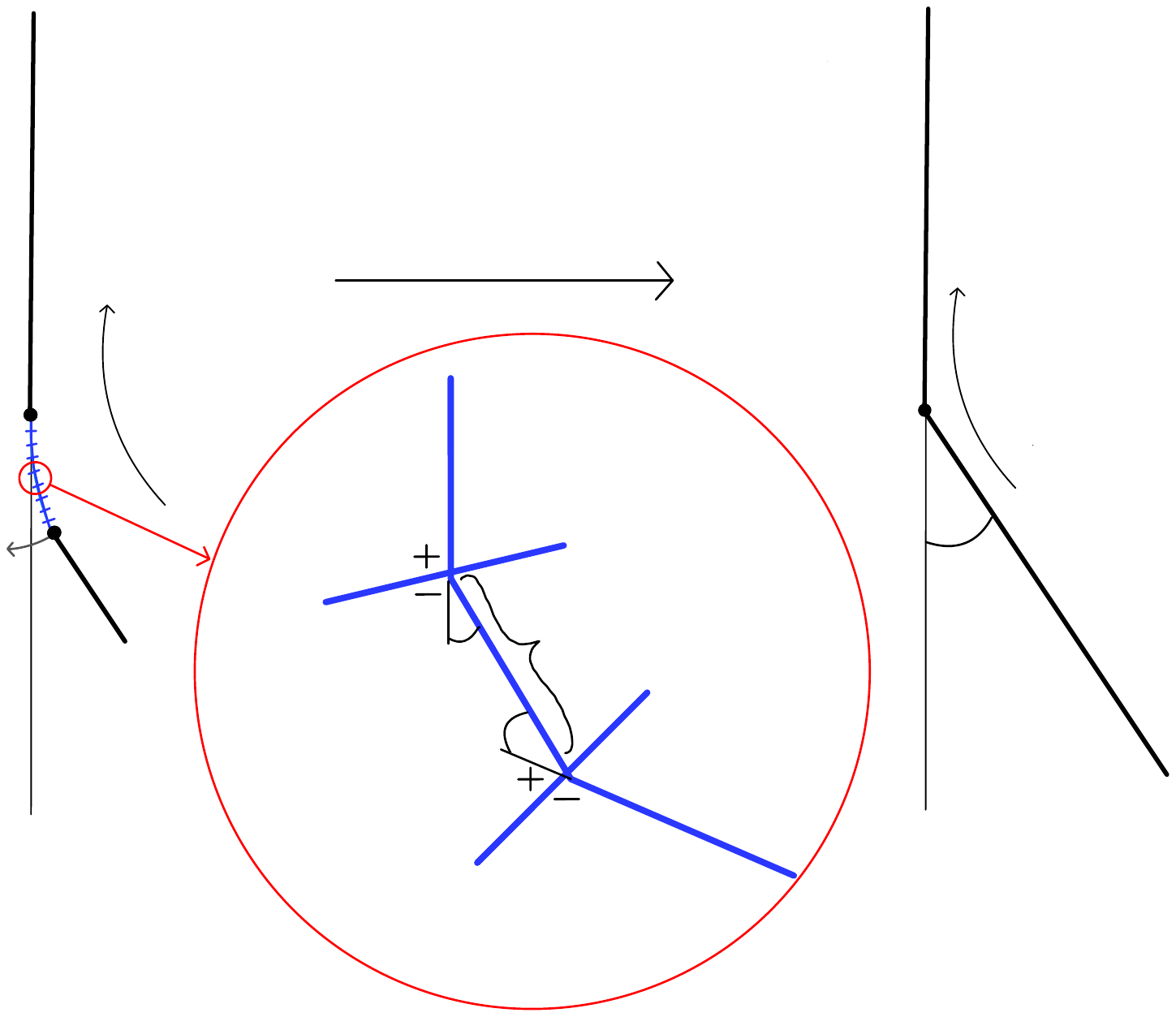}};

\node at (4.3,2) [fontscale=1]{$\tau$};
\node at (-4.8,2) [fontscale=1]{$\tau$};
\node at (3.2,1) [fontscale=1]{$\partial \Sigma$};
\node at (-6.4,1) [fontscale=1]{$\partial \Sigma$};

\node at (4,-0.7) [fontscale=1]{$\alpha_0$};

\node[align=center] at (-5,-0.9) [fontscale=2]{$\Sigma$};
\node[align=center] at (5.3,-0.9) [fontscale=2]{$\Sigma$};
\node at (-1,2.7) [fontscale=1]{$\epsilon \to 0$};
\node at (-6.8,0.2) [fontscale=1]{$\partial \Sigma_{\text{rough}}$};
\node at (-6.8,-0.5) [fontscale=1]{$\tau=0$};
\node at (-6.6,4) [fontscale=1]{$\partial \mathcal{B}^{(+)}$};
\node at (3,4) [fontscale=1]{$\partial \mathcal{B}^{(+)}$};

\node at (-0.4,-1.35) [fontscale=1]{$\epsilon$};
\node at (-1.3,-1.7) [fontscale=1]{$\delta\alpha$};
\node at (-1.2,-2.2) [fontscale=1]{$\delta\alpha$};

\end{tikzpicture}
\caption{\small This figure shows how to get from an infinitesimal angle to a finite angle via successive kinks. In the limit that the proper size $\epsilon$ of each segment is taken to 0, we have that $\alpha=N\delta\alpha$, where N is the number of kinks, which will be taken to $\infty$ while keeping $\alpha$ fixed. At each kink we can perform the infinitesimal kink analysis, with the notion of $(+)$ and $(-)$ being updated at each step. Working to leading order in $\epsilon$ we can equate $A^{(-)}$ with $A^{(+)}$ of each segment.}\label{rough_fig}
\end{figure}

Up to now everything holds for an infinitesimal kink angle. We now build up the case of finite angle $\alpha_0$ in the following way. We imagine regulating a finite angle kink by introducing an interpolating manifold $\partial\Sigma_\text{rough}$ comprised of many infinitesimal kinks of angle $\delta\alpha$, as shown in Figure \ref{rough_fig}. At each such kink the relations derived above hold. Let the proper size of each segment be $\epsilon$, which is also infinitesimal. Working to leading order in $\epsilon$ we can equate $A^{(-)}$ with $A^{(+)}$ within each segment, where the notions of $(+)$ and $(-)$ keep being updated along $\partial\Sigma_\text{rough}$. From this perspective, equations \eqref{junk1} and \eqref{junk2} can now be turned into a system of coupled first order ODEs, which holds at every intermediate value of the angle:
 
\begin{eqnarray}
   \frac{d}{d\alpha}\Pi^{\tau\tau}&=& \frac{\sqrt{\text{det(g)}}}{16\pi G_N}\;K,\\
   \frac{d}{d\alpha}K&=& -\frac{16\pi G_N}{\sqrt{\text{det}(g)}}\;\Pi^{\tau\tau}.
\end{eqnarray}
Integrating these up to the final angle $\alpha_0$ we can obtain the solution for the discontinuous jump across the finite angle kink of the original setup. The boundary conditions are:
\begin{eqnarray}
    \Pi^{\tau\tau}(\alpha=0)&=&\Pi^{\tau\tau(+)},\\
    K(\alpha=0)&=&K^{(+)}.
\end{eqnarray}
Solving for $\Pi^{\tau\tau(-)}(\alpha)$ and $K^{(-)}(\alpha)$ with the above boundary conditions, we can re-arrange to get the following junction conditions:\footnote{These junction conditions were originally derived in \cite{Wall_CEB} as a method for proving a covariant version of the holographic entropy bound.}
\begin{eqnarray}\label{junction}
    \Pi^{\tau\tau(+)}&=&\cos{(\alpha_0)}\Pi^{\tau\tau(-)}+\frac{\sqrt{\text{det(g)}}}{16\pi G_N}\sin(\alpha_0)K^{(-)} \label{juncsol1},\\
    K^{(+)}&=&-\frac{16\pi G_N}{\sqrt{\text{det}(g)}}\sin(\alpha_0) \Pi^{\tau \tau(-)} +\cos{( \alpha_0)} K^{(-)}\label{juncsol2}.
\end{eqnarray}
As the metric on the boundary $\partial \mathcal{B}^{(+)}$ is time translation invariant (boundary of asymptotically AdS spacetime), $K^{(+)}=0$, which imposes a constraint on the field theory operators $\Pi^{\tau\tau(-)}$ and $\alpha_0$ via \eqref{juncsol2}.

We conclude that the field theory observable
\begin{equation}\label{sqrt}
    \overline{\Pi^{\tau\tau}}:=\sqrt{\left(\frac{16\pi G_N}{\sqrt{\text{det}(g)}}\Pi^{\tau \tau}\right)^2+K^2}
\end{equation} 
is constant across the junction, independently of the angle $\alpha_0$. %This operator is well-defined provided $\Pi^{\tau\tau}$ is self-adjoint. 
By equating the values of $\overline{\Pi^{\tau\tau}}$ on either side of the junction and integrating over the $S^{d-1}$ boundary, we obtain an equality between two field theory operators, one of which is the boundary generator of time evolution, $\text{H}_\text{QFT}$ (because $K^{(+)}=0$), and the other is given the suggestive symbol $\text{H}_\text{ADM}$:
\begin{equation}\label{qft=adm}
    \text{H}_\text{QFT}=\text{H}_\text{ADM}.
\end{equation}
They are both operators acting on the field theory Hilbert space. If we now take the proper size of our boundary $\partial\mathcal{B}$ to be infinite, this is equivalent to taking the limit of vanishing deformation in the $T^2$-deformed field theory. In this limit, the boundary theory looks like a CFT, up to an infinite counterterm. Thus, $\text{H}_\text{QFT}$ becomes the same as $\text{H}_\text{CFT}$, up to this counterterm, which simply carries out an infinite shift in the energy spectrum of the theory.  Hence we find:
\begin{equation}\label{cft=adm}
    \boxed{\text{H}_\text{CFT}=\text{H}_\text{ADM}}
\end{equation}

It is crucially important to realize that the analysis above was purely field-theoretic, simply making use of properties of $T^2$-deformed theories.

Classically, the observable on the RHS of \eqref{sqrt} resembles the form of the ADM Hamiltonian in Einstein gravity. This is shown in the Appendix. Upon canonical quantization we can upgrade it to an operator acting on $\mathcal{H}_{\text{QG}}$, expressed in terms of the canonical variables ($g_{ab},\Pi^{ab})$ where now $\Pi^{ab} = -i\delta/\delta g_{ab}$. The question remains whether this operator is the generator of time evolution on $\mathcal{H}_{\text{QG}}$. We can now show this easily via our duality map.

On the junction $\partial\Sigma$ lives a boundary CFT state $\ket{\psi,t}$. On the bulk slice $\Sigma$ ending on $\partial\Sigma$ we have the quantum gravity state $\ket{\Psi,t}$, as given by the duality map. We now show that the WDW states satisfy the Schr\"{o}dinger equation with the ADM Hamiltonian generating time translations. We start by acting with the ADM Hamiltonian on the WDW state, which we are allowed to do since it is an operator defined on $\Sigma$, and we make use of our duality map \eqref{map1}:
\begin{eqnarray}
    \text{H}_{\text{ADM}} \Psi[g,t] &=& \int d\{\chi\}\,   (\text{H}_{\text{ADM}}\,Z[g,\{\chi\}])\; \psi[\{\chi\},t]\\
    &=& \int d\{\chi\}\,   (\text{H}_{\text{CFT}}\,Z[g,\{\chi\}])\; \psi[\{\chi\},t]\\
    &=& \int d\{\chi\}\,   Z[g,\{\chi\}]\; \left(\text{H}_{\text{CFT}}\,\psi[\{\chi\},t]\right)\\
    &=& \int d\{\chi\}\,   Z[g,\{\chi\}]\; i\frac{\partial}{\partial t}\left(\psi[\{\chi\},t]\right)\\
    &=& i\frac{\partial}{\partial t}\Psi[g,t].
\end{eqnarray}
In going to the second line we used the equality of field theory operators that we derived above. In going to the third line we used the self-adjointness of the CFT Hamiltonian. To do this we notice that the expression is nothing but an inner product on the boundary theory. In particular, 
\begin{equation}
   \int d\{\chi\}\,   (\text{H}_{\text{CFT}}\,Z[g,\{\chi\}])\; \psi[\{\chi\},t]=\braket{\text{H}_{\text{CFT}}\,\phi}{\psi(t)}=\braket{\phi}{\text{H}_{\text{CFT}}\,\psi(t)}, 
\end{equation}
where we have defined the state $\phi[\chi]:=Z[g,\{\chi\}]$.

So we see that our duality map maps the boundary and bulk generators of time evolution to each other. In other words, the dynamics is consistent on both sides of the duality. Thus, the quantum gravity theory inherits a Schr\"{o}dinger evolution generated by the ADM Hamiltonian. This can be seen as a derivation of one of the key aspects of the AdS/CFT dictionary.

\subsection{Compatibility with other Bulk Reconstruction Approaches}\label{compatible}

A word is necessary about the compatibility of the dynamics defined in this section and section \ref{dictionary} with the usual approaches to bulk reconstruction.  There is a large literature about how to perturbatively reconstruct AdS fields using CFT data, e.g.~\cite{Balasubramanian:1998sn,Balasubramanian:1998de,Banks:1998dd,Son:2002sd,Satoh:2002bc,Herzog:2002pc,Marolf:2004fy,Lawrence:2006ze,Skenderis:2008dg,Hamilton:2005ju,Hamilton:2006az,Kabat:2011rz,Kabat:2012hp,Heemskerk:2012mn,Heemskerk:2012np,Kabat:2013wga,Morrison:2014jha}.  The general approach to reconstruction in these papers rests on two basic premises:
\begin{enumerate}
\item The asymptotic value of any bulk field $\varphi$ limits (after a suitable rescaling) to a CFT primary $\cal O$,
\item 
The bulk fields $\varphi$ obey the bulk dynamical equations of motion.
\end{enumerate}
Here the first premise is necessary for there to be a nontrivial dictionary at the boundary $\partial {\cal B}$, while the second premise is required to say anything about fields a finite distance in the interior of the bulk ${\cal B}$.  (In Lorentzian signature it is also necessary to impose suitable initial and final boundary conditions \cite{Herzog:2002pc,Marolf:2004fy,Lawrence:2006ze,Skenderis:2008dg}, which can be derived from the Euclidean signature dictionary.)

In our approach both of these premises are guaranteed to be satisfied.  Premise 1 follows because the timelike CFT boundary $\partial {\cal B}$ is described by the $\mu \to 0$ limit of the $T^2$-deformed theory; therefore every field which exists at finite distances in the bulk (finite $\mu$) will have some limiting analogue at $\mu = 0$.  Premise 2 follows because, as shown in \ref{thegeneraldeformation}, the Cauchy slice theory obeys the Hamiltonian constraint equation ${\cal H}(x) = 0$, which encodes all the dynamics of the theory and thus implies the bulk equations of motion. (We will also get the correct boundary conditions at infinity, so long as we obtain the right ADM boundary term $H_{\text{ADM}}$, as argued in \ref{time} and \ref{admH=cftH} above).\footnote{In addition, the proper relation between the Euclidean and Lorentzian signature follows if the $T^2$ theory matches onto the correct quantum gravity contour in AdS, an issue raised but not completely addressed in section \ref{contour}.}

Since these assumptions are verified in our approach, anything that follows from them should also be the case, and hence in the abstract our approach \emph{must} agree with these reconstruction approaches, assuming the AdS/CFT duality is itself consistent.  However, we leave detailed checks of this agreement to future work.

There is another family of approaches to bulk reconstruction based on holographic entropy surfaces \cite{Faulkner:2013ica,lashkari2014gravitational,lashkari2016canonical,Swingle:2014uza,Czech:2012bh,Jafferis:2014lza,Jafferis:2015del,Dong:2016eik,Kim:2016ipt,Faulkner:2017vdd,Cotler:2017erl,Hayden:2018khn,Chen:2019gbt}, and information theoretic quantities such as relative entropy or modular flow in the corresponding entanglement wedges.  Checking compatibility with this approach would probably require the study of gluing rules for the $T^2$ deformed theory on curved manifolds \cite{Wall_CEB}, which goes beyond the scope of this paper (except for our motivational remarks about tensor networks in section \ref{TN}).  Relatedly, it would be interesting to see if the bulk/boundary maps defined in section \ref{dictionary} could be extended to obtain the entanglement wedge subregion form of the duality \cite{Czech:2012bh,Wall:2012uf,Headrick:2014cta,Almheiri:2014lwa}, where only part of the CFT and bulk is specified.

\subsection{Emergence of Bulk Unitarity}\label{unitarity}

As shown in the previous section, the boundary theory living on Cauchy slices is nonunitary.  One might worry that the dual bulk theory will therefore also be nonunitary.  Fortunately, this will not be the case.  In order for the bulk time evolution to be unitary, two basic properties are needed: (i) positivity of norms, and (ii) preservation of norms under time evolution, i.e. a self-adjoint Hamiltonian.  And we can show that both of these properties are satisfied.  

We can consider unitarity either with respect to a dynamical inner product (after imposing the constraint equations) or with respect to a kinematic inner product (prior to imposing the Hamiltonian constraint equations).  The theory turns out to be unitary in \emph{both} senses, but for different reasons.

%Going beyond the semiclassical regime, it is clear that the $T^2$ quantum gravity will in fact be unitary in this sense.  In addition to the usual AdS/CFT argument based on the boundary CFT, this follows from the fact that a partition function of a field theory living on a manifold $\Sigma$ with only one boundary $\partial \Sigma$, always gives rise to a pure state on the boundary.  In this case, $\Sigma$ may be a Cauchy slice that goes behind a the horizon of a black hole formed from collapse (see Fig. \cite{}), and $\partial \Sigma$ is the place where this Cauchy slice intersects the CFT boundary.}

%\cite{Wall:2021bxi}

%There are 3 distinct senses of unitarity which might be considered: (i) positivity of the norm, (ii) preservation of the norm under time-evolution (i.e. the Hamiltonian is self-adjoint) and (iii) pure states do not evolve to mixed states, even in the presence of black holes.  In fact, the theory of $T^2$ deformed AdS Cauchy slices is unitary in all three senses.

\subsubsection*{Positive Dynamical Norm}

In the context of AdS/CFT, the violations of reflection positivity when the Cauchy slice $\Sigma$ is sewn together from pieces, do not spoil the existence of a probabilistic interpretation for the AdS spacetime \emph{taken as a whole}.  As long as the CFT at infinity has a positive inner product, the bulk Hilbert space also does because (as we showed in section \ref{dictionary}) the two Hilbert spaces agree at the dynamical level:
\bea\label{match}
\mathcal{H}_\text{QG} = \mathcal{H}_\text{CFT}.
\eea

\subsubsection*{Self-Adjointness with the Dynamical Inner Product}

The bulk quantum gravity theory should also be unitary in the sense that the Hamiltonian H is self-adjoint: $\text{H}=\text{H}^{\dagger}$.  Now the definition of the dagger symbol depends on the choice of inner product.\footnote{An inner product defines an adjoint as follows: for any operator ${\cal O}$, ${\cal O}^\dagger$ is obtained by both raising and lowering ${\cal O}^*$ using the inner product.}  If we use the inner product on dynamical states which satisfy the WDW constraint equations, then ${\cal H}(x) = 0$ trivially (since time evolution is pure gauge) so we need concern ourselves only with the ADM boundary term at spatial infinity.  As argued above in sections \ref{time}-\ref{admH=cftH}, this boundary term $\mathrm{H_{ADM}}$ equals the CFT Hamiltonian $\mathrm{H_{CFT}}$, so if we start with a unitary CFT then $\mathrm{H_{ADM}}$ is clearly self-adjoint. Thus $\text{H} = \text{H}^{\dagger_I}$ where $I$ is the dynamical inner product of the theory.

As is well-known, the relation $\mathrm{H_{ADM}} = \mathrm{H_{CFT}}$ already seems to imply that information is not lost inside of black holes, because the boundary CFT is unitary.  (Indeed, Marolf has argued that, at the quantum level, boundary unitarity follows solely from the existence of a boundary Hamiltonian $\mathrm{H_{ADM}}$, without appealing to the AdS/CFT dictionary \cite{Marolf:2008mf,Marolf:2008mg}.)  However, it is difficult to see how the boundary unitarity manifests in the bulk to ensure that the Hawking radiation is unitary.

The $T^2$ deformation helps to clarify the issue of what is happening in the bulk.  In addition to the unitary boundary Hamiltonian argument, there is a second reason why it is obvious that the $T^2$ deformed theory does not lose information inside of black holes.  

The reason is that the partition function of a field theory living on a manifold $\Sigma$ with only one boundary $\partial \Sigma$ always gives rise to a pure state on the boundary.  Consider a black hole that forms from collapse from a star initially in a pure state.  If $\Sigma$ is a Cauchy slice that goes inside of the horizon, then the collapsing matter that fell into the black hole will determine the sources $J$ on $\Sigma$.  (If this matter is in a pure quantum superposition, we will need to take a coherent superposition of different $J$ values.)  By the linearity of the bulk to boundary map, the resulting boundary CFT state on $\partial \Sigma$ will necessarily be pure.  %See Fig. \ref{} for a description of this process.\textcolor{red}{Need Figure}

\subsubsection*{Self-Adjointness with the Kinematic Inner Product}

We would also like to see that the Hamiltonian constraint ${\cal H}(x)\Psi = 0$ is in some sense self-adjoint.  But here, we cannot use the dynamical inner product, as all physical states have ${\cal H}(x)\Psi = 0$.  We can, however ask if ${\cal H}(x)$ is self-adjoint with respect to some \emph{kinematic} inner product $\langle \Psi_2|\Psi_1\rangle_K$, defined on the kinematic space $K$ of wavefunctions $\Psi[g]$,\footnote{We defined the kinematic space $K$ in section \ref{HS} as a preliminary towards constructing the Hilbert space, which we were able to do without reference to a kinematic inner product.} prior to imposing the constraint equations.

This kinematic unitarity property ${\cal H}(x) = {\cal H}^{\dagger_K}\!(x)$ is needed to guarantee that the various terms in the Hamiltonian constraint have the right relative phases so that ${\cal H}(x)$ looks real in Lorentzian signature.  (The overall phase of ${\cal H}(x)$ is not very meaningful since the meaning of ${\cal H}(x)\Psi = 0$ is the same regardless, but the phase determines what we mean by a Lorentzian sign for the lapse $N$.)  Without this, there is no reason to expect that the bulk theory reduces to a unitary QFT in the weak gravity limit.  

(This type of kinematic unitarity should definitely \emph{not} be confused with unitarity of the black hole information puzzle, since it refers to unitary evolution from one Cauchy slice to the next.  Indeed, it is in intellectual tension with the unitarity of the black hole S-matrix \cite{Unruh:2017uaw}.  Na\"{i}vely, one might think that this would imply that the Hawking radiation is mixed after tracing out the interior degrees of freedom.  But as we have stated above, on a $T^2$ Cauchy slice this is not so.)

Note that the bulk adjoint $\dagger_K$ we are discussing reverses the order of operators \emph{and sources}, so it does not preserve the normal-ordering prescription defined in section \ref{largeN}.  For the same reason $\dagger_K$ is \emph{not} the same as the adjoint in the boundary theory.  Recall that the usual Euclidean field theory adjoint depends on a foliation and reverses the direction of imaginary time, but does not reverse the order of sources and operators.  This contrasts with $CPT$ which means the same thing on both sides of the duality. (That is, after selecting a particular Cauchy slice $\Sigma$ on the gravity side.  The spontaneous breaking of $CPT$ in the boundary is dual to the non-$CPT$ invariance of the embedding of $\Sigma$ in the bulk $\cal B$; despite this $CPT$ remains a good symmetry of the bulk dynamics, considered in abstraction of a particular slice $\Sigma$.)

The existence of a kinematic norm, with respect to which ${\cal H}(x)$ is self-adjoint, follows as an emergent consequence from the fact that $CPT$ is not \emph{explicitly} broken in the $T^2$ deformed theory.\footnote{We are ignoring questions related to boundaries of the space of metrics, so self-adjointness of ${\cal H}(x)$ really just means that it is Hermitian.  A proper treatment of such boundaries would require an analysis of nonperturbative quantum gravity effects such as singularities.}  To illustrate this phenomenon, let us first consider a Hamiltonian which is quadratic (and derivative-free) in the momenta, such as \eqref{Hscalar}-\eqref{matterHC}.  In that particular example, $C$ acts trivially (all fields are real); $P$ is obviously satisfied (there are no $\epsilon$ permutation symbols); while time reversal $T$ acts by complex conjugation, which includes sending $\Pi_{ab} \to -\Pi_{ab}$, $\Pi_\Phi \to -\Pi_\Phi$.  Since all terms in this Hamiltonian have an even number of $\Pi$'s, $CPT$ therefore guarantees that their coefficients are real.  It follows that, up to (divergent) operator-ordering ambiguities,\footnote{Such operator-ordering ambiguities in general involve lower powers of $\Pi$, and are therefore be implicitly dealt with by the arguments below.} this Hamiltonian ${\cal H}(x)$ is self-adjoint using a standard inner product with respect to which $\Pi_{ab}=\Pi_{ab}^\dagger$, $\Pi_\Phi = \Pi_\Phi^\dagger$, and $g_{ab} = g_{ab}^\dagger$.

The construction of such a standard inner product is not entirely trivial.  If we write the kinematic inner product in the obvious path integral manner as\footnote{We could have included in our measure an integral over spatial diffeos, as was done in \eqref{gamma}, in order to get an intermediate Hilbert space which is invariant under spatial diffeomorphisms but not lapses.  But this would have complicated the discussion of whether ${\cal H}(x)$ is self-adjoint, because $x$ would no longer have an invariant meaning.}
\bea\label{KIP}
\int Dg\,D\Phi\:\Psi^*_2[g,\Phi] \Psi_1[g,\Phi],
\eea
then it follows that $\Pi_{ab}^\dagger = \Pi_{ab}^{\phantom{\dagger}} +i(\delta / \delta g^{ab})Dg +i(\delta / \delta g^{ab})D\Phi$,
where the extra terms do not obviously vanish since covariant path integral measures typically depend on the metric (cf.~\eqref{metric^2}).  We assume that, after a suitable regulation procedure, these extra terms can be eliminated via counterterms.  

Alternatively, we can eliminate such divergences (formally) by defining $\Pi^{ab}$ as the \emph{covariant} functional derivative $-i\nabla_{g_{ab}(x)}$ with respect to a
Wheeler-De Witt metric $G^{A|B}$ on superspace:
\bea
G^{ab|cd} = \frac{1}{\sqrt{g}}\left(\frac{g^{ac}g^{bd}+g^{ad}g^{bc}}{2} - \frac{1}{d-1} g^{ab}g^{cd}\right),\qquad G^{\Phi|\Phi} = \frac{1}{\sqrt{g}}, \qquad G^{ab|\Phi} = 0.
\eea
By taking our path integral measure to be (again formally, since a UV regulator is required to make sense of this expression):
\bea
Dg D\Phi= 
\prod_x d^n\!g \;d\Phi\; \sqrt{\det G},
\eea
where $n = d(d+1)/2$ is the number of components of $g$.  It is now manifest that $\Pi_{ab}$, $\Pi_\Phi$, and all terms in ${\cal H}(x)$ including the kinetic term $G^{A|B}\Pi_{A}\Pi_{B}$ are self-adjoint, since $G^{A|B}$ is covariantly constant (as is ${\cal D}_a(x)$).  Furthermore, the inner product \eqref{KIP} is now manifestly positive.

(It therefore differs from the inner product introduced by DeWitt on superspace \cite{DeWitt:1967yk}, which is of indefinite sign.  The motivation of that (Klein-Gordon-like) inner product depended on the fact that $G^{A|B}(x)$ has Lorentzian signature at each point $x$, and therefore there is a (hyper)-analogue of Cauchy slices in superspace on which this Klein-Gordon norm is conserved.  In a 3rd quantized framework, this Klein-Gordon norm basically counts the number of universes which expand past a specified scale factor, and is non-positive since it counts contracting universes negatively.  We will have no use for that norm in this article.)

Now if we add additional terms to ${\cal H}(x)$ that are linear in the $\Pi$'s, $CPT$ would imply these terms have imaginary coefficient.  At a first glance, this looks non-Hermitian.  However, as we have seen in section \ref{example}, such terms can be eliminated by an imaginary canonical transformation \eqref{canonical} of the form ${\cal H}(x) = e^{-C}\tilde{{\cal H}}(x)e^{C}$ (with $C = C^\dagger$) which shifts the value of $\Pi$'s by a constant.  If one Hamiltonian ${\cal H}(x)$ is self-adjoint with respect to the inner product $\langle \Psi_2 | \Psi_1\rangle$, then the other Hamiltonian $\tilde{\cal H}(x)$ is self-adjoint with respect to the modified inner product
\bea
\widetilde{\langle \Psi_2 | \Psi_1\rangle} = \langle \Psi_2 |e^{-2C} |\Psi\rangle.
\eea
It follows that for a generic CPT-invariant Hamiltonian of this form, we can define an inner product with respect to which it is Hermitian.  Since $e^{2C}$ is positive, this does not spoil the positivity of kinematic inner product, so we also have $\forall \Psi: \langle \Psi | \Psi \rangle \ge 0$.

More generally, consider an arbitrary $CPT$-invariant Hamiltonian constraint $\tilde{\cal H}(x)$ which satisfies the the ADM closure condition \eqref{closure}.  On standard grounds we expect that generically there will also exist a bulk spacetime Lagrangian formulation of this theory with $CPT$-invariant action $I_\text{grav}$, at least to any finite order in perturbation theory.  Since closure implies local Lorentz invariance (section \ref{thegeneraldeformation}), $I_\text{grav}$ should also be Lorentz invariant.  Now in Lorentzian signature, unitarity corresponds to all terms in $I_\text{grav}$ being real,\footnote{Technically this is only true if we stipulate that the measure factors in the gravitational path integral are chosen correctly, but let us assume this has been done.  While reality of the action does not rule out the kind of non-unitarity associated with negative norm states, we can rule out such negative norm states by simply checking that the fields appearing in the low-energy bulk action are standard types of matter fields, which have positive norm.} and such terms are necessarily $CPT$-invariant by the usual $CPT$ theorem \cite{Greaves:2012iq}.  Any imaginary nonunitary terms in the action would therefore be odd under $CPT$ (since $i \to -i$ under $T$) and hence (by a sort of converse to the $CPT$ theorem) a $CPT$ invariant action must necessarily also be unitary in the sense of preserving a norm.\footnote{For some special cases of this argument, see \cite{Kuchar:1974es,Donnelly:2013tia}.}

The above argument is unaffected by the fact that $CPT$ is spontaneously broken when the slice $\Sigma$ is embedded in Lorentzian signature.  This is because $CPT$ is broken by the choice of \emph{solution} to ${\cal H}(x)\Psi = 0$.  However, the \emph{algebraic form} of the constraint ${\cal H}(x)$ is independent of this choice, and hence remains $CPT$-invariant.  This is all that is needed to argue for bulk unitarity.\pagebreak[4]

\subsubsection*{Flowchart}

In order to clarify the conceptual foundations of our approach, we have drawn a chart (Figure \ref{flowchart}) to show how various bulk properties emerge from the premises of our holographic model:

\vspace{25pt}

\tikzset{
block/.style = {rectangle, draw, fill=black!20, 
  text width=6em, text centered, rounded corners, minimum height=6em,node distance=4cm},
block2/.style = {rectangle, draw, fill=black!0, 
  text width=7em, text centered, rounded corners, minimum height=4.2em,node distance=4cm},
block3/.style = {rectangle, draw, fill=black!20, 
  text width=6em, text centered, rounded corners, minimum height=4em,node distance=3cm},
block4/.style = {rectangle, draw, fill=black!0, 
  text width=8em, text centered, rounded corners, minimum height=6em,node distance=3cm},
block5/.style = {rectangle, draw, fill=black!0, 
  text width=7em, text centered, rounded corners, minimum height=4.2em,node distance=3.5cm},
line/.style = {draw, -latex'},
lined/.style = {draw}
}
\begin{figure}[h] \label{emerge}
    \centering
    \begin{tikzpicture}[node distance = 2cm, auto]
    % Place nodes
    \node [block] (init) {No \textbf{explicit} CPT breaking};
    \node [block5, above of=init] (blli) {Bulk local Lorentz invariance};
    \node [block5, below of=init] (scpt) {Spontaneous CPT breaking (for some $g_{ab}$)};
    \node [block2, right of=blli] (bu){Bulk kinematic unitarity: $\exists K$:
    
    \vspace{0.05cm}
    ${\cal H}(x)\!=\!{\cal H}^{\dagger_K}\!(x)$};
    \node [block2,right of=init] (btnu){Boundary theory \textbf{not} reflection positive};
    \node [block2,right of=scpt] (bls){Bulk Lorentzian signature};
    \node[block, left of=blli] (cft){Conformal invariance of starting CFT};
    \node[block, left of=init] (diff){Deformation closes nontrivially (up to ${\cal D}^a$ terms)};
    \node[block, left of=scpt] (quad){
    Deformation quadratic in $\Pi^{ab}$};
    \node[block3, above of=cft] (uscft) {Unitarity of starting CFT};
    \node[block4, above of=bu]
    (bdu) {Bulk dynamical unitarity:
        
    \vspace{0.05cm}
    $\langle \Psi | \Psi \rangle_I \ge 0$  and
    ${\text H}^{\phantom{\dagger_I}}_\text{ADM}={\text H}^{\dagger_I}_\text{ADM}$};
   
    % Draw edges
    \path [lined,dashed] (init) -- (scpt);
    \path [line] (scpt)--(btnu);
    \path [line] (scpt)--(bls);
    \path [lined,dashed] (btnu) -- (bu);
    \path[line] (blli)--(bu);
    \path[line] (init)--(bu);
    \path[line] (cft)--(blli);
    \path[line] (diff)--(blli);
%    \path[line] (quad)--(blli);
    \path[line] (quad)--(scpt);
    \path [line] (uscft)--(bdu);
    \path [lined,dashed] (bdu) -- (bu);
    
\end{tikzpicture}

\vspace{1cm}
    \caption{Chart showing how various properties of the $T^2$ deformed theory relate to properties of the gravity theory. The shaded boxes highlight properties that we impose on the field theory as assumptions. The unshaded boxes correspond to emergent properties of either the field theory or the gravity theory, which arise as consequences. The arrows show implications. The dashed lines illustrate the contrast between two subtly distinct properties.}
    \label{flowchart}
\end{figure}
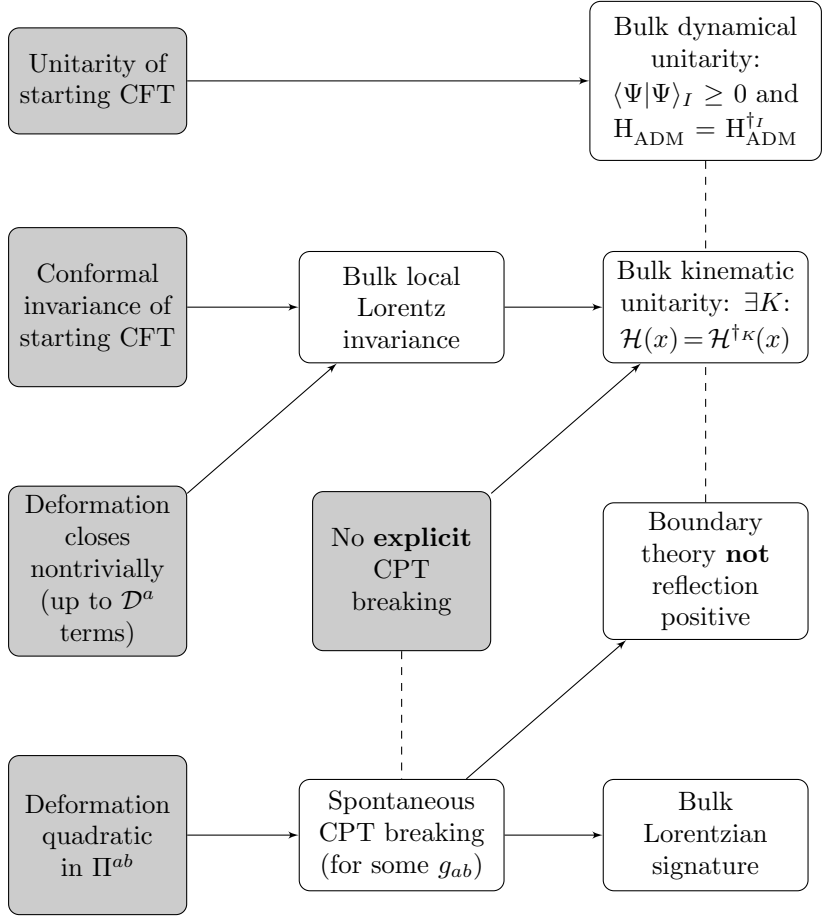

\vspace{25pt}

\section{Discussion}\label{dis}

% \textcolor{blue}{Points to mention:
% \begin{itemize}
%     \item contour problem
%     \item understanding second Freidel solution
%     \item uniqueness (or not) of deformation
%     \item dealing explicitly with finite N complications
%     \item applying this to cosmology
%     \item explicit toy models of spontaneous CPT breaking
%     \item constructing explicitly the deformation that gets bulk supergravity
%     \item UV completion
%     \item tensor networks
%     \item background independence
%     \item massive bulk fields and nonrenormalizable couplings
%     \item a more rigorous proof of positivity of the dynamical norm from the gravitational path integral
% \end{itemize}
% }

\subsection{A New AdS/CFT Dictionary}

The goal of this paper was to reformulate the holographic principle in terms of a field theory living on a bulk Cauchy slice $\Sigma$.  The field theory on $\Sigma$ is defined by the irrelevant $T^2$ deformation.   Hence, it flows to the usual boundary CFT in the IR, but it looks different at short distances.  Unlike the usual formulation of holography, here it is \emph{time} rather than space that is the emergent dimension.  This emergent time dimension corresponds to dynamics in the bulk.

But the distinguishing mark of a holographic theory is not the ability to reconstruct data from a codimension-1 surface---that's already true classically---rather, it's the ability to reconstruct a spacetime from data on a codimension-2 surface that is surprising!  In our work, this property arises naturally from the fact that (just like any other partition function) the numerical evaluation of $Z[\Sigma]$ requires specifying some boundary data $\{\chi \} $ which lives on $\partial \Sigma$.  Since $\partial \Sigma$ is also a slice of the usual boundary CFT (which lives at spatial infinity), the boundary conditions for the Cauchy slice partition function are naturally identified with CFT states.  

This allowed us to uncover a new entry in the AdS/CFT dictionary: a map between states in the Hilbert space of the usual holographic CFT, and states of the bulk canonical quantum gravity theory. 

%The latter are states living on bulk Cauchy slices, not radial states as usually considered in formulations of finite-cutoff AdS/CFT. So they are genuine elements of the Hilbert space of quantum gravity.

%dual field theory lives, not only on the boundary of the spacetime $\partial\mathcal{B}$, but also on a

%We believe this approach can be used to define a general bulk quantum gravity \emph{non-perturbatively} (after UV completing the field theory), for both signs of the cosmological constant and for arbitrary bulk matter content.

The dictionary entries are as follows: Given any (asymptotically hyperbolic) bulk spatial metric $g$, Eq. \eqref{ketmap} gives the corresponding CFT boundary state $\psi$.  Conversely, given a CFT boundary state $\psi$, Eq. \eqref{map1} gives the corresponding bulk state $\Psi[g]$ which satisfies the Wheeler-DeWitt constraint equations.  As this is the key result of our paper, we repeat the formulae below:
\bea\label{2maps}
\psi_\text{CFT}[\{\chi\}] =Z[g_0,\{\chi\}],\qquad
\Psi_\text{WDW}[g]= \int d\{\chi\}\,  Z[g,\{\chi\}] \, \psi_\text{CFT}[\{\chi\}],
\eea
where $g_0$ is a specific choice of spatial metric (and topology) on $\Sigma$.
%Throughout the paper, the topology of $\Sigma$ on which the metric $g$ is defined, was not taken to be fixed but was an argument of the state $\Psi[g]$ and the partition function $Z[g,\{\chi\}]$ implicitly (see footnote \ref{superpositionoftopologies}). 
%So the bulk state we have in our dictionary is a superposition of metric and topologies (which goes beyond the formalism of canonical quantum gravity). 
These maps are explicit in the sense that, if you know how to calculate $T^2$-deformed partition functions $Z$, the maps almost trivially follow!  In previous approaches to the AdS/CFT dictionary, in order to determine the CFT dual $\psi$ to a given spacetime $\cal M$, you first have to figure out how to construct $\cal M$ by boundary CFT path integral constructions (e.g. Euclidean CFT constructions, throwing stuff in from the boundary etc.).\footnote{See \cite{Skenderis:2008dh,Marolf:2017kvq} for attempts to specify Cauchy slice data in the bulk subject to these limitations.}  This left it somewhat unclear whether it is always possible to reconstruct all the data behind an eternal black hole from the dual CFT \cite{Engelhardt:2014gca, Marolf:2012xe}.  But now, using our new entry in the dictionary, one can find the dual to an arbitrary spatial geometry $g$.  This clarifies the manner in which information is able to escape from the black hole, since on any given Cauchy slice $\Sigma$, the $T^2$ deformed theory propagates all information to the boundary (cf.~section \ref{unitarity}).\footnote{The maximum entropy capacity of the $T^2$ deformed theory to transmit information will be discussed in a forthcoming article \cite{Wall_CEB}.  Even if this information capacity is exceeded, the boundary state manifestly \emph{remains pure}, but some postselection must happen on bulk data as suggested in \cite{Brown:2019rox,Engelhardt:2021qjs,Bak:2021qbo}.  Despite this postselection, the evolution of the entire spacetime remains unitary, because the boundary CFT is unitary.}

If we compose the {\sf bulk$\to$boundary} and {\sf boundary$\to$bulk} maps, we get a map which takes any metric $g_0$ to a wavefunctional $\Psi[g]$.  Assuming our proposed generalization of the holographic principle (the GHP defined in section \ref{sectionGHP}) is valid, we showed in section \ref{dictionary} that this {\sf bulk$\to$bulk} map is equivalent to the usual gravitational path integral.  (In other words, the CFT inner product is mapped to the dynamical inner product in the bulk.)  Hence, a holographic state automatically encodes local gravitational dynamics in the bulk.  From the perspective of the holographic duality, this is not surprising since local gravitational dynamics is pure gauge, and therefore the boundary CFT state at $\partial \Sigma$ doesn't pick out any specific time slice $\Sigma$.  Rather it encodes them all.

This means that the CFT state encodes the bulk information in a background-independent way, since the bulk Cauchy slice $\Sigma$ is simply to be thought of as a codimension-1 manifold anchored to the Cauchy slice of the holographic theory, and not to be thought of as being embedded into some bulk spacetime \textit{a priori}. (Such an interpretation can only emerge in the semiclassical regime of the bulk theory.)

We also confirmed the duality for nonzero boundary time evolution, by deriving an equality \eqref{cft=adm} between the ADM Hamiltonian and the CFT Hamiltonian operators: 
\begin{equation}
\text{H}_\text{CFT}[\partial {\cal B}]=\text{H}_\text{ADM}[\Sigma].
\end{equation}
We derived this equality in section \ref{admH=cftH} in the large $N$ limit (on the field theory side of the duality), but we expect it can be shown more generally.  We also derived it in section \ref{time} using the assumption that the GHP is valid.

\subsection{Summary of Additional Results}

In order for the maps above to make sense, it is important for both sides of the dictionary (boundary and bulk) to be as well-defined as possible.  The initial sections of this paper were devoted to outlining, as precisely as we could, the procedures for constructing both sides, although there are various points which would benefit from further analysis.

\subsubsection*{Boundary side} 
In section \ref{thegeneraldeformation} we gave a prescription to construct a $T^2$-like deformation operator which gives rise to an arbitrary Hamiltonian constraint equation ${\cal H} = 0$, including arbitrary matter fields in arbitrary dimension.  (We require only that the usual ADM constraint closure conditions hold.)  Our prescription requires the imposition of certain anomaly matching conditions for the starting CFT, which turn out to encode the usual bulk-boundary relations, e.g. relations between the central charges and $G_N$, or the relations between mass $m$ and operator dimensions $\Delta$.  The starting CFT cannot be specified beyond this, at least at this level of analysis. However, the form of the $T^2$ deformation is strongly constrained; for any given Hamiltonian constraint $\cal H$ there appears to be only one possible deformation which gives rise to it.  As an example, we worked out the case of gravity coupled to a scalar field in \ref{example}.

In section \ref{largeN}, we also took some steps towards defining the $T^2$ deformation even away from the $N = \infty$ limit, by a careful definition of the normal-ordering prescription which is intended to preserve locality on sub-AdS scales.  Although formally we wrote down the quantum version of our deformation for $N$ large but finite, we were unable to totally remove a role for a small UV cutoff parameter $\epsilon$, and hence technically our boundary theory is only defined perturbatively in a $1/N$ expansion.  It would be good to explore these quantum renormalization issues further, particularly their role in defining sub-AdS scales, and better understanding the role of the Planck scale.

$T^2$ deformed theories have a slew of unusual properties for the field theory, e.g. lack of reflection-positivity, spontaneous CPT breaking (section \ref{exotic}).  Nevertheless, in section \ref{unitarity}, we explained why the bulk dynamics will nevertheless remain unitary, in several physically important senses.  It would be interesting to explore these effects in concrete Lagrangians that spontaneously break CPT. 

%In section \ref{sectionGHP}, we proposed a general form of the duality between $T^2$-deformed field theories and bulk quantum gravity. 

\subsubsection*{Bulk side}
On the other side of the duality, after defining the Lorentzian signature gravitational path integral in \ref{Lorentzian} we argue that this transition amplitude satisfies some key properties in \ref{prop}, although further remains to be done to establish positivity and finiteness (after smearing) of the gravitational inner product.  In section \ref{HS} we showed how to use the gravitational path integral to construct the dynamical Hilbert space in quantum gravity.  (Since this construction does not use holography in any way, it may be of broader interest to the quantum gravity community.)

Along the way, we also reviewed some of the usual problems and issues with the quantum gravity path integral in sections \ref{sectionGHP} and \ref{QG Inner Product}.  In particular, in sections \ref{contour}-\ref{challenge} we discussed the contour, factorization, and nonrenormalizability problems.  And in section \ref{lapse} we argued for the importance of including histories with both positive and negative lapse $N$ in the transition amplitude.

%Nevertheless, even if these results are only valid near semiclassical saddles, it is important to define the Wheeler-De Witt regime as accurately and precisely as possible in order to be able to talk about a holographic dictionary to gravity at all.

\subsection{Towards a UV completion}\label{UV}

Since it is not clear that the quantum gravity path integral will make sense nonperturbatively, we proposed in section \ref{defofbulktheory} that in the end it is the $T^2$ deformed boundary theory that will ultimately be the definition of quantum gravity.  However, for this to work, the boundary theory must itself be UV completed.

In cases where the $T^2$-deformation is not exactly solvable, it is not clear that the deformation is fully defined in the UV.  This is for the usual field theory reason that an irrelevant term becomes dangerous in the deep infrared.  (Although the pure $T\overline{T}$ deformation in $d = 2$ is exactly solvable, at least in flat spacetime, this is probably an artifact of the fact that $D = 3$ pure gravity has no local degrees of freedom.)

If we think of the irrelevant $T^2$ coupling in a Wilsonian manner, it is natural to hypothesize that, at finite $N$, the IR model may arise from some better-behaved field theory in the deep UV.  One possibility is that it might come from a discrete model, e.g. a literal tensor network.  

Alternatively, it might come from a (nonunitary, CPT invariant) UV fixed point CFT, which is perturbed in the UV by some \emph{relevant} operator.  If this theory flows to a (unitary, large $N$) CFT in the IR, then the model will be defined everywhere along its renormalization group flow trajectory.  From the perspective of somebody observing the RG flow at low energies{\bf---}but not all the way at the IR fixed point!{\bf---}the theory would look like the IR large $N$ CFT, deformed by an \emph{irrelevant} deformation.

If this irrelevant deformation is the $T^2$-deformation, then one would have a nonperturbative definition of this deformation, but now defined by flowing `downstream' from the UV rather than `upstream' from the IR.  One would simply need to check that the $T^2$-description of the theory is valid over a sufficiently large range of distance scales to justify its use as an approximately local bulk description at sub-AdS scales, and one would have a model of quantum gravity.

Although the UV theory must be nonunitary in order to give rise to the $T^2$-deformation, it is critical to assume the UV theory is still CPT invariant.  Without this assumption, there is no good Wilsonian reason to obtain a Hamiltonian constraint in the IR whose parameters are real, corresponding to a unitary bulk evolution.  Note that this property is retained even when CPT is spontaneously broken (as described in section \ref{CPT}), since the Hamiltonian constraint \eqref{ham} is not itself broken, i.e. it takes the same algebraic form in both sectors of the broken phase.  See section \ref{unitarity} for more discussion on this point.

This would be tantamount to defining a nonperturbative quantum gravity model in the bulk, which is valid at arbitrarily short distances (even below the Planck scale!).  Apart from the specification of boundary conditions at the spatial boundary (which need not be asymptotically AdS) this holographic quantum gravity model would automatically be \emph{background free}, because the background sources of the local field theory (like $g_{ab}$) are reinterpreted as \emph{fields} on the quantum gravity side.  For this reason, the bulk quantum gravity theory would necessarily have no undetermined global coupling constants, apart from dynamically evolving fields; a property already familiar in string theory.  Taken to the extreme, this suggests that there is only a single unique theory of holographic quantum gravity.\footnote{If this perspective is correct, then even if the original CFT were holographically dual to some superficially non-stringy type of quantum gravity, it should probably be considered a different vacuum sector of the same overarching unified theory.  For whatever differences it has from string theory, might themselves be regarded as a mere difference of state! In this picture, ``Quantum gravity'' would then just be the state space of all possible CPT-invariant (but non-unitary) partition functions, perhaps satisfying some additional axioms yet to be determined.  The dynamical inner product would simply be the natural one coming from sewing open partition functions to each other, but a probabilistic interpretation (using the Born rule) would be viable only in regions bounded by a ``holographic screen'', which is a codimension-2 surface satisfying whatever constraints are necessary for this inner product to be positive \cite{Wall_CEB,Wall:2021bxi}.}

If the UV is indeed described by a scale-invariant fixed point, the dual gravity model would presumably also be scale-invariant at short distances.  So this would presumably be a holographic dual of the Asymptotic Safety Scenario \cite{Niedermaier:2006ns}.  Even more attractively, one could suppose that the UV model might be constructed to be a \emph{trivial} theory with $Z = 1$ (prior to turning on the relevant coupling).  This is conceivable since the theory is nonunitary; an interesting example of such a theory is Yang-Mills with a $U(N|N)$ supergroup \cite{Vafa:2001qf}, in which the contributions of the fermionic gluons exactly cancel the contributions from the bosonic gluons.\footnote{This model should not be confused with the usual type of local supersymmetry, because the fermionic symmetries of this model are scalars, not spinors.  As a result this theory violates spin-statistics and so is nonunitary, but we already know from section \ref{exotic} that the $T^2$ theory is nonunitary, so we are looking for a nonunitary UV completion.} Since the fields at short distances have no physical effects, this could be loosely called a ``RG flow from Nothing''.  This would presumably be a holographic dual to the Induced Gravity Scenario \cite{Visser:2002ew}.

An interesting hypothetical scenario to consider is what we should think if we discover two \emph{different} UV completions of the $T^2$ Cauchy slice theory at the Planck scale (with the \emph{same} starting CFT) which differ from each other only by highly irrelevant terms in the IR.  How would we decide between them? In fact, we wouldn't need to.  So long as these two UV models flow to numerically the same partition function $Z[g]$ at large distance scales, the two models would actually give rise to the same boundary states, and thus should be regarded as dual descriptions of each other!\footnote{Of course, the $T^2$ theory also has the same IR behavior as the original CFT, so there is a genuine sense in which the $T^2$ model is \emph{also} dual to the original CFT.  However, in that case, the ``duality'' between the two models involves a nontrivial flow of time evolution into the bulk, which means that the implementation of the duality is nonlocal and highly nontrivial (since it is equivalent to the bulk dynamics).  An approximately local bulk duality would presumably require, at minimum, that the difference between the two regulators be more irrelevant than the $T^2$ deformation itself.}  

The usual intuition from particle physics models{\bf---}that knowledge of the IR physics is not sufficient to know what physics is like in the deep UV{\bf---}is not applicable here, because we are discussing an intrinsically holographic description in which all information flows to the conformal boundary of AdS.  However, it is important to note that the IR theories must agree at the level of the detailed IR statistical microstates, not merely at the level of a coarse-grained thermodynamic description.  This detailed information might be very difficult for a low energy bulk observer to obtain.  (For example, there are probably lots of distinct CFTs which are dual to pure gravity at low energies, all of which would correspond to distinct quantum gravity backgrounds in the bulk.)

That being said, the invariant information in the partition function is characterized by more information than just the spectrum of light local operators.  In particular, the existence and properties of (suitably smeared) spatially large Wilson loop observables should be insensitive to the method used to regulate the $T^2$ deformation at short distances.  (Such Wilson loop observables should appear in any large $N$ gauge theory when the gauge field is a 1-form.)  Since in AdS/CFT, Wilson loop observables are dual to string fields in the bulk \cite{Maldacena:1998im,Rey:1998ik,Drukker:1999zq}, we expect that whether or not a particular bulk quantum gravity vacuum contains \emph{stringy} excitations is objectively determinable from the Cauchy slice partition function (or indeed from the starting CFT) irrespective of the precise UV completion of the $T^2$ flow.  Similarly, higher-dimensional surface operators \cite{Gaiotto:2014kfa} in the boundary theory should be associated with higher dimensional membranes in the bulk.  So to be clear, our statement that the UV completion doesn't matter refers specifically to the irrelevant boundary $T^2$ deformation---we are not claiming that it is possible to eliminate the phenomena of string theory/M-theory from the \emph{bulk} side of the AdS/CFT duality.\footnote{It would be interesting, but likely extremely difficult, to extend the deformation flow defined in section \ref{Hdual} to the bulk Hamiltonian constraint of string field theory.  Finding a UV completion of the $T^2$ deformation would sidestep this problem.}

\subsection{Holographic Cosmology}

In this paper we have mostly restricted our attention to Cauchy slices that go to an asymptotically AdS boundary (or in some cases, a very slightly $T^2$-regulated timelike boundary).  However, it is possible to extend the Cauchy slice holography formalism to other types of spacetimes, e.g. cosmological spacetimes.  In \cite{Araujo-Regado:2022jpj} this philosophy was applied to the case of closed slices in asymptotically dS cosmologies.  (A very brief discussion of the dS/CFT case was mentioned in section \ref{example}.)

If this idea makes sense in cosmological settings, the field theory partition function $Z$ would have to represent the amplitude to arise from some specific initial condition!  This means we would obtain a unification of the dynamical laws, with the initial conditions of the universe!  

Indeed, if the Cauchy slice $\Sigma$ has no boundary, our dictionary seems to suggest that there is only a \emph{single} quantum gravity state that can be encoded holographically via a $T^2$-deformed field theory. This follows from the fact that in this case the field theory Hilbert space is trivial (i.e.~1-dimensional), since $\partial\Sigma=\emptyset$. The absence of a boundary also implies the lack of a physical dynamical evolution, which is also consistent with a trivial Hilbert space. This suggests additional philosophical puzzles in the case of a holographic description of the quantum gravity of a closed universe.  The idea that it does not make sense to have a Hilbert space for the entire universe and that states must somehow ``relational'' was discussed in  \cite{Baez:1992uf,Crane:1995qj,Smolin:1995vq,Rovelli:1995fv}.

In this respect, the proposal is similar to the Hartle-Hawking \cite{Hartle:1983ai, Hawking:1983hj} or Vilenkin \cite{Vilenkin:1982de,Vilenkin:1984wp,Vilenkin:1986cy} ``no-boundary'' proposals, in which the gravitational path integral is closed on one end, so as to define a unique global no-boundary state $\Psi_\text{NB}$.  We would then seek a relation of the form:
%The statement of the duality would then be
\begin{equation}\label{1}
\Psi_\text{NB}[g] = Z[g].
\end{equation}
where $Z$ is the partition function of some field theory.\footnote{As an early example of such a proposed state, the 3d Chern-Simons partition function was used to construct the Kodama solution \cite{Kodama:1990sc} to the Ashtekar form of the 3+1 dimensional constraint equations.  This construction is analogous to Cauchy slice holography in that it relies on a $\Psi = Z$ relation like \eqref{Psi=Z}.  This Kodama state was, however, shown to be physically unrealistic by Witten \cite{Witten:2003mb} due to one helicity of graviton having negative energy.  For modifications of the Kodama construction to address this concern, see \cite{Randono:2006rt,Randono:2006ru,Magueijo:2020qcj}.}

Similar holographic cosmology models were proposed in \cite{Strominger:2001pn, Anninos:2011ui,McFadden:2009fg,McFadden:2010na,Maldacena:2011nz,Hertog:2011ky,Hartle:2012tv,Hawking:2017wrd}.\footnote{For a somewhat different construction involving branes to get a re-collapsing cosmology with $\Lambda < 0$, see \cite{Antonini:2022blk}.}  However, in most of these proposals, the holographic theory is defined to live on the geometry of future conformal infinity $\cal{I}+$ (or the \emph{effective} $\cal{I}+$ at the end of inflation), and one learns about earlier times via the usual (single-trace) holographic RG.  Our approach would instead suggest turning on a $T^2$ deformation of the theory, in order to get a holographic model of cosmology on a finite-time Cauchy slice.\footnote{Another approach to obtaining a holographic cosmology using $T^2$ deformation is dS/dS holography \cite{gorbenko2019ds}.  Here the holographic boundary is timelike, so it is more similar to radial $T^2$-deformations.}  If this model can be UV completed, then the UV model may be a description which does not input any particular assumptions about the late-time fate of the Universe, or the spatial boundary conditions (if any).  This would be a significant step towards constructing a background-independent formulation of the holographic principle.

\vspace{4pt}
{\centering
\noindent\rule{8cm}{0.8pt}
\par}
%\newpage
%\medskip
\vspace{-6pt}

{\paragraph{Acknowledgements:} \small This work was supported in part by AFOSR grant FA9550-19-1-0260 “Tensor Networks and Holographic Spacetime”, STFC
grant ST/P000681/1 “Particles, Fields and Extended Objects”, and an Isaac Newton Trust Early Career grant.  R.K. is also supported by a Trinity Henry Barlow scholarship, and G.A.-R. is also supported by a Harding Distinguished
Postgraduate Scholarship.  We are grateful for helpful conversations with Amr Ahmadain, Raphael Bousso, Jeremy Butterfield, Steve Carlip, William Donnelly, Laurent Freidel, Tom Hartman, Ted Jacobson, Chethan Krishnan, Jorrit Kruthoff, Juan Maldacena, Don Marolf, Filipe Miguel, Prahar Mitra, Masamichi Miyaji, Jose Sa, Arvind Shekar, Eva Silverstein, Vasudev Shyam, Ronak Soni, Brian Swingle, Tadashi Takayanagi, Bilyana Tomova, Joao Melo, Herman Verlinde, Manus Visser, Houwen Wu and Balt van Rees.       
}

\begin{appendices}
\section{ADM Hamiltonian on Cauchy slices}

The ADM boundary Hamiltonian is given by a integral over $\partial\Sigma$ of the time component of the boundary conjugate momentum:
\begin{equation}
    \text{H}_\text{ADM}=\int d\Omega\, \Pi^{\tau\tau(+)},
\end{equation}
where the usual boundary lapse factor is 1 in our coordinate system \cite{Hayward:1992ix}.

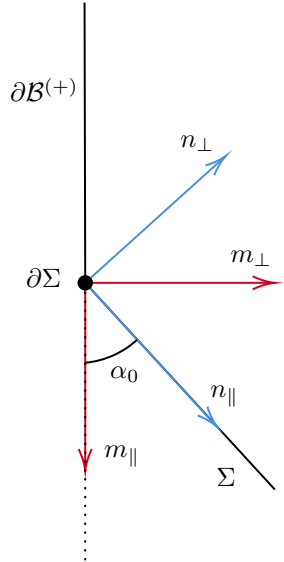
\begin{figure}[h]
\centering
\tikzset{every picture/.style={line width=0.75pt}} %set default line width to 0.75pt        

\begin{tikzpicture}[x=0.75pt,y=0.75pt,yscale=-1,xscale=1]
%uncomment if require: \path (0,452); %set diagram left start at 0, and has height of 452

%Straight Lines [id:da6210932552273652] 
\draw    (271,230) -- (283.24,243.48) -- (365.68,334.02) ;
%Straight Lines [id:da07108976679652246] 
\draw [color={rgb, 255:red, 208; green, 2; blue, 27 }  ,draw opacity=1 ][fill={rgb, 255:red, 208; green, 2; blue, 27 }  ,fill opacity=1 ]   (271,230) -- (363.54,230) ;
\draw [shift={(365.54,230)}, rotate = 180] [color={rgb, 255:red, 208; green, 2; blue, 27 }  ,draw opacity=1 ][line width=0.75]    (10.93,-3.29) .. controls (6.95,-1.4) and (3.31,-0.3) .. (0,0) .. controls (3.31,0.3) and (6.95,1.4) .. (10.93,3.29)   ;
%Straight Lines [id:da2952631766219159] 
\draw [color={rgb, 255:red, 208; green, 2; blue, 27 }  ,draw opacity=1 ][fill={rgb, 255:red, 208; green, 2; blue, 27 }  ,fill opacity=1 ]   (271,230) -- (271,322.54) ;
\draw [shift={(271,324.54)}, rotate = 270] [color={rgb, 255:red, 208; green, 2; blue, 27 }  ,draw opacity=1 ][line width=0.75]    (10.93,-3.29) .. controls (6.95,-1.4) and (3.31,-0.3) .. (0,0) .. controls (3.31,0.3) and (6.95,1.4) .. (10.93,3.29)   ;
%Straight Lines [id:da9119165460763199] 
\draw [color={rgb, 255:red, 74; green, 144; blue, 226 }  ,draw opacity=1 ][fill={rgb, 255:red, 208; green, 2; blue, 27 }  ,fill opacity=1 ]   (271,230) -- (339.78,167.42) ;
\draw [shift={(341.26,166.08)}, rotate = 137.7] [color={rgb, 255:red, 74; green, 144; blue, 226 }  ,draw opacity=1 ][line width=0.75]    (10.93,-3.29) .. controls (6.95,-1.4) and (3.31,-0.3) .. (0,0) .. controls (3.31,0.3) and (6.95,1.4) .. (10.93,3.29)   ;
%Straight Lines [id:da024448294345794208] 
\draw [color={rgb, 255:red, 74; green, 144; blue, 226 }  ,draw opacity=1 ][fill={rgb, 255:red, 208; green, 2; blue, 27 }  ,fill opacity=1 ]   (271,230) -- (335.58,301.1) ;
\draw [shift={(336.92,302.58)}, rotate = 227.75] [color={rgb, 255:red, 74; green, 144; blue, 226 }  ,draw opacity=1 ][line width=0.75]    (10.93,-3.29) .. controls (6.95,-1.4) and (3.31,-0.3) .. (0,0) .. controls (3.31,0.3) and (6.95,1.4) .. (10.93,3.29)   ;
%Straight Lines [id:da5677305850820533] 
\draw    (270.66,89.03) -- (271,230) ;
\draw [shift={(271,230)}, rotate = 89.86] [color={rgb, 255:red, 0; green, 0; blue, 0 }  ][fill={rgb, 255:red, 0; green, 0; blue, 0 }  ][line width=0.75]      (0, 0) circle [x radius= 3.35, y radius= 3.35]   ;
%Straight Lines [id:da8679234808806198] 
\draw  [dash pattern={on 0.84pt off 2.51pt}]  (271,230) -- (270.98,371.02) ;
%Shape: Arc [id:dp15039445604801527] 
\draw  [draw opacity=0] (296.9,258.59) .. controls (293.71,261.63) and (289.75,264.33) .. (285.2,266.41) .. controls (280.39,268.6) and (275.48,269.84) .. (270.88,270.18) -- (276.71,247.79) -- cycle ; \draw   (296.9,258.59) .. controls (293.71,261.63) and (289.75,264.33) .. (285.2,266.41) .. controls (280.39,268.6) and (275.48,269.84) .. (270.88,270.18) ;  

% Text Node
\draw (232,125.4) node [anchor=north west][inner sep=0.75pt]    {$\partial \mathcal{B}^{( +)}$};
% Text Node
\draw (240,221.4) node [anchor=north west][inner sep=0.75pt]    {$\partial \Sigma $};
% Text Node
\draw (282,270.4) node [anchor=north west][inner sep=0.75pt]    {$\alpha _{0}$};
% Text Node
\draw (317,154.4) node [anchor=north west][inner sep=0.75pt]    {$n_{\perp }$};
% Text Node
\draw (332,280.4) node [anchor=north west][inner sep=0.75pt]    {$n_{\parallel }$};
% Text Node
\draw (279,310.4) node [anchor=north west][inner sep=0.75pt]    {$m_{\parallel }$};
% Text Node
\draw (342,212.4) node [anchor=north west][inner sep=0.75pt]    {$m_{\perp }$};
% Text Node
\draw (335,321.4) node [anchor=north west][inner sep=0.75pt]    {$\Sigma $};

\end{tikzpicture}
\caption{\small This shows a bulk Cauchy slice $\Sigma$ in the neighbourhood of its junction with the boundary at $\partial\Sigma$. The submanifold $\partial\Sigma$ is a codimension-two surface and so has a two-dimensional space orthogonal to it. Two different bases of this space are shown.}\label{normals_fig}
\end{figure}

Now we express this operator in terms of the extrinsic curvature of the boundary, as embedded in the bulk space:
\begin{equation}
    \Pi^{\tau\tau(+)}=\frac{\sqrt{\text{det}(g)}}{16\pi G_N}\;\frac{1}{2} g^{ij}\left(\mathcal{L}_{m_\bot}g_{ij}\right),
\end{equation}
where $\mathcal{L}_{m_\bot}g_{ij}$ is the Lie derivative of the metric induced on $\partial\Sigma$, $g_{ij}$, in the direction of the normal to $\partial\mathcal{B}$, as shown in Figure \ref{normals_fig}. $\text{det(g)}$ is the determinant of the boundary metric evaluated at $\partial\Sigma$.

We now consider the vector $V$ living on the tangent space at $\partial\Sigma$ defined via:
\begin{equation}
    V=\left(\Pi^{\tau\tau(+)}\right) m_\bot + (0) m_\parallel,
\end{equation}
where the set $\{m_\bot,m_\parallel\}$ is an orthonormal basis of the two-dimensional space orthogonal to $\partial\Sigma$. $m_\bot$ is the normal to $\partial\mathcal{B}$ and $m_\parallel$ is the normal to $\partial\Sigma$ as embedded in $\partial\mathcal{B}$. We see that the norm of this vector is $|V|=\Pi^{\tau\tau(+)}$. If we express the same vector in a new orthonormal basis $\{n_\bot,n_\parallel\}$ it will instead have components given by:
\begin{eqnarray}
    V&=&\frac{\sqrt{\text{det}(g)}}{16\pi G_N}\Bigg\{\left(\cos^2{\alpha_0}\frac{1}{2} g^{ij}\left(\mathcal{L}_{n_\bot}g_{ij}\right) +\sin{\alpha_0}\cos{\alpha_0}\frac{1}{2} g^{ij}\left(\mathcal{L}_{n_\parallel}g_{ij}\right)\right)n_\bot\\ &+&\left(\sin^2\alpha_0 \frac{1}{2} g^{ij}\left(\mathcal{L}_{n_\parallel}g_{ij}\right) +\sin\alpha_0\cos\alpha_0\frac{1}{2} g^{ij}\left(\mathcal{L}_{n_\bot}g_{ij}\right)\right)n_\parallel\Bigg\},
\end{eqnarray}
where the new basis is also shown in Figure \ref{normals_fig}. $n_\bot$ is the normal to $\Sigma$ and $n_\parallel$ is the normal to $\partial\Sigma$ as embedded in $\Sigma$. Relating the Lie derivatives to the conjugate momentum on $\Sigma$ and to the extrinsic curvature of $\partial\Sigma$ we get:
\begin{eqnarray}
    V&=&\left(\cos^2{\alpha_0}\;\Pi^{\tau\tau(-)} +\sin{\alpha_0}\cos{\alpha_0}\;\frac{\sqrt{\text{det}(g)}}{16\pi G_N}K^{(-)}\right)n_\bot\\ &+&\left(\sin^2\alpha_0\; \frac{\sqrt{\text{det}(g)}}{16\pi G_N}K^{(-)} +\sin\alpha_0\cos\alpha_0\;\Pi^{\tau\tau(-)}\right)n_\parallel,
\end{eqnarray}
where the $(\pm)$ convention is the same as in the main text. The norm, as expressed in the components of the new basis, is:
\begin{equation}
    |V|=\Pi^{\tau\tau(+)}=\cos\alpha_0\;\Pi^{\tau\tau(-)}+\sin\alpha_0\;\frac{\sqrt{\text{det}(g)}}{16\pi G_N}K^{(-)}.
\end{equation}

Therefore, the ADM Hamiltonian can equally be written as:
\begin{equation}\label{gravityadm}
    \text{H}_\text{ADM}=\int d\Omega^{d-1} \left( \cos\alpha_0\;\Pi^{\tau\tau(-)}+\sin\alpha_0\;\frac{\sqrt{\text{det}(g)}}{16\pi G_N}K^{(-)} \right).
\end{equation}
This relation is true in the coordinate system we have picked and with the choice of the boundary metric, which makes $K^{(+)}=0$, but the bulk metric is arbitrary (other than satisfying the Dirichlet boundary condition).

Equation \eqref{gravityadm} is just a re-expression of the ADM Hamiltonian only using gravity computations. Although this looks identical to equation \eqref{junction}, the difference is that \eqref{junction} is a quantum field theory result (to be interpreted as operator relation in the field theory) while equation \eqref{gravityadm} is a gravity result. Upon canonical quantization, equation $\eqref{gravityadm}$ tells us how to write the quantum operator $\text{H}_\text{ADM}$ in terms of the phase-space variables of quantum gravity $(g_{ab}, \Pi^{ab})$ on the Cauchy slice $\Sigma$. In particular, the RHS involves functional derivatives with respect to the metric on the Cauchy slice $\Sigma$: $\Pi^{\tau\tau(-)} = -i\frac{\delta}{\delta g_{\tau\tau}}$ (where $\tau$ runs along $\Sigma$ radially) and the extrinsic curvature of $\partial\Sigma$ as embedded in $\Sigma$, which are operators that can act on the WDW states $\Psi[g]$.  

%\begin{figure}[h]
%\centering
%\includegraphics[width=.4\textwidth]{normals.pdf}
%\caption{\small This shows a bulk Cauchy slice $\Sigma$ in the neighbourhood of its junction with the boundary at $\partial\Sigma$. The submanifold $\partial\Sigma$ is a codimension-two surface and so has a two-dimensional space orthogonal to it. Two different bases of this space are shown.}\label{normals_fig}
%\end{figure}

\end{appendices}

\vspace{1cm}
\color{black}
\noindent\rule[0.25\baselineskip]{\textwidth}{1pt}

\bibliographystyle{ieeetr}
\bibliography{references}

\end{document}